\documentclass{article}
\pdfoutput=1
\usepackage{jheppub}

\usepackage{braket}
\usepackage[table]{xcolor}
\usepackage{mathrsfs}
\usepackage[english]{babel}

\def\nn{\nonumber}

\def\pd{\partial}
\def\cA{{\cal A}}
\def\cB{{\cal B}}
\def\cQ{{\cal Q}}
\def\cD{{\cal D}}
\def\cC{{\cal C}}
\def\cE{{\cal E}}

\def\cI{{\cal I}}
\def\cH{{\cal H}}
\def\cJ{{\cal J}}
\def\cL{{\cal L}}
\def\cM{{\cal M}}
\def\cO{{\cal O}}
\def\cN{{\cal N}}
\def\cG{{\cal G}}

\def\cR{{\cal R}}
\def\cT{{\cal T}}
\def\cU{{\cal U}}
\def\cV{{\cal V}}

\def\cP{{\cal P}}
\def\cS{{\cal S}}

\def\bfp{{{\bf p}}}
\def\bfk{{{\bf k}}}
\def\bfq{{{\bf q}}}
\def\bfl{{{\bf l}}}
\def\bfr{{{\bf r}}}
\def\bfx{{{\bf x}}}
\def\bfy{{{\bf y}}}

\def\bfz{{{\bf z}}}

\def\mfa{{\mathfrak{a}}}
\def\mfb{{\mathfrak{b}}}

\def\mfm{{\mathfrak{m}}}
\def\mfp{{\mathfrak{p}}}

\def\mfH{{\mathfrak{H}}}
\def\mfI{{\mathfrak{I}}}

\def\exd{{\hbox{d}}}

\def\bea{\begin{eqnarray}}
\def\eea{\end{eqnarray}}
\def\be{\begin{equation}}
\def\ee{\end{equation}}

\def\ssA{{\scriptscriptstyle A}}

\def\ssH{{\scriptscriptstyle H}}
\def\ssI{{\scriptscriptstyle I}}

\def\ssR{{\scriptscriptstyle R}}
\def\ssS{{\scriptscriptstyle S}}

\def\ssU{{\scriptscriptstyle U}}
\def\ssV{{\scriptscriptstyle V}}

\def\MF{{\scriptscriptstyle MF}}

\def\pref#1{(\ref{#1})}
\newcommand{\roughly}[1]{\mathrel{\raise.3ex\hbox{$#1$\kern-0.85em
\lower1ex\hbox{$\sim$}}}}
\newcommand{\lsim}{\roughly<}

\def\csch{{\rm csch}}

\def\la{{\langle}}
\def\ra{{\rangle}}
\def\lla{{\langle\hspace{-0.6mm}\langle}}
\def\rra{{\rangle\hspace{-0.6mm}\rangle}}

\def\avg#1{{ \lla \, #1 \, \rra}}

\def\ol#1{\overline{#1}}

\def\sb{ {\scriptstyle(\cdot)} }
\def\abare{\lambda}
\def\Tr{\mathrm{Tr}}

\def\TrA{{\mathrm{Tr}}_+}
\def\TrB{{\mathrm{Tr}}_-}
\def\TrAB{{\mathrm{Tr}}}
\def\vac{\mathrm{vac}}
\def\Vm{{\ol {V}}}

\def\d{\mathrm{d}}

\def\smath#1{\text{\scalebox{.85}{$#1$}}}
\def\sfrac#1#2{\smath{\frac{#1}{#2}}}

\numberwithin{equation}{section}

\def\vac{\mathrm{vac}}

\def\smath#1{\text{\scalebox{.85}{$#1$}}}
\def\sfrac#1#2{\smath{\frac{#1}{#2}}}

\numberwithin{equation}{section}
\date{June 2021}

\author[a,b]{C.P. Burgess,}

\author[c]{R. Holman}

\author[a,b]{and G. Kaplanek}

\affiliation[a]{Department of Physics \& Astronomy, McMaster University, 1280 Main Street West, Hamilton ON, Canada.}
\affiliation[b]{Perimeter Institute for Theoretical Physics, 31 Caroline Street North, Waterloo ON, Canada.}
\affiliation[c]{Minerva Schools at KGI,
1145 Market Street, San Francisco, CA 94103, USA.}

\title{Quantum Hotspots: Mean Fields, Open EFTs, Nonlocality and Decoherence Near Black Holes}
 
\date{}
 
\abstract{ Effective theories describing black hole exteriors resemble open quantum systems inasmuch as many unmeasurable degrees of freedom beyond the horizon interact with those we can see. A solvable Caldeira-Leggett type model of a quantum field that mixes with many unmeasured thermal degrees of freedom on a shared surface was proposed in {\tt arXiv:2106.09854} to provide a benchmark against which more complete black hole calculations might be compared. We here use this model to test two types of field-theoretic approximation schemes that also lend themselves to describing  black hole behaviour: Open EFT techniques (as applied to the fields themselves, rather than Unruh-DeWitt detectors) and mean-field methods. Mean-field methods are of interest because the effective Hamiltonians to which they lead can be nonlocal; a possible source for the nonlocality that is sometimes entertained as being possible for black holes in the near-horizon regime. Open EFTs compute the evolution of the field state, allowing discussion of thermalization and decoherence even when these occur at such late times that perturbative methods fail (as they often do). Applying both of these methods to a solvable system identifies their domains of validity and shows how their predictions relate to more garden-variety perturbative tools.}

\begin{document}
\maketitle

\section{Introduction and Discussion}
\label{sec:intro}

Quantum fields in black-hole backgrounds have long been known to behave in surprising ways \cite{Bekenstein:1971hc, Hawking:1974rv}, even at energies well below whatever new physics ultimately describes gravity at its most foundational level. Although the nature of the approximations being made when studying these effects was initially puzzling, this quantum-gravitational interplay has been integrated into the broader framework of theoretical physics within the formalism of effective field theories (EFTs) \cite{Weinberg:1978kz, Donoghue:1994dn} (for reviews see \cite{, Burgess:2003jk, Goldberger:2007hy, Porto:2016pyg, Donoghue:2017ovt, EFTBook}). 

In the meantime black hole physics grew up, with the discovery of gravitational waves \cite{LIGO} confronting EFT calculations \cite{Goldberger:2004jt, Goldberger:2005cd, Porto:2005ac, Kol:2007bc, Kol:2007rx, Gilmore:2008gq, Porto:2008jj, Damour:2009vw, Emparan:2009at, Damour:2009wj, Levi:2015msa} of black hole properties with experimental measurements. This has stimulated much work and has underlined some of the unique challenges posed when working with black holes in an EFT framework. One of these challenges asks how the EFT should handle the large numbers of gapless and dissipative degrees of freedom \cite{Goldberger:2005cd, Goldberger:2019sya} associated with the black hole's entropy. 

It would be useful to compare black-hole calculations with similar ones for well-understood solvable systems that share as many of these features as possible, and to this end ref.~\cite{Hotspot} proposed a solvable Caldeira-Leggett style \cite{FeynmanVernon, CaldeiraLeggett} model in which an `external' massless quantum field $\phi$ interacts with many unseen gapless thermal fields, but only on a surface meant as a poor man's model of the event horizon. Following \cite{Hotspot}, in what follows we call such a hot localized source a `hotspot'. 

In this paper we use this model to explore two types of approximate tools that both lend themselves to black-hole applications and capture different aspects of black-hole exceptionalism. (A companion paper \cite{Kaplanek:2021xxx} computes the response an Unruh-DeWitt (qubit) detector that couples to the external field $\phi$ at a fixed distance from the hotspot.) Applying approximate tools to a solvable model allows explicit identification of their domain of validity, which can be useful for applications to more realistic systems for which a full solution is not known. 

The two approximations explored here are late-time Open-EFT methods\footnote{Refs.~\cite{Kaplanek:2021xxx, Kaplanek:2019dqu, Kaplanek:2019vzj, Kaplanek:2020iay} also explore the use of Open EFT techniques, but do so for the much simpler case where late-time predictions are only sought for an Unruh-DeWitt qubit detector \cite{Unruh:1976db, DeWitt:1980hx}, rather than for the entire $\phi$ field. (See \cite{Decoherence1, Decoherence2, Decoherence3, Decoherence4, Decoherence5, Decoherence6, Decoherence7, Decoherence8, Decoherence9, Burgess:2014eoa, Martin:2018zbe} for related discussions of field decoherence in cosmology.)} and mean-field expansions. Ref.~\cite{Hotspot} solves the system dynamics exactly, but does so within the Heisenberg picture and so obscures how the $\phi$ field state evolves once couplings to the hotspot are switched on. Open-EFT methods are designed to extract this state evolution, in principle allowing access to questions such as whether (and how quickly) the hotspot decoheres the external $\phi$ field. Furthermore it does so with a domain of validity that allows it to treat phenomena (like thermalization) that occur at times sufficiently late that naive perturbative methods generically fail.

Mean-field methods provide a framework within which an effective Hamiltonian description is possible even while including open-system effects (see for example \cite{EFTBook} for practical examples treated in the same framework used here). Such a Hamiltonian description need not be guaranteed for generic systems. Furthermore, the effective mean-field Hamiltonian is often nonlocal and/or non-Hermitian, making it natural to ask whether mean-field methods might provide relatively mundane origins for exotic non-Wilsonian behaviour in the vicinity of black hole horizons; exotic behaviour that is often speculated to exist near black holes \cite{Hawking:1976ra, Giddings:2006sj, Skenderis:2008qn, Almheiri:2012rt, Almheiri:2013hfa, Banks:1994ph, Mathur:2009hf}.

Our arguments and results are laid out as follows.  \S\ref{sec:setup} starts by briefly reviewing the hotspot model given in \cite{Hotspot} and summarizes the results computed there that are relevant for later comparisons. The model's main variables are a massless field, $\phi$, (meant to represent observable degrees of freedom on the near side of the horizon) plus a thermal bath (also with massless fields, $\chi^a$) meant to model the dissipative effects of beyond-the-horizon physics. These fields `interact' locally by mixing only on a surface meant to represent the horizon itself (though the interaction surface is not an actual geometrical local horizon).  

Once the model is set up \S\ref{sec:NZ} formulates the evolution of the external field $\phi$ after tracing out the thermal degrees of freedom. This section does so by deriving eq.~\pref{NZbeforeMarkov}; a Nakajima-Zwanzig evolution equation \cite{Nak, Zwan} -- a common open-systems tool -- for the reduced density matrix of the field ${\phi}$. We find a limit in which this equation takes a Markovian form -- \pref{SchPicMarkovian} -- and we solve for the field's density matrix in this Markovian limit, doing so perturbatively in the system couplings and working in the field basis where the solution is a Gaussian state. Using this result we compute the equal-time correlations $\langle {\phi}(t,\bfx) {\phi}(t,\bfy) \rangle$ -- with the result \pref{eqtimecorrMarkovianLT} -- and identify the domain of validity of Markovian methods by comparing this to the correlator given in \S\ref{sec:setup}. 

\S\ref{sec:NZ} closes by computing a measure of the $\phi$ state's purity after the $\chi^a$ fields are integrated out, and computes the decoherence rate as a function of the $\phi$-$\chi$ coupling parameters and hotspot temperature. For each mode the rate of departure from an initially pure free-vacuum state is controlled by the time scale $\tilde g^2/\beta$ where $\beta$ is the inverse hotspot temperature and $\tilde g$ is a measure of the $\phi$-$\chi$ couplings. 

Finally, \S\ref{sec:meanfieldH} defines what a mean-field Hamiltonian is, and what this definition means -- {\it c.f.} eq.~\pref{MFHeffHS} -- for the field $\phi$ in the hotspot model. The result is in general nonlocal on the interaction surface  and in time (with characteristic nonlocality scales given by the surface radius, $\xi$, and the inverse temperature, $\beta$), but is typically local in the radial `off-horizon' direction. This mean-field Hamiltonian is used to compute once again the system correlation function $\langle {\phi}(t,\bfx) {\phi}(s,\bfy) \rangle$ for comparison with the results of \S\ref{sec:setup}, showing that validity of mean-field methods requires some couplings (like the $\phi$ self-coupling $\lambda$) to dominate the temperature-dependent combination $\tilde g^2/\beta$.

Taken together, the respective approximate methods described in \S\ref{sec:NZ} and \S\ref{sec:meanfieldH} hold in complementary regimes of parameter space, both of which are subsets of the broader domain of perturbative methods found in \cite{Hotspot}.

\section{The hotspot reloaded}
\label{sec:setup}

\begin{figure}[h]
\begin{center}
\includegraphics[width=60mm,height=60mm]{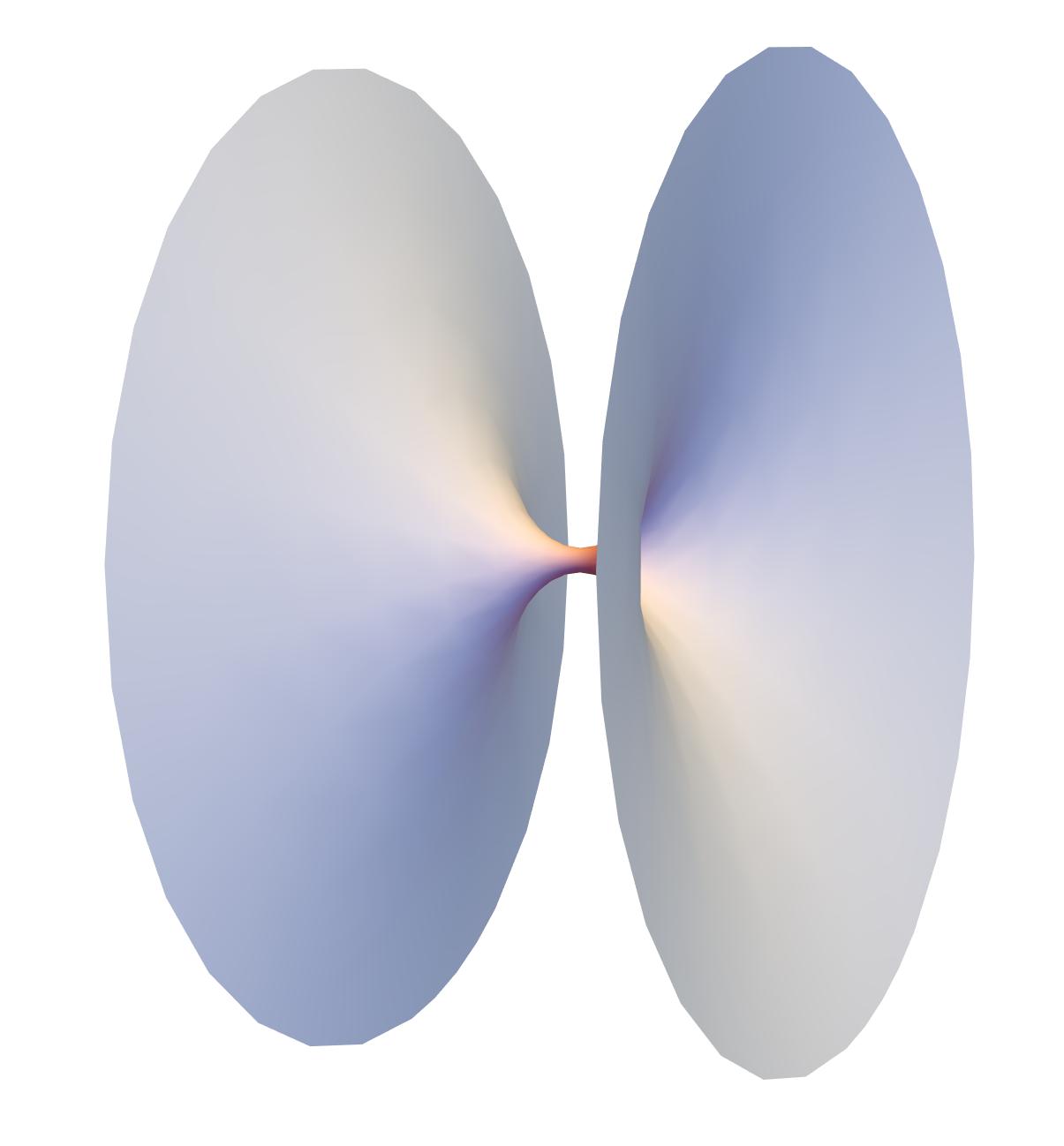} 
\caption{The two spatial branches, $\cR_+$ and $\cR_-$, in which the field $\phi$ and the $N$ fields $\chi^a$ repsectively live. In practice the two regions are idealized as flat (though curvature can in principle also be included) with the spherical interaction region $\cS_\xi $ identified. The two types of fields only couple to one another within $\cS_\xi $ (which effectively becomes the world-line of a point in the special case that $\xi $ is much smaller than all other scales of interest). Figure taken from \cite{Hotspot}.} \label{fig:FunnelFig} 
\end{center}
\end{figure}

This section briefly summarizes the hotspot-model setup as given in \cite{Hotspot}. The fields involved consist of an observable sector --- a single real massless scalar field, $\phi$ --- and an unmeasured disspative and gapless environment --- $N$ real massless scalar fields, $\chi^{a}$ prepared in a thermal state. These two systems live in different spatial regions ($\cR_\pm$) that only intersect in a small localized domain: a sphere $\cS_\xi $ of radius $\xi $ enclosing the origin that is identified for the two spaces (see Fig.~\ref{fig:FunnelFig}). We choose here not to follow the gravitational back-reaction of these fields, and so treat $\cR_\pm$ as independent flat spatial slices. A surface $\cS_{\xi\pm}$ encircles the origin within each of these spaces, with $\cS_{\xi\pm}$ identified to obtain the interaction surface $\cS_\xi$. The $\chi^a$ fields are meant to represent unmeasured degrees of freedom internal to the black hole with which external fields can interact.

Interactions on $\cS_\xi$ are taken to be bilinear mixings of the fields, of the form $\cL_{\rm int} = - g_a \, \chi^a \, \phi$, plus possible quadratic self-interactions of the fields as required by any renormalization-group flows.\footnote{Such flows arise due to renormalizations of the Coulomb-like divergences that appear even at the classical level near the interaction surface \cite{Goldberger:2001tn, deRham:2007mcp, Burgess:2008yx, Bayntun:2009im, PPEFT}.} With black holes in mind we consider observables that depend only on the $\phi$ field and do not directly measure any of the $\chi^a$'s. 

The action for the model is 
\be \label{Naction}
S_{\ssU\ssV} = - \frac{1}{2} \int_{\cR_+^t} \exd^{4}x \; \partial_{\mu} \phi \, \partial^{\mu} \phi  -\frac12   \int_{\cR_-^t} \exd^4x\; \delta_{ab} \, \partial_{\mu} \chi^{a} \partial^{\mu} \chi^{b} -     \int_{\cS_\xi^t} \exd^3 x \; \left[  G_a\,  \chi^{a}  \phi + \frac{G_\phi}2 \, \phi^2 \right] \,,
\ee
where $\cR_\pm^t$ denote spacetime regions whose spatial slices are $\cR_\pm$, and the integration for the interaction is over the world-tube, $\cS_\xi^t$ swept out by the surface $\cS_\xi$ over time. The couplings $G_a$ and $G_\phi$ have dimension mass, and $G_\phi$ is included because it can be required to renormalize divergences that arise due to the presence of the $\phi$-$\chi$ mixing $G_a$. When performing calculations we usually specialize to the case where $G_a = G$ is $a$-independent and define the combination
\be \label{GtildeDef}
  \widetilde G^2  := \delta^{ab} G_a G_b = N G^2 \,.
\ee

As described in \cite{Hotspot} this hotspot model comes in two versions, depending on whether or not the radius $\xi$ is regarded to be an ultraviolet scale.\footnote{The same distinction also arises for black-hole EFTs, for which the black hole event horizon can either be regarded as being shrunk to a point -- as in world-line point-particle EFTs \cite{Goldberger:2004jt, Goldberger:2005cd, Porto:2005ac, Kol:2007bc, Kol:2007rx, Gilmore:2008gq, Porto:2008jj, Damour:2009vw, Emparan:2009at, Damour:2009wj, Levi:2015msa} -- or can be regarded as being macroscopic \cite{Burgess:2018pmm, Rummel:2019ads}. This latter type of EFT can be used to summarize situations where it is ordinary GR \cite{Price:1986yy, Thorne:1986iy, Damour:1978cg, Parikh:1997ma, Donnay:2019jiz} or exotic physics \cite{Cardoso:2016rao, Abedi:2016hgu, Holdom:2016nek, Cardoso:2017cqb, Bueno:2017hyj, Mark:2017dnq, Conklin:2017lwb, Berti:2018vdi, Zhou:2016hsh} whose UV near-horizon physics is being summarized.} When $\xi$ is a UV scale the 2-sphere $\cS_\xi$ degenerates to a point and the above description gets replaced by an effective action organized in powers of $\xi$. The leading $\phi$-dependent interactions in this effective theory become
\be \label{Sint}
S_{\rm int-eff} \simeq   - \int   \exd t \; \left[ g_a \, \chi^{a}(t,\mathbf{0}) \,  \phi(t,\mathbf{0})  + \frac{\lambda}{2} \, \phi^2(t,\mathbf{0}) \right] \,,  
\ee
where the integration is over the proper time of the interaction point. At leading order the effective couplings $\lambda$ and $g_a$ are related to the couplings in \pref{Naction} by 
\be \label{gavsGa}
   g_a = 4\pi \xi^2 G_a \quad \hbox{and} \quad
   \lambda = 4 \pi \xi^2 G_\phi 
\ee
and so have dimensions of length. In this limit it is the quantity 
\be \label{tildegdef}
   \tilde g^2 = \delta^{ab} \, g_a g_b = N g^2
\ee
that plays the same role as did \pref{GtildeDef} when $\xi$ was not small.

\subsection{Time evolution}
\label{subsec:schropic}

The Hamiltonian for this system can be written $H = H_0 + H_{\rm int}$ where ${H}_0 :=  {\cH}_+ \otimes \cI_- + \cI_+ \otimes  {\cH}_-$ is the free Hamiltonian, with $\cH_\pm$ and $\cI_\pm$ the Hamiltonian and identity operators acting separately within the $\phi$- and $\chi$-sectors of the Hilbert space:
\be
 {\cH}_{+}  := \frac12  \int_{\cR_+} \exd^3\bfx \; \Bigl[  {\mfp}^2 + \big( \boldsymbol{\nabla}  {\phi} \big)^2 \Bigr]    \quad\hbox{and}\quad
 {\cH}_{-}    :=   \frac12 \int_{\cR_-} \exd^{3}x \;\Bigl[ \delta^{ab} {\Pi}_a{\Pi}_b + \delta_{ab}   \boldsymbol{\nabla} {\chi}^{a} \cdot   \boldsymbol{\nabla} {\chi}^{b} \Bigr] \,,
\ee
and the canonical momenta are defined by $\mfp  := \partial_t  {\phi}$ and $\Pi_a  := \delta_{ab} \, \partial_t  {\chi}^b$. The interaction Hamiltonian is similarly
\be \label{Hint(t)}
 {H}_{\mathrm{int}}  = \int_{\cS_\xi} \exd^2x \; \left[ G_a \, {\phi}  \otimes  {\chi}^a + \frac{G_\phi}2 \, \phi^2 \otimes \cI_-  \right]\,,
\ee
which in the point-hotspot limit reduces to
\be \label{Hint(t)pt}
 {H}_{\mathrm{int}}(t) \simeq    g_a \, {\phi}(t,\boldsymbol{0})  \otimes  {\chi}^a(t,\boldsymbol{0}) + \frac{\lambda}2 \, \phi^2(t,\boldsymbol{0}) \otimes \cI_-   \,.
\ee
 
Reference \cite{Hotspot} solves explicitly for the time-evolution for this model within the Heisenberg picture, with most of the explicit results given for the point-hotspot limit (for which $\xi \to 0$). The remainder of this section quotes a few of the results found for later comparisons.

\subsubsection{System state}

In the Heisenberg picture states do not evolve with time, and correspond to the initial state of the system in the Schr\"odinger or interaction pictures. Correlation functions in \cite{Hotspot} are computed assuming the $\phi$ and $\chi^a$ fields are initially uncorrelated, with
\be\label{initialstate}
\rho_0 = \rho_+ \otimes \rho_-  \,,
\ee
where the $\phi$ sector is in its standard Minkowski vacuum, $\ket{\vac}$, and the $\chi^a$ sector is in a thermal state, $\varrho_\beta$, with temperature $T = 1/\beta$:
\be \label{initialstate+-}
\rho_+ = \ket{\vac} \bra{\vac}  \quad \hbox{and} \quad
\rho_- = \varrho_{\beta} := \frac{e^{ - \beta \cH_{-} } }{ \TrB[ e^{ - \beta \cH_{-}} ] }  \,.
\ee
Here $\cH_-$ is the $\chi^a$-sector bulk Hamiltonian and the subscript `$-$' on the trace indicates that it is only taken over the $\chi$ sector. 

\subsubsection{Operator evolution}

In Heisenberg picture the entire burden of time evolution falls on the field operators,
which evolve according to
\be  \label{Hpicfield}
 {\phi}_{\ssH}(t,\bfx)   :=   {U}^{-1}(t,0) \big[  {\phi}_{\ssS}(\bfx) \otimes  \cI_- \big]  {U}(t,0) \quad \hbox{and} \quad
 {\chi}^{a}_{\ssH}(t,\bfx)   :=  {U}^{-1}(t,0) \big[ \cI_+ \otimes  {\chi}^{a}_{\ssS}(\bfx) \big]  {U}(t,0) \,,
\ee
where the full time-evolution operator is ${U}(t,t')   =   \cT\exp\left( - i \int_{t'}^t \exd s \; {H} (s) \right)$ and $\cT$ denotes time-ordering. 

In differential form, a generic Heisenberg-picture operator $A_\ssH(t)$ satisfies $\partial_t A_{\ssH}(t)  = - i \Bigl[ A_{\ssH}(t) , H_{\ssH}(t) \Bigr]$, which in particular implies the fields $\phi_\ssH$ and $\chi^a_\ssH$ satisfy the equations of motion that follow from the action \pref{Naction}: $\Box \phi = \Box \chi^a = 0$ for all points exterior to $\cS_\xi$. For points on $\cS_\xi$ the presence of the interaction implies the equations of motion instead impose the boundary condition
\be
      \partial_r \phi \Bigr|_{r \to \xi} =  \Bigl( G_a \chi^a + G_\phi \,\phi \Bigr)_{r \to \xi} \,,
\ee
(and similarly for $\chi^a$). In the $\xi \to 0$ limit of a point source both the bulk equation and boundary condition are efficiently summarized by the equations
\be  \label{heis1}
( - \partial_t^2 + \nabla^2 ) \phi _{\ssH}(t,\bfx) = \delta^{3}(\bfx) \bigg[ \lambda \phi _{\ssH}(t,\mathbf{0}) +  g_a \chi_{\ssH}^a(t,\mathbf{0}) \bigg]   
\ee
and
\be   \label{heis2}
\delta_{ab} ( - \partial_t^2 + \nabla^2 ) \chi^{b}_{\ssH}(t,\bfx) = \delta^{3}(\bfx) \; g_a \phi _{\ssH}(t,\mathbf{0}) \,.
\ee

It is these last equations that are solved explicitly in \cite{Hotspot}, with the result used to compute the time-evolution of the system's correlation functions. These integrations are performed assuming the couplings $G_a$ and $G_\phi$ (or $g_a$ and $\lambda$) turn on suddenly at $t = 0$ and remain time-independent thereafter. These solutions generically diverge as $r \to \xi$ (or $\bfr \to 0$ in the point-hotspot limit) and because of this these equation must be regulated, such as by evaluating the boundary conditions at a small distance $\xi + \epsilon$ from the singular point. The divergences associated with taking $\epsilon \to 0$ are ultimately absorbed into renormalizations of couplings like $G_\phi$ \cite{Goldberger:2001tn, deRham:2007mcp, Burgess:2008yx, Bayntun:2009im, PPEFT}

\subsection{Correlation functions}

The main result of \cite{Hotspot} is the calculation of the correlation functions for the fields, whose results we now quote for future use.

\subsubsection{$\phi$ correlation functions}

The $\phi$-field Wightman function is defined by
\be \label{CorrelationStart}
W_\beta(t,\bfx; t',\bfx') := \Tr\Bigl[ \phi _{\ssH}(t,\bfx) \phi _{\ssH}(t',\bfx') \rho_0 \Bigr]  \,,
\ee
where $\rho_0$ is the state given in \pref{initialstate}, with explicit formulae given in the small-hotspot limit ($\xi \to 0$).

The result computed to leading nontrivial order in $\tilde{g}^2$ and $\lambda$ turns out to be given by
\bea \label{Wpertfull}
&\ & W_\beta(t,\bfx; t',\bfx') \simeq \frac{1}{4\pi^2 \big[ - (t - t' - i \delta)^2 + |\bfx - \bfx'|^2 \big]}\nn \\ 
&\ & \qquad + \frac{\lambda}{16\pi^3} \bigg( \frac{\Theta(t-|\bfx|)}{|\bfx|} \frac{1}{(t-t'-|\bfx| - i \delta)^2 - |\bfx'|^2} + \frac{\Theta(t'-|\bfx'|)}{|\bfx'|} \frac{1}{(t-t'+|\bfx'| - i \delta)^2 - |\bfx|^2} \bigg) \nn \\
& \ & \qquad - \frac{\tilde{g}^2 \Theta(t-|\bfx|) \Theta(t' - |\bfx'|)}{64 \pi^2 \beta^2 |\bfx| |\bfx'| \sinh^2 \left[ \frac{\pi}{\beta}( t - |\bfx| - t' + |\bfx'| - i \delta ) \right]}  \\
&\ & \qquad + \frac{\tilde{g}^2}{32 \pi^4} \bigg( - \frac{\Theta(t-|\bfx|)}{|\bfx|} \frac{t-t'-|\bfx|}{\big[ (t-t'-|\bfx| - i \delta)^2 - |\bfx'|^2 \big]^2} + \frac{\Theta(t'-|\bfx'|)}{|\bfx'|} \frac{t-t'+|\bfx'|}{\big[ (t-t'+|\bfx'| - i \delta)^2 - |\bfx|^2 \big]^2} \bigg) \nn \\
&\ & \qquad + \frac{\tilde{g}^2}{64\pi^4} \bigg( \frac{\delta(t - |\bfx|)}{|\bfx| \big[ - (t' + i \delta)^2 - |\bfx'|^2 \big]} + \frac{\delta(t' - |\bfx'|)}{|\bfx'| \big[ - (t - i \delta)^2 - |\bfx|^2 \big]} \bigg) \qquad \hbox{(perturbative)}\,, \nn
\eea
where the delta functions and step functions describe the transients due to the switch-on of couplings at $t = |\bfx| = 0$. The infinitesimal $\delta \to 0^+$ is taken to zero at the end of the calculation.

Of most interest here is the form for the correlation function inside the future light-cone of this switch-on ({\it i.e.} for $t>|\bfx|$ and $t' > |\bfx'|$), for which this perturbative expression becomes
\bea  \label{pertcorr}
W_\beta(t,\bfx; t',\bfx') & \simeq & \frac{1}{4\pi^2 \big[ - (t - t' - i \delta)^2 + |\bfx - \bfx'|^2 \big]} + \frac{\lambda}{16\pi^3 |\bfx| |\bfx'|} \bigg[ \frac{|\bfx| + |\bfx'|}{(t - t' - i \delta)^2 - (|\bfx + |\bfx'|)^2 } \bigg]\nn \\
& \ & \qquad - \frac{ \tilde{g}^2 }{64 \pi^2 \beta^2 |\bfx| |\bfx'| \sinh^2 \left[ \frac{\pi}{\beta}( t - |\bfx| - t' + |\bfx'| - i \delta ) \right]} \\
&\ & \qquad + \frac{\tilde{g}^2}{32 \pi^4} \bigg( - \frac{1}{|\bfx|} \frac{t-t'-|\bfx|}{\big[ (t-t'-|\bfx| - i \delta)^2 - |\bfx'|^2 \big]^2} + \frac{1}{|\bfx'|} \frac{t-t'+|\bfx'|}{\big[ (t-t'+|\bfx'| - i \delta)^2 - |\bfx|^2 \big]^2} \bigg) \nn\\
&& \qquad \qquad \qquad\qquad \qquad \qquad \qquad \qquad \qquad \qquad  \qquad\quad \hbox{(inside light cone, perturbative)}\nn
\eea

The exact correlator is also computed in \cite{Hotspot} and the result is compared there to this perturbative limit, verifying that the full dependence on $\lambda$ agrees with RG-improved resummations using point-particle EFT boundary-condition based methods \cite{PPEFT}. Because the full result is not needed in what follows it is not repeated here, beyond observing that the perturbative result emerges from the full one once it is expanded in powers of   
\be \label{PertParams}
   \frac{\tilde g^2}{ 16 \pi^2 \epsilon \tau} \left( 1 + \frac{\lambda}{4\pi \epsilon} \right)^{-1} \ll 1 \quad \hbox{and} \quad
     \frac{\tilde g^2}{ 16 \pi^2\epsilon \beta} \left( 1 + \frac{\lambda}{4\pi  \epsilon} \right)^{-1} \ll 1 \,.
\ee

For later comparisons it is useful to focus on the equal-time special case of these formulae, for which $t' = t$. Of particular interest when comparing with other approximation schemes is the equal-time limit of the perturbative result \pref{pertcorr}, which is
\bea  \label{largeNcorrteqs0}
W_\beta(t,\bfx; t,\bfx') & \simeq & \frac{1}{4\pi^2 |\bfx - \bfx'|^2} - \frac{\lambda}{16\pi^3 |\bfx| |\bfx'| ( |\bfx| + |\bfx'| )}  \\
& \ & \qquad \qquad  - \frac{\tilde{g}^2}{64 \pi^2 \beta^2 |\bfx| |\bfx'| \sinh^2 \left[ \frac{\pi}{\beta}( |\bfx| - |\bfx'| ) \right]} + \frac{\tilde{g}^2}{16 \pi^4 ( |\bfx|^2 - |\bfx'|^2 )^2} \ . \nn \\
&& \qquad \qquad \qquad\qquad \qquad \qquad \qquad \qquad \qquad \qquad  \quad \hbox{(inside light cone, perturbative)}\nn
\eea

\subsubsection{$\chi^a$ correlation functions}

Reference \cite{Hotspot} also gives the explicit form for the $\chi^a$ free-field thermal correlator. Evaluated at spacetime points $x = (t, \mathbf{x})$ and $x' = (t', \mathbf{x}')$ the result (at large $N$) is 
\bea \label{thermalcorrelatorab}
\langle \chi^a(t,\bfx) \chi^b(t',\bfx') \rangle_{\beta}  &=&  \frac{\delta^{ab}}{8 \pi \beta |\mathbf{x} - \mathbf{x}'|}\left\{ \coth\left[  \dfrac{\pi}{\beta} \left( t - t' + |\mathbf{x} - \mathbf{x}'| - i \delta \right) \right] \right. \\
&& \qquad\qquad \qquad\qquad \qquad \left. - \coth\left[ \dfrac{\pi}{\beta} \left( t - t' - |\mathbf{x} - \mathbf{x}'| - i \delta \right) \right] \right\} \,, \notag
\eea
in agreement with standard formulae \cite{Thermal}. In this expression the limit $\delta \to 0^{+}$ is to be taken at the end of the calculation. 

\section{Open EFT late-time field evolution}
\label{sec:NZ} 

The solution to the hotspot problem given in \cite{Hotspot} is provided in the Heisenberg picture, and a drawback of this picture is that it obscures how the system's state evolves. This is unfortunate because it makes it difficult to compare with much of the literature on open quantum systems, which is often phrased in terms of the system's reduced density matrix (see for example \cite{EFTBook}).

In this section we aim to make this comparison more transparent, by solving the hotspot problem using the Schr\"odinger-picture (in practice we use the interaction picture once we resort to perturbative methods), computing in particular the reduced density matrix $\sigma_{\ssS}(t)$
\be
\sigma_{\ssS}(t) := \TrB[ \rho_{\ssS}(t) ] \,,
\ee
for the $\phi$-field (defined as the partial trace of the full Schr\"odinger-picture density matrix $\rho_{\ssS}(t)$ over the $\chi^a$ sector).

The cumbersome nature of the Schr\"odinger picture for field theories prevented us from solving exactly for $\sigma_{\ssS}(t)$ even within the hotspot model (though we have no reason to believe that this cannot be done). So we instead compute this evolution perturbatively in the hotspot couplings, also restricting for simplicity to the case of a point-like hotspot ($\xi \to 0$). A major drawback of using perturbation theory, however, is that perturbative methods intrinsically break down at late times, seeming to put beyond reach a reliable calculation of phenomena like decoherence or thermalization. We therefore also adapt Open EFT techniques \cite{EFTBook, Burgess:2014eoa, Agon:2014uxa, Burgess:2015ajz, Braaten:2016sja, Kaplanek:2019dqu, Kaplanek:2019vzj, Kaplanek:2020iay} to verify that they allow the perturbative result to be resummed to extend the perturbative domain of validity to very late times.\footnote{Open EFT techniques were also applied to the hotspot in ref.~\cite{Kaplanek:2021xxx}, but only to obtain the late-time thermalization behaviour of an Unruh-DeWitt detector \cite{Unruh:1976db, DeWitt:1980hx} that sits at rest displaced from the hotspot.} Doing so also allows the testing of these tools in this relatively unfamiliar quantum-field setting.

\subsection{Open EFT evolution}

In principle the time evolution of the system's state is given within the Interaction picture by the Liouville equation,
\be   \label{Liouville}
\frac{\partial \rho}{\partial t} \ = \ - i \big[ H_{\rm int}(t), \rho(t) \big]
\ee
where $H_{\mathrm{int}}(t)$ is given by eq.~\pref{Hint(t)}, and where $\rho(t)$ is the interaction-picture state
\be
\rho(t) = e^{+ i H_0 t} \rho_{\ssS}(t) \, e^{- i H_0 t}
\ee
where $H_0= \cH_{+} \otimes \cI_{-} + \cI_{+} \otimes \cH_{-}$ is the free Hamiltonian for the combined system. It is easy to see from the above that the Schr\"odinger-picture reduced density matrix $\sigma_{\ssS}(t)$ is related to the interaction-picture reduced density matrix $\sigma(t)$ through the relation
\be \label{sigmaSvssigma}
\sigma(t) = e^{+ i \cH_{+} t} \sigma_{\ssS}(t) \,e^{- i \cH_{+} t } \ .
\ee
When solving these equations we assume the uncorrelated initial state $\rho(0) = \rho_0 = \rho_+ \otimes \rho_-$ given in (\ref{initialstate}) where $\rho_+ = \ket{\vac} \bra{\vac}$ and $\rho_{-} = \varrho_{\beta}$ is the thermal configuration for the $\chi$ sector.

\subsubsection{Nakajima-Zwanzig equation}

In principle the evolution of $\sigma(t)$ is given by taking the trace of \pref{Liouville} over all unmeasured degrees of freedom (in this case the fields $\chi^a$). In perturbation theory one usually first formally solves \pref{Liouville} and then takes the trace of the result, leading to
\be
   \sigma(t) = \sigma(0) -i \int_0^t \exd s\; \Tr_- \Bigl[ H_{\rm int}(s) \,, \rho_0 \Bigr] + (-i)^2 \int_0^t \exd s_1 \int_0^{s_1} \exd s_2 \; \Tr_- \Bigl[ H_{\rm int}(s_1) \,, \Bigl[ H_{\rm int}(s_2) \,, \rho_0 \Bigr] \Bigr] + \cdots
\ee
where $\Tr_-$ denotes the partial trace only over the $\chi^a$ sector. The drawback of this expression is the relatively complicated dependence of its right-hand side on the full system's state. Because the right-hand side refers explicitly to the initial state $\rho_0$ successive terms in the series generically grow without bound for large $t$, causing the perturbative approximation to fail at late times and precluding accessing issues like thermalization and late-time decoherence. 

The better route for late-time purposes is to take the trace of the differential relation \pref{Liouville} and to eliminate from this the dependence of the right-hand side on any unmeasured degrees of freedom (for a review of the steps given below see for example \cite{EFTBook}). The good news is that because of the linearity of \pref{Liouville} this can be done in great generality, with the resulting evolution equation for $\sigma(t)$ known as the Nakajima-Zwanzig equation \cite{Nak, Zwan}. Although this is a textbook derivation, we now describe it in some detail since it is not often applied to quantum fields (as we do here) in the relativity literature.
 
The logic of the derivation proceeds as follows. One first defines the super-operator $\mathscr{P}$ acting on operators in the Hilbert space by
\be \label{ProjSupOp}
\mathscr{P}(\cO) \ = \ \TrB\left[ \cO \right] \otimes \rho_- \ , 
\ee
where $\rho_- = \varrho_{\beta}$ is the time-independent initial thermal density matrix for the fields $\chi^a$. This is a projection super-operator because it satisfies $\mathscr{P}^2 = \mathscr{P}$, as therefore must its complement $\mathscr{R} = 1 - \mathscr{P}$. The definition \pref{ProjSupOp} is defined so that it projects the full density matrix $\rho(t)$ onto the reduced density matrix $\sigma(t)$, 
\be  \label{sigmarelation}
\mathscr{P}\big[ \rho(t)\big] =  \sigma(t) \otimes \rho_- \ ,
\ee
and so $\mathscr{R}[\rho(t)]$ can be regarded as describing all of the unmeasured parts of the full density matrix.

Our goal is therefore to rewrite the Liouville equation as a coupled set of evolution equations for the mutually exclusive quantities $\mathscr{P}\big[ \rho(t)\big]$ and $\mathscr{R}\big[ \rho(t)\big]$, and then eliminate $\mathscr{R}\big[ \rho(t)\big]$ from these by solving its equation as a function of $\mathscr{P}\big[ \rho(t)\big]$.  To this end write the interaction-picture Liouville equation \pref{Liouville} in terms of a linear Liouville super-operator,
\be
\partial_t \rho  =  \mathscr{L}_t( \rho  ) \quad \quad \quad \mathrm{where} \quad \quad \quad \mathscr{L}_t( \rho  ) = - i [ H_{\rm int}(t) , \rho \, ] \ ,
\ee
and project the result using the operators $\mathscr{P}$ and $\mathscr{R}$. Since $\mathscr{P} + \mathscr{R} = 1$ this leads to 
\be  \label{Peq}
\mathscr{P}( \partial_{t} \rho  ) = \mathscr{P} \mathscr{L}_{t} \mathscr{P}(\rho  ) + \mathscr{P} \mathscr{L}_{t} \mathscr{R}(\rho  ) 
\ee
and
\be  \label{Req}
  \mathscr{R}( \partial_{t} \rho   ) = \mathscr{R} \mathscr{L}_{t} \mathscr{P}(\rho  ) + \mathscr{R} \mathscr{L}_{t} \mathscr{R}(\rho  ) \,.
\ee

The unmeasured degrees of freedom are eliminated by formally solving eq.~\pref{Req}:
\be 
\mathscr{R}[ \rho (t) ]  =  \cG(t,0) \mathscr{R}[ \rho(0) ] + \int_0^{t} \exd s\ \mathcal{G}(t,s) \mathscr{R} \mathscr{L}_{t} \mathscr{P}[ \rho (s) ]  
\ee
with
\be
  \mathcal{G}(t,s)  =  1 + \sum_{n=1}^{\infty} \int_{s}^{t} \exd s_1 \ \cdots \int_{s}^{s_{n-1}} \exd s_{n}\ \mathscr{R} \mathscr{L}_{s_1} \cdots \mathscr{R} \mathscr{L}_{s_{n}}  \,,
\ee
and inserting the result into \pref{Peq}. This yields the desired self-contained evolution equation for $\mathscr{P}[\rho(t)]$:
\be  \label{NZfull} 
 \mathscr{P}[ \partial _{t} \rho(t) ]  =  \mathscr{P} \mathscr{L}_{t} \mathscr{P}[ \rho (t) ]  + \mathscr{P} \mathscr{L}_{t} \cG(t,0) \mathscr{R}[ \rho(0) ] + \int_0^{t} \exd s\ \mathcal{K}(t,s)[ \rho (s) ] \quad 
\ee
with kernel
\be
  \mathcal{K}(t,s)  =  \mathscr{P} \mathscr{L}_{t} \cG(t,s) \mathscr{R} \mathscr{L}_{s} \mathscr{P} \,.
\ee
For uncorrelated initial states of the form $\rho(0) = \rho_+ \otimes \rho_-$ the second term on the right-hand side of \pref{NZfull} vanishes because $\mathscr{P}\big[ \rho(0) \big] = \rho(0)$ and so $\mathscr{R}\big[ \rho(0) \big] = 0$.

In what follows we wish to use \pref{NZfull} but work only to second order in $H_{\rm int}(t)$, which means expanding out the kernel $\mathcal{K}(t,s)$ to second order in $\mathscr{L}_t$. At this order we can therefore take $\cG(t,s) \simeq 1$ in $\mathcal{K}(t,s)$, which becomes
\be
\mathcal{K}(t,s)  \simeq  \mathscr{P} \mathscr{L}_{t} \mathscr{R} \mathscr{L}_{s} \mathscr{P} \,,
\ee
and so (\ref{NZfull}) simplifies to
\bea \label{NZ2ndOrder}
 \mathscr{P}[ \partial_{t} \rho(t) ]  & \simeq &  \mathscr{P} \mathscr{L}_{t} \mathscr{P}[ \rho(t) ] + \int_0^{t} \exd s\ \mathscr{P} \mathscr{L}_{t} \mathscr{R} \mathscr{L}_{s} \mathscr{P} [ \rho(s) ] \,.
\eea
Writing this out explicitly using the definitions of $\mathscr{P}$, $\mathscr{R}$ and $\mathscr{L}_{t}$ then gives the more explicit form
\bea
\frac{\partial \sigma}{\partial t} & \simeq & - i \; \TrB\Bigl\{  \Bigl[ H_{\rm int}(t) , \sigma(t) \otimes \rho_{-} \Bigr] \Bigr\} \\
&\ & \quad  - \int_0^{t} \exd s\ \TrB\left\{ \bigg[ H_{\rm int}(t) \,,\, \Bigl[ H_{\rm int}(s) \,,\, \sigma(s) \otimes \rho_{-}  \Bigr] - \TrB\Bigl( \Bigl[ H_{\rm int}(s), \sigma(s) \otimes \rho_{-} \Bigr] \Bigr) \otimes \rho_-\bigg] \right\} \,.\nn
\eea

To apply this expression to the hotspot fields expand $H_{\rm int}(t)$ in terms of a basis of operators with the factorized form
\be
   H_{\rm int}(t) = \cA^\ssA(t) \otimes  \cB_\ssA(t)  \, ,
\ee
where $\cA^\ssA$ acts only in the $\phi$ sector and $\cB_\ssA$ acts only in the $\chi^a$ sector. With this choice \pref{NZ2ndOrder} simplifies to become
\bea
&&\frac{\partial \sigma}{\partial t}  \simeq  - i   \; \Bigl[ \cA_a(t) \,,\, \sigma(t) \Bigr] \avg{ \cB^a(t) } \\
&& \qquad\quad + \int_0^{t} \exd s\ \bigg( \Bigl[  \cA^a(s) \, \sigma(s) \,, \cA^b(t) \Bigr] \avg{ \cB_b(t)\cB_a(s) } +  \Bigl[ \cA^b(t) \,, \sigma(s) \cA^a(s) \Bigr] \avg{ \cB_a(s)\, \cB_b(t) } \bigg) \nn \\
&& \qquad\qquad - \int_0^{t} \exd s\ \bigg( \Bigl[ \cA^a(s)\sigma(s) \,, \cA^b(t) \Bigr] + \Bigl[ \cA^b(t) \,, \sigma(s) \cA^a(s) \Bigr] \Bigr] \bigg) \avg{ \cB_b(t) }\avg{\cB_a(s) }  \,. \nn
\eea
where $\avg{ \cO } := \TrB\, [ \rho_- \,\cO ] = \TrB \, [ \varrho_\beta \cO]$ is the thermal trace for operators acting purely in the $\chi$ sector. 

For the point-hotspot system the interaction Hamiltonian given in \pref{Hint(t)pt} has the form $H_{\mathrm{int}}(t) = g_a \phi(t,\mathbf{0}) \otimes \chi^{a}(t,\mathbf{0}) + \frac{1}{2} \lambda \phi^2(t,\mathbf{0}) \otimes \cI_{-}$ (in the Interaction picture) and so using $\avg{\chi^a} = 0$ the second-order Nakajima-Zwanzig equation reduces to
\bea
&&\frac{\partial \sigma}{\partial t}  \simeq - \frac{i \lambda}{2} \Bigl[ \phi^2(t,\mathbf{0}) , \sigma(t) \Bigr] +   g_a g_b \int_0^{t} \exd s\ \bigg( \avg{ \chi^b(t,\mathbf{0}) \chi^a(s,\mathbf{0}) } \Bigl[  \phi(s,\mathbf{0}) \, \sigma(s) \,, \phi(t,\mathbf{0}) \Bigr] \\
&& \qquad \qquad \qquad \qquad  \qquad \qquad \qquad \qquad \qquad \qquad \qquad \qquad  + \avg{ \chi^a(s,\mathbf{0})\, \chi^b(t,\mathbf{0}) } \Bigl[  \phi(t,\mathbf{0}) \,, \sigma(s)  \phi(s,\mathbf{0}) \Bigr] \bigg) \nn 
\eea
We write the thermal correlation function for two $\chi$ fields as 
\be \label{ThermalWexp}
  g_a g_b  \avg{ \chi^b(t,\mathbf{0}) \chi^a(s,\mathbf{0}) }  =:  \tilde g^2 \mathscr{W}(t -s)  =  - \frac{\tilde g^2}{4 \beta^2 } \; \csch^2\left[ \frac{\pi }{\beta}( t - s - i \delta ) \right]
\ee
where $\delta = 0^+$ goes to zero at the end of any calculation and the first equality defines the function $ \mathscr{W}(t)$ with $\tilde g^2 :=   \delta^{ab}\,  g_a g_b$ as given in \pref{tildegdef}, while the second equality uses the explicit form for the $\bfx' \to \bfx$ limit of the correlation function given in \pref{thermalcorrelatorab}. 

Finally, after a change of integration variables $s \to t - s$ we arrive at the form for the Nakajima-Zwanzig equation whose properties are explored below:
\bea\label{NZbeforeMarkov} 
&& \frac{\partial \sigma}{\partial t} \simeq - \frac{i \lambda}{2} \Bigl[ \phi^2(t,\mathbf{0}) , \sigma(t) \Bigr] \\
&& \qquad \qquad \qquad + g^2 \int_0^{t} \exd s\ \bigg(  \mathscr{W}(s) \Bigl[  \phi(t-s,\mathbf{0}) \, \sigma(t-s) \,, \phi(t,\mathbf{0}) \Bigr] +  \mathscr{W}^{\ast}(s) \Bigl[  \phi(t,\mathbf{0}) \,, \sigma(t-s)  \phi(t-s,\mathbf{0}) \Bigr] \bigg) \ . \nn
\eea

\subsubsection{Markovian limit}

Since the correlation function $ \mathscr{W}(s)$ is sharply peaked about $s=0$ and falls off exponentially fast like $ \mathscr{W}(s) \propto e^{ - {2 \pi s}/{\beta} }$ for $s \gg \beta$, the integral simplifies if the rest of the integrand varies more slowly in the region where $\mathscr{W}$ varies quickly. In such a case the integral is well-approximated by expanding the rest of the integrand in powers of $s$, using
\bea  \label{TaylorMarkov}
\phi(t - s) \sigma(t-s)  \simeq  \phi(t,\mathbf{0})\, \sigma(t)  - s \big[ \partial_t \phi(t,\mathbf{0}) \,\sigma(t)  +  \phi(t,\mathbf{0}) \partial_t\sigma(t) \big] + \ldots
\eea
beneath the integral sign in (\ref{NZbeforeMarkov}). Notice that this assumes {\it both} $\phi$ and $\sigma$ vary slowly, and so its justification requires both that $\sigma(t)$ should be slowly varying and that we work in an effective description that keeps only those modes of $\phi$ whose energies satisfy $E \ll 1/\beta$. This becomes relevant when choosing momentum cutoffs for later integrals, since this Markovian derivation requires $\Lambda \ll 1/\beta$. Part of the discussion to follow aims to identify more precisely the region of parameter space for which this Markovian approximation is valid (which we find by asking when the subleading terms in the series \pref{TaylorMarkov} are small).

Keeping only the leading-order term in the Taylor-series (\ref{TaylorMarkov}) yields the approximate equation of motion
\be \label{NZafterMarkov} \frac{\partial \sigma}{\partial t} \simeq - \frac{i \lambda}{2} \Bigl[ \phi^2(t,\mathbf{0}) , \sigma(t) \Bigr]  + \tilde g^2 \mathscr{C} \Bigl[  \phi(t,\mathbf{0}) \, \sigma(t) \,, \phi(t,\mathbf{0}) \Bigr] +  \tilde g^2 \mathscr{C}^* \Bigl[  \phi(t,\mathbf{0}) \,, \sigma(t)  \phi(t,\mathbf{0}) \Bigr]   \,. 
\ee
where the coefficient is given by
\be \label{mathscrCdef}
  \mathscr{C}(t) := \int_0^{t} \exd s\, \mathscr{W}(s) 
\ee
and the approximate equality requires $t\gg \beta$ so that the integration includes the strong peaking of $\mathscr{W}(s)$ (with exponential fall-off) noted above. Although the coefficient $\mathscr{C}(t)$ is in principle a function of $t$, in practice the narrowly peaked form of $\mathscr{W}(t)$ ensures it approaches a constant exponentially quickly once $t \gg \beta$. The value of this constant can be evaluated explicitly by taking the upper integration limit to infinity and evaluating the resulting integral using expression \pref{ThermalWexp} for $\mathscr{W}(t)$:
\be \label{NZdivergence}
  \mathscr{C}(t)  \simeq  \mathscr{C}_\infty =  \int_0^{\infty} \exd s\, \mathscr{W}(s) 
 = - \frac{1}{4\beta^2}   \int_0^{\infty} \frac{\exd s}{\sinh^2 \big[ \tfrac{\pi}{ \beta} (s - i \delta) \big] } 
\simeq \frac{1}{4\pi\beta} - \frac{i}{4\pi^2\beta} \left[ \frac{\beta}{\delta} + \mathcal{O}\left( \sfrac{\delta}{\beta} \right) \right] \,,
\ee
for $t \gg \beta$. The divergence as $\delta \to 0$ is a reflection of the divergence of the integrand as $s \to 0$. 

It is sometimes useful to convert \pref{NZafterMarkov} to the Schr\"odinger picture, as is done by noting that 
\be \label{SchroVsInt}
e^{- i \cH_{+} t} \frac{\partial \sigma}{\partial t} e^{+ i \cH_{+}t}  =  \frac{\partial \sigma_{\ssS}}{\partial t} + i [\cH_{+} , \sigma_{\ssS}(t)] \ . 
\ee
which follows from the relation (\ref{sigmaSvssigma}), giving
\be \label{IntermedNZeq}
\frac{\partial \sigma_{\ssS}}{\partial t}   \simeq  - i \Bigl[\cH_{+} + \sfrac{\lambda}{2}\phi_{\ssS}^2(\mathbf{0}) , \sigma_{\ssS}(t)\Bigr] +\tilde  g^2 \mathscr{C}   \Bigl[  \phi_{\ssS}(\mathbf{0}) \, \sigma_{\ssS}(t) , \phi_{\ssS}(\mathbf{0}) \Bigr]  + \tilde g^2 \mathscr{C}^* \Bigl[ \phi_{\ssS}(\mathbf{0}) \, , \sigma_{\ssS}(t)  \phi_{\ssS}(\mathbf{0})  \Bigr]  \,.  
\ee
Using this in \pref{IntermedNZeq} finally gives
\bea \label{SchPicMarkovian} 
\frac{\partial \sigma_{\ssS}}{\partial t} & \simeq & - i  \Bigl[\cH_{+} + \sfrac{\lambda}{2}\phi_{\ssS}^2(\mathbf{0}) , \sigma_{\ssS}(t)\Bigr] \nn\\
&& \ \quad + \frac{\tilde g^2}{4\pi} \left( \frac{1}{\beta} - \frac{i}{\pi \delta} \right) \Bigl[  \phi_{\ssS}(\mathbf{0}) \, \sigma_{\ssS}(t) \,, \phi_{\ssS}(\mathbf{0}) \Bigr] + \frac{\tilde g^2}{4\pi} \left( \frac{1}{\beta} + \frac{i}{\pi \delta} \right) \Bigl[  \phi_{\ssS}(\mathbf{0}) \,, \sigma_{\ssS}(t) \, \phi_{\ssS}(\mathbf{0}) \Bigr] \\
&=&  - i  \Bigl[\cH_{+} + \sfrac{\lambda_{\rm ren}}{2}\phi_{\ssS}^2(\mathbf{0}) , \sigma_{\ssS}(t)\Bigr] + \frac{\tilde g^2}{4\pi \beta} \Bigl( \Bigl[  \phi_{\ssS}(\mathbf{0}) \, \sigma_{\ssS}(t) \,, \phi_{\ssS}(\mathbf{0}) \Bigr] +  \Bigl[  \phi_{\ssS}(\mathbf{0}) \,, \sigma_{\ssS}(t)  \,\phi_{\ssS}(\mathbf{0}) \Bigr] \Bigr) \,, \nn
\eea
which shows that the divergence can be absorbed into the renormalization
\be
  \lambda_{\mathrm{ren}} \ := \ \lambda - \frac{\tilde g^2}{2\pi^2 \delta} \,.
\ee

\subsubsection{Evolution equation in a field basis} 

It is easiest to solve an equation like \pref{SchPicMarkovian} in a basis that diagonalizes the interaction Hamiltonian, and in this instance this suggests using a basis of field eigenstates defined as the basis that diagonalizes the Schr\"odinger-picture field operator $\phi(0,\bfx) =\phi_{\ssS}(\bfx)$:
\be
\phi_{\ssS}(\bfx) \ket{\varphi\sb}  = \varphi(\bfx)\ket{\varphi\sb} 
\ee
where the eigenvalue $\varphi(\bfx)$ is a real-valued function of position. 

The wave-functional of the free vacuum $\ket{\hbox{vac}}$ in this reprsentation is given by a gaussian \cite{Weinberg:1995mt}
\bea    \label{overlap} 
\bra{ \varphi \sb} \hbox{vac} \ra& = & \sqrt{ \mathcal{N}_0}\;  \exp\left[ - \frac{1}{2} \int \exd^3 \bfx  \int d^3\bfy\ \mathcal{E}(\bfx,\bfy) \varphi(\bfx) \varphi(\bfy)  \right] \\
&=& \sqrt{\mathcal{N}_0 }\; \exp\left[ - \frac{1}{2} \int \frac{\exd^3p}{(2\pi)^3} \; E_p \; \varphi_\bfp \varphi_{-\bfp} \right] \nn
\eea
where the kernel $\mathcal{E}(\bfx,\bfy)$ is given by
\be   \label{freekernel}
\mathcal{E}(\bfx,\bfy) = \int \frac{d^3\bfp}{(2\pi)^3} \ E_p \ e^{i\bfp \cdot (\bfx - \bfy)} \,,
\ee
where $E_p :=  |\bfp|$ and the normalization factor $\mathcal{N}_0$ is determined by using the normalization condition $\langle \hbox{vac} | \hbox{vac} \rangle=1$.

In the field basis the reduced density matrix is a time-dependent functional of the basis field configurations $\varphi_1(\bfx)$ and $\varphi_2(\bfx)$, with components
\be
\sigma_{\ssS}[\varphi_1, \varphi_2; t] := \la \varphi_1\sb | \sigma_{\ssS}(t) | \varphi_2\sb \ra
\ee
which the Markovian equation \pref{SchPicMarkovian} implies must satisfy 
\bea
\frac{\partial \sigma_{\ssS}[\varphi_1, \varphi_2; t]}{\partial t} & \simeq & - i \bra{\varphi_1\sb} \Bigl[\cH_{+} , \sigma_{\ssS}(t)\Bigr] \ket{\varphi_2\sb} \\
&& \qquad -\bigg[  \frac{\tilde g^2}{4 \pi \beta} \big( \varphi_1(\mathbf{0}) - \varphi_2(\mathbf{0}) \big)^2 + \frac{i\lambda_{\rm ren}}{2} \big( \varphi_1^2(\mathbf{0}) - \varphi^2_2(\mathbf{0}) \big) \bigg] \sigma_{\ssS}[\varphi_1, \varphi_2; t] \ . \notag
\eea
We henceforth drop the subscript `ren' on the renormalized coupling parameter $\lambda$. Evaluating the commutator term using
\be
\bra{ \varphi \sb} \cH_{+} \ket{ \Psi } \ = \ \frac{1}{2} \int \exd^3 \bfx \bigg[  -  \frac{\delta^2}{\delta \varphi(\bfx)^2 } + \big| \boldsymbol{\nabla} \varphi(\bfx) \big|^2 \bigg] \la \varphi\sb | \Psi \ra \ ,
\ee
for any state $|\Psi \rangle$, the equation of motion for $\sigma_\ssS$ finally becomes
\bea \label{Markeq} 
&&\frac{\partial \sigma_{\ssS}[\varphi_1, \varphi_2; t]}{\partial t}   \simeq  - \frac{i}{2} \int \exd^3 \bfx \bigg[  -  \frac{\delta^2}{\delta \varphi_1(\bfx)^2 } + \big| \boldsymbol{\nabla} \varphi_1(\bfx) \big|^2 + \frac{\delta^2}{\delta \varphi_2(\bfx)^2 } - \big| \boldsymbol{\nabla} \varphi_2(\bfx) \big|^2 \bigg] \sigma_{\ssS}[\varphi_1, \varphi_2; t] \nn \\
&& \qquad\qquad\qquad\qquad - \frac{\tilde g^2}{4 \pi \beta} \big( \varphi_1(\mathbf{0}) - \varphi_2(\mathbf{0}) \big)^2 \sigma_{\ssS}[\varphi_1, \varphi_2; t] - \frac{i\lambda}{2} \big( \varphi_1^2(\mathbf{0}) - \varphi^2_2(\mathbf{0}) \big) \sigma_{\ssS}[\varphi_1, \varphi_2; t] \ .\nn \\
\eea
As is easily verified, when $\lambda = \tilde g = 0$ this equation has as a solution
\bea\label{FreeVacSchrodPic}
   \sigma_\ssS(t,\varphi_1,\varphi_2) &=& \langle \varphi_1 \sb | \hbox{vac} \rangle \, \langle \hbox{vac} | \varphi_2\sb \rangle \nn\\
  & = & \cN_0 \exp\left( - \frac{1}{2} \int \exd^3 \bfx \int \exd^3 \bfy\; \mathcal{E}(\bfx-\bfy) \big[ \varphi_1(\bfx) \varphi_{1}(\bfy)  + \varphi_2(\bfx) \varphi_{2}(\bfy) \big]  \right) 
\eea
where the second equality uses \pref{overlap} and the kernel $\mathcal{E}(\bfx-\bfy)$ is as defined in equation (\ref{freekernel}). 

\subsection{Solutions for the reduced density matrix}

We next solve eq.~\pref{Markeq} for the $\phi$-sector density matrix in the presence of the hotspot interactions.

\subsubsection{Gaussian ansatz}

Keeping in mind that the hotspot `interactions' are all bilinear in the fields we seek solutions to \pref{Markeq} subject to the more general Gaussian ansatz
\bea \label{firstAnsatz}
\sigma_{\ssS}[\varphi_1, \varphi_2; t] & = & \cN(t) \exp\bigg( - \frac{1}{2} \int \exd^3 \bfx \int \exd^3 \bfy\; \bigg\{ \mathcal{A}_1(\bfx,\bfy ; t ) \,\varphi_1(\bfx) \varphi_{1}(\bfy) + \mathcal{A}_2(\bfx,\bfy ; t ) \, \varphi_2(\bfx) \varphi_{2}(\bfy) \notag \\
& \ & \qquad \qquad \qquad \qquad \qquad \qquad \quad \quad \quad \quad + 2 \cB(\bfx, \bfy ; t) \, \varphi_1(\bfx) \varphi_{2}(\bfy) \bigg\} \bigg) \,,
\eea
with the kernels $\cA_{1}$, $\cA_{2}$ and $\cB$ to be determined. Note that we can without loss of generality assume the symmetry
\be \label{AjSYM}
\cA_j(\bfx,\bfy;t) = \cA_j(\bfy,\bfx;t) \ ,
\ee
and that hermiticity of the reduced density matrix -- $\sigma_{\ssS}^{\ast}[\varphi_1, \varphi_2; t] = \sigma_{\ssS}[\varphi_2, \varphi_1; t]$ -- implies $\cN^{\ast}(t) = \cN(t)$ and
\be  \label{cBsym1}
\cA_{1}^{\ast}(\bfx, \bfy ; t) = \cA_{2}(\bfx,\bfy ; t) \quad \hbox{and} \quad
\cB^{\ast}(\bfx, \bfy ; t) = \cB(\bfy,\bfx ; t) \,.
\ee

Notice that $\sigma_\ssS = | \Psi \rangle \, \langle \Psi |$ for a gaussian pure state $\langle \varphi \sb | \Psi \rangle \propto \exp\left[ - \frac12 \, K(\bfx,\bfy) \, \varphi(\bfx) \varphi(\bfy)\right]$ only if  
\be
   \cA_1(\bfx,\bfy) = \cA^*_2(\bfx,\bfy) = K(\bfx,\bfy) \quad\hbox{and} \quad 
   \cB(\bfx,\bfy) = 0 \,.
\ee
with the free vacuum \pref{FreeVacSchrodPic} corresponding to the choice $K(\bfx,\bfy) = \cE(\bfx-\bfy)$. Since the first of these is an automatic consequence of \pref{cBsym1}, this shows that $\cA_1$ and $\cA_2$ can be regarded as the deformations of the ground state away from the free result due to the interactions, while having $\cB \neq 0$ corresponds to the interaction causing the initially pure state to become mixed.

The kernels are obtained by plugging the ansatz \pref{firstAnsatz} into \pref{Markeq} and equating the coefficients of the different functional forms on both sides of the equation. The details are worked out in Appendix \ref{App:KernelPosSpace}, with the results simply quoted here. Equating the coefficients of terms independent of $\varphi_i$ implies
\be \label{Neq}
\frac{1}{\cN} \frac{\pd \cN}{\pd t} \ = \  - \frac{i}{2} \int \exd^3 \bfx \Bigl[ \cA_1(\bfx,\bfx;t) -  \cA_2(\bfx,\bfx;t) \Bigr] \,.
\ee
This expression can also be derived from the condition that $\TrA \sigma_\ssS(t) = 1$ for all times. The coefficient of $\varphi_1(\bfx)  \varphi_1(\bfy)$ similarly gives
\bea \label{11eq}
 \frac{\pd \cA_{1}(\bfx, \bfy ; t)}{\pd t} & = & - i \nabla_\bfx^2 \delta^3(\bfx - \bfy) + \left(  \frac{\tilde g^2}{2\pi \beta} + i \lambda \right) \delta^3(\bfx) \delta^3(\bfy)  \\
& \ & \quad  \quad  \quad  \quad  \quad  \quad + \int \exd^3 \bfz \; \Bigl[ - i \cA_{1}(\bfz,\bfx ; t) \cA_{1}(\bfz,\bfy ; t) + i \cB(\bfx,\bfz ; t) \cB(\bfy,\bfz ; t) \Bigr] \,, \notag
\eea
while the coefficient of $\varphi_2(\bfx)  \varphi_2(\bfy)$ leads to
\bea \label{22eq}
 \frac{\pd \cA_{2}(\bfx, \bfy ; t)}{\pd t} & = &  i \nabla_\bfx^2 \delta^3(\bfx - \bfy) + \left(  \frac{\tilde g^2}{2\pi \beta} - i \lambda \right) \delta^3(\bfx) \delta^3(\bfy)  \\
& \ & \quad  \quad  \quad  \quad  \quad  \quad + \int \exd^3 \bfz \; \Bigl[  i \cA_{2}(\bfx,\bfz ; t) \cA_{2}(\bfy,\bfz ; t) - i \cB(\bfz,\bfx ; t) \cB(\bfz,\bfy ; t) \Bigr] \,.\notag
\eea
Finally, the coefficient of $\varphi_1(\bfx) \varphi_2(\bfy)$ gives
\bea  \label{12eq}
\frac{\pd \cB(\bfx, \bfy ; t)}{\pd t} & = & - \frac{\tilde g^2}{2\pi \beta} \delta^3(\bfx) \delta^3(\bfy)  \\
& \ & \quad  \quad  \quad  \quad  \quad  \quad  + \int \exd^3 \bfz \; \Bigl[ - i  \cA_{1}(\bfz,\bfx ; t) \cB(\bfz,\bfy ; t) + i \cB(\bfx,\bfz ; t) \cA_{2}(\bfz,\bfy ; t) \Bigr]  \,.\notag
\eea

The implications of these equations are simpler to see in momentum space, so we define
\bea
\cA_j(\bfx, \bfy ; t) & = &  \int \frac{\exd^{3} \bfk}{(2\pi)^{3}} \int \frac{\exd^{3} \bfq}{(2\pi)^{3}} \; A_{j}(\bfk, \bfq ; t)\;  e^{ + i \bfk \cdot \bfx} \, e^{- i \bfq \cdot \bfy}\\
\hbox{and} \quad
\cB(\bfx, \bfy ; t) & = &  \int \frac{\exd^{3} \bfk}{(2\pi)^{3}}  \int \frac{\exd^{3} \bfq}{(2\pi)^{3}} \; B(\bfk, \bfq ; t) \; e^{ + i \bfk \cdot \bfx} \, e^{- i \bfq \cdot \bfy} \,, \nn
\eea
for which the symmetry $\cA_j(\bfx,\bfy;t) = \cA_j(\bfy,\bfx;t)$ of (\ref{AjSYM}) implies 
\bea \label{AjSYMmom}
A_{j}(\bfk, \bfq ; t) \ = \ A_{j}(- \bfq, -\bfk ; t) \ . 
\eea

In terms of these equation (\ref{Neq}) becomes
\bea \label{NeqMom}
\frac{1}{\cN} \frac{\pd \cN}{\pd t} \ = \  - \frac{i}{2} \int \frac{\exd^{3} \bfk}{(2\pi)^{3}}  \;  \Bigl[ A_{1}(\bfk, \bfk ; t) - A_{2}(\bfk, \bfk ; t) \Bigr]
\eea
while equations (\ref{11eq}) through  (\ref{12eq}) become
\bea\label{A1momeq}
\frac{\pd A_{1}(\bfk, \bfq ; t)}{\pd t} & = & i (2\pi)^3 |\bfk|^2 \delta^3(\bfk - \bfq) + \left( \frac{\tilde g^2}{2\pi \beta} + i \lambda \right)  \\
& \ & \quad  \quad  \quad + \int \frac{\exd^{3} \bfp}{(2\pi)^{3}} \Bigl[ - i A_{1}(\bfk, \bfp ; t) A_{1}(\bfp, \bfq ; t) + i B(\bfk, \bfp ; t ) B(-\bfq, -\bfp ; t) \Bigr] \,, \notag
\eea
\bea \label{A2momeq}
\frac{\pd A_{2}(\bfk, \bfq ; t)}{\pd t} & = & - i (2\pi)^3 |\bfk|^2 \delta^3(\bfk - \bfq) + \left( \frac{\tilde g^2}{2\pi \beta} - i \lambda \right) \\
& \ & \quad  \quad  \quad + \int \frac{\exd^{3} \bfp}{(2\pi)^{3}} \Bigl[ i A_{2}(\bfk, \bfp ; t) A_{2}(\bfp, \bfq ; t) - i B(-\bfp, -\bfk ; t ) B(\bfp, \bfq ; t) \Bigr] \notag
\eea
and  
\be\label{Beqmomentumspace}
\frac{\pd B(\bfp, \bfq ; t)}{\pd t} = - \frac{\tilde g^2}{2\pi \beta} + \int \frac{\exd^{3} \bfk}{(2\pi)^{3}} \Bigl[ - i A_1(\bfp, \bfk ; t ) B(\bfk, \bfq ; t) + i B(\bfp, \bfk  ; t) A_2(\bfk, \bfq ; t) \Bigr] \,.
\ee
These are the equations we solve in the next few sections. Notice in particular that \pref{12eq} or \pref{Beqmomentumspace} implies $\tilde g^2/\beta \neq 0$ is an obstruction to $B(\bfq,\bfx,t) = 0$ being a solution.

\subsubsection{Perturbative solution}

As is easily verified, for $\lambda = \tilde g = 0$ these above equations are solved by 
\be
   B(\bfp,\bfq,t) = 0 \quad \hbox{and} \quad
   A_j(\bfk, \bfq) \ = \ (2\pi)^3 |\bfk| \; \delta^3(\bfk - \bfq) \,,
\ee
corresponding to the vacuum solution of \pref{FreeVacSchrodPic}. We next seek solutions that are perturbatively close to this vacuum solution, as should be possible for small $\tilde g$ and $\lambda$. 

To this end we write
\be
A_{j}(\bfk, \bfq ; t) = (2\pi)^3 |\bfk| \; \delta^3(\bfk - \bfq) +  \mfa_{j}(\bfk, \bfq ; t)  + \ldots \quad \hbox{and} \quad
B(\bfk, \bfq ; t)  =  \mfb(\bfk, \bfq ; t)  + \ldots \,,
\ee
and linearize eqs.~\pref{NeqMom} through  \pref{Beqmomentumspace} in the perturbations $\mfa_j$ and $\mfb$. The resulting evolution equations decouple, to become
\be
\frac{\partial \mfa_1(\bfk, \bfq; t)}{\pd t} = \frac{\tilde g^2}{2\pi\beta} + i \lambda - i ( |\bfk| + |\bfq| ) \, \mfa_1(\bfk, \bfq; t) \,,
\ee
\be
\frac{\partial \mfa_2(\bfk, \bfq; t)}{\pd t} = \frac{\tilde g^2}{2\pi\beta} - i \lambda + i ( |\bfk| + |\bfq| )\, \mfa_2(\bfk, \bfq; t)
\ee
and
\be \label{mfbEvo}
\frac{\partial \mfb(\bfk, \bfq; t)}{\pd t} = - \frac{\tilde g^2}{2\pi\beta}   - i \big( |\bfk| - |\bfq| \big)  \, \mfb(\bfk,\bfq;t) \,.
\ee
These are to be solved subject to the initial conditions 
\be
\mfa_1(\bfk,\bfq ; 0) = \mfa_2(\bfk,\bfq; 0) = \mfb(\bfk,\bfq; 0) = 0 \,,
\ee
since the scalar $\phi$ is starts off in its vacuum state. 

The solutions to these initial-value problems are given by
\be \label{mfa1kk}
\mfa_1(\bfk,\bfq ; t) = \left( \lambda - \frac{i \tilde g^2 }{2\pi\beta} \right) \frac{1 - e^{- i \big( |\bfk| + |\bfq| \big) t}}{|\bfk| + |\bfq|} \,,
\ee
\be \label{mfa2kk}
\mfa_2(\bfk,\bfq ; t) =  \left( \lambda + \frac{i \tilde g^2}{2\pi\beta} \right) \frac{1 - e^{+ i \big( |\bfk| + |\bfq| \big) t}}{|\bfk| + |\bfq|} 
\ee
and
\be \label{mfbkk}
\mfb(\bfk,\bfq ; t)  =  \left( \frac{i \tilde g^2}{2\pi\beta} \right) \frac{1 - e^{- i ( |\bfk| - |\bfq| ) t} }{|\bfk| - |\bfq|} \,.
\ee
In the limit $\bfk = \bfq$ (or in the limit of small $t$) this last solution simplifies to
\be
 \mfb(\bfk,\bfq ; t)  \to
  - \frac{\tilde g^2t}{2\pi\beta} \qquad \hbox{when } \quad \bfk \to \bfq \,.
\ee

As remarked earlier --- and explored in more detail in \S\ref{ssec:Decoherence} --- nonzero $\mfb$ is a signature of $\sigma_\ssS$ becoming a mixed state, and so \pref{mfbkk} shows that this only happens for nonzero $\tilde g^2/\beta$. Furthermore when $\tilde g^2/\beta$ is nonzero there can be no static solution with $\partial_{t} \mfb =0$, (and in particular no solution with $\mfb = 0$) and $|\mfb(\bfk,\bfk; t)|$ monotonically increases. The coupling $\lambda$, by contrast, just deforms the ground state but leaves it pure.

The normalization $\cN(t)$ is found in a similar way. Using the above solution in \pref{NeqMom} allows it to be written
\bea \label{Neqveck}
\frac{\pd_t \cN}{\cN} & = & - \frac{i}{2} \int \frac{\exd^3 \bfk}{(2\pi)^3} \Bigl[ \mfa_{1}(\bfk, \bfk ; t) -  \mfa_{2}(\bfk, \bfk ; t) \Bigr] \nn \\
& = & \int \frac{\exd^3 \bfk}{(2\pi)^3|\bfk|} \bigg[ \lambda \cos(|\bfk| t ) \sin(|\bfk| t ) - \frac{\tilde g^2}{2\pi\beta} \sin^2(|\bfk|t) \bigg] \,.
\eea
The integral on the right-hand side diverges in the ultraviolet, which we regulate using a momentum cutoff $|\bfk| < \Lambda$, leading to the result
\be \label{NtSol}
\cN(t)  =  \exp\left( C_0 - \frac{\tilde g^2\Lambda^2 t}{16 \pi^3 \beta} -  \frac{\lambda}{16 \pi^2 t} \sin( 2 \Lambda t ) + \frac{\tilde g^2}{16 \pi^3 \beta t } \sin^2(\Lambda t) \right)  \,, 
\ee
where $C_0$ is the integration constant. In terms of the initial condition $\cN(0) = \cN_0$, where $\cN_0$ is the normalization of the free-vacuum state, we have
\be
\cN_0 =  \exp\left(C_0 -  \frac{\lambda \Lambda}{8 \pi^2}  \right)  \,, 
\ee
and so  
\be \label{cNtsol}
\cN(t)  =  \cN_0 \; \exp\bigg\{ \frac{\Lambda}{8 \pi^2} \bigg[ - \frac{\tilde g^2  \Lambda t }{2 \pi \beta } \left( 1 - \frac{\sin^2(\Lambda t)}{(\Lambda t)^2} \right) + \lambda \left(1 - \frac{ \sin\left( 2 \Lambda t \right)}{ 2 \Lambda t }\right)  \bigg] \bigg\} \ . 
\ee
The significance of the divergences in the time-dependence of $\cN(t)$ is discussed further in \S\ref{ssec:Decoherence}.

\subsection{Equal-time $\phi$-correlator}

This section uses the reduced density matrix for $\phi$ computed in the previous section to calculate the $\langle \phi \, \phi \rangle$ two-point function. Comparison of the result with the Wightman function given in \S\ref{sec:setup} provides a check on the domain of validity of the Nakajima-Zwanzig late-time evolution.  

The correlator evaluated at $t = t'$ is convenient because it has a simple representation in terms of the reduced density matrix. This is because the only evolution operators that appear are the ones that convert between Heisenberg, interaction and Schr\"odinger pictures:
\be
\Tr \Bigl[ {\phi}_\ssH(t,\bfx) \,{\phi}_\ssH(t,\bfx') \,\rho_0 \Bigr]  = \TrA\Bigl[ {\phi}(t,\bfx)\, {\phi}(t,\bfx')\, \sigma(t) \Bigr]  =  \TrA\Bigl[ {\phi}_{\ssS}(\bfx) \,{\phi}_{\ssS}(\bfx') \,\sigma_{\ssS}(t) \Bigr] \,,
\ee
where $\TrA$ in the second two terms denotes a trace only over the $\phi$ sector of the Hilbert space.

The calculation based on the field-representation of the Schr\"odinger-picture reduced density matrix evaluates the trace using a partition of unity written as a functional integral over the field eigenstates,
\bea \label{EqTimePathInt}
\TrA \Bigl[  \phi_{\ssS}(\mathbf{x}) \phi_{\ssS}(\mathbf{x}') \sigma_{\ssS}(t) \Bigr] & = & \int \cD \varphi \; \bra{\varphi \sb } \phi_{\ssS}(\mathbf{x})\, \phi_{\ssS}(\mathbf{x}') \,\sigma_{\ssS}(t) | \varphi \sb \rangle \\
 & = & \int \cD \varphi \; \varphi(\mathbf{x}) \, \varphi(\mathbf{x}') \,  \sigma_{\ssS}[\varphi, \varphi;t] \,,  \nn
\eea
with $\sigma_\ssS[\varphi_1, \varphi_2;t]$ given in terms of the kernels $\cA_j$ and $\cB$ as in \pref{firstAnsatz}. Our focus here is in particular on the $\lambda$ and $\tilde g$ dependent parts of the density matrix, since Appendix \ref{sec:freeequalcorrelator} verifies that the above functional integral correctly reproduces the free-field Wightman function inasmuch as it shows that in the limit $\lambda = \tilde g = 0$ eq.~\pref{EqTimePathInt} reproduces the usual expression
\be
\bra{ \mathrm{vac} } \phi(t,\bfx) \phi(t,\bfx') \ket{ \mathrm{vac} }  =  \frac{1 }{4 \pi^2 |\bfx - \bfx'|^2 } \,.
\ee

Eq.~\pref{EqTimePathInt} shows that only the diagonal part of the reduced density matrix is required to compute the equal-time Wightman function. Explicitly, this is given by
\be
\sigma_{\ssS}[ \varphi, \varphi ;t] = \cN(t) \exp\bigg( - \frac{1}{2} \int \exd^3 \bfx \int \exd^3 \bfx'\; \cM(\bfx, \bfx' ; t) \varphi(\bfx) \varphi(\bfx')\bigg) \,, 
\ee
where
\be \label{MvsAB}
\cM(\bfx, \bfx'; t) := 2 \mathrm{Re} \Bigl[\mathcal{A}_1(\bfx,\bfx' ; t ) +   \cB(\bfx, \bfy ; t) \Bigr] \,,
\ee
and we use the symmetry \pref{cBsym1}. Evaluating the gaussian integrals then implies 
\bea  \label{cMinvcorrelator}
\TrA[  \phi_{\ssS}(\mathbf{x}) \phi_{\ssS}(\mathbf{x}') \sigma_{\ssS}(t) ] & = &\cN(t) \int \cD \varphi \; \varphi(\mathbf{x}) \varphi(\mathbf{x}') \;e^{- \frac{1}{2} \int \mathrm{d}^3\bfz_1 \int \rm{d}^3 \bfz_2\; \cM(\bfz_1,\mathbf{z}_2; t) \varphi(\bfz_1) \varphi(\bfz_2) } \nn \\
& = & \cM^{-1}(\bfx, \bfx' ; t) \,,
\eea
where $\cM^{-1}(\bfx, \bfx' ; t)$ is the inverse of $\cM(\bfx, \bfx' ; t)$, in the sense that $\int \exd^{3} \bfz\; \cM^{-1}(\bfx, \bfz ; t) \cM(\bfz, \bfx' ; t)  =  \delta^3(\bfx - \bfx' )$. The components $\cM^{-1}(\bfx,\bfx')$ are computed explicitly in Appendix \ref{MinvCalc} using \pref{MvsAB} and the perturbative solutions for $\cA_j$ and $\cB$ given earlier. Using the result in \pref{cMinvcorrelator} then gives
\bea \label{corrwithmfians}
 \TrA[  \phi_{\ssS}(\mathbf{x}) \phi_{\ssS}(\mathbf{x}') \sigma_{\ssS}(t) ]
& = &  \frac{1}{4\pi^2|\bfx - \bfx'|^2}  - \frac{\lambda}{16 \pi^3 (|\bfx|^2 - |\bfx'|^2)}   \bigg[  \frac{1}{ |\bfx'|} \,  \Theta(t - |\bfx'| ) - \frac{1 }{ |\bfx|}\, \Theta(t - |\bfx| ) \bigg]  \nn \\
&&\qquad\qquad\qquad\qquad + \frac{\tilde g^2}{32 \pi^3 \beta |\bfx| |\bfx'|} \;   \Theta(t-|\bfx|) \, \delta(|\bfx|-|\bfx'|) \,.
\eea
Specializing to the forward light cone of the switch-on of couplings --- {\it i.e.}~to $t>|\bfx|$ and $t > |\bfx'|$ --- the above becomes 
\be \label{eqtimecorrMarkovianLT}
\TrA[  \phi_{\ssS}(\mathbf{x}) \phi_{\ssS}(\bfx') \sigma_{\ssS}(t) ] \simeq  \frac{1}{4\pi^2|\bfx - \bfx'|^2} - \frac{\lambda}{16 \pi^3 |\bfx| \; |\bfx'| (|\bfx| + |\bfx'| )} + \frac{\tilde{g}^2}{32 \pi^3 \beta |\bfx| |\bfx'|} \;  \delta(|\bfx| - |\bfx'|)\ .  
\ee

\subsubsection*{Comparison with Heisenberg picture}

Expression  \pref{eqtimecorrMarkovianLT} is to be compared with the Wightman function computed in Heisenberg picture, whose expansion at linear order in $\lambda$ and $\tilde g^2$ is quoted in \pref{pertcorr} and whose equal-time limit is given in \pref{largeNcorrteqs0}, reproduced for convenience here:
\bea \label{pertcorrequal}
W_\beta(t,\bfx; t,\bfx') & \simeq & \frac{1}{4\pi^2 |\bfx - \bfx'|^2} - \frac{\lambda}{16\pi^3 |\bfx| |\bfx'| ( |\bfx| + |\bfx'| )}  \\
& \ & \qquad \qquad - \frac{\tilde{g}^2}{64 \pi^2 \beta^2 |\bfx| |\bfx'| \sinh^2 \left[ \frac{\pi}{\beta}( |\bfx| - |\bfx'| ) \right]} + \frac{\tilde{g}^2}{16 \pi^4 ( |\bfx|^2 - |\bfx'|^2 )^2} \,. \nn
\eea
Although the free and the $\lambda$-dependent terms here agree with those in \pref{eqtimecorrMarkovianLT}, the same is in general not true for the $\tilde g^2$-dependent terms. This need not be a problem since they should only be expected to agree within the domain of validity of both, and \pref{eqtimecorrMarkovianLT} is derived under more restrictive assumptions. 

Recall in particular that the derivation within the Schr\"odinger picture approximated the Nakajima-Zwanzig equation \pref{NZbeforeMarkov} using a Markovian limit in which $\sigma(t - s)\phi(t-s,\mathbf{0}) \simeq \sigma(t) \phi(t,\mathbf{0})$ is used under the integral sign. Doing this assumes both $\sigma_\ssS$ and $\phi$ vary very slowly on the time-scale $\beta$ for which the thermal $\chi^a$ correlator was sharply peaked (see the discussion surrounding eq.~\pref{TaylorMarkov}). This is only valid if the UV cutoff, $\Lambda$, for the $\phi$-field modes satisfies $\beta \Lambda \ll 1$, since only in this EFT is the field $\phi$ sufficiently slowly varying. 

Within this regime the term involving the hyperbolic function has a microscopic width and so approaches a delta function. To see this explicitly it is easier to work in momentum space, in which case \pref{pertcorrequal} is given by the following mode sum
\bea
W_\beta(t,\bfx; t,\bfx') & \simeq & \frac{1}{4\pi^2 |\bfx - \bfx'|} \int_0^\infty \exd p \; \sin(p |\bfx - \bfx'|  ) - \frac{\lambda}{16 \pi^3 |\bfx| |\bfx'|} \int_0^\infty \exd p \; e^{- i p |\bfx'| } \sin\big[ p ( |\bfx| + |\bfx'| ) \big] \nn \\ 
& & \qquad   + \frac{\tilde{g}^2}{64 \pi^4  |\bfx| |\bfx'|} \int_0^\infty \exd p\; p \left\{ \; \cos\big[ p ( |\bfx| + |\bfx'| ) \big] + \frac{2 \cos \big[ p ( |\bfx| - |\bfx'|)\big]}{e^{ \beta p }  - 1 } \; \right\} \,.
\eea
Because $p < \Lambda \ll \beta^{-1}$ we can expand the temperature-dependent last term using $\beta p \ll 1$, 
\be
 \frac{\tilde{g}^2}{64 \pi^4  |\bfx| |\bfx'|} \int_0^\infty \exd p\;  p\left\{  \cos\big[ p ( |\bfx| + |\bfx'| ) \big] +  \bigg[ \frac{2}{\beta p} - 1 + \frac{\beta p}{6} + \cO(p^3 \beta^3 ) \bigg]  \cos \big[ p ( |\bfx| - |\bfx'|)\big] \; \right\} \,,
\ee
and perform the momentum integrals term-by-term, giving
\be
 \frac{\tilde{g}^2}{64 \pi^4 \beta |\bfx| |\bfx'|} \left\{ - \frac{1}{( |\bfx| + |\bfx'| )^2} + \bigg[ \frac{2\pi}{\beta} \delta( |\bfx| - |\bfx'| ) +\frac{1}{( |\bfx| - |\bfx'| )^2}  - \frac{\beta}{6} \delta''(|\bfx| - |\bfx'| ) + \ldots \bigg] \right\} \,.
\ee
Using this in the Wightman function gives
\bea
W_\beta(t,\bfx; t,\bfx') & \simeq & \frac{1}{4\pi^2 |\bfx - \bfx'|^2} - \frac{\lambda}{16 \pi^3 |\bfx| |\bfx'| \big( |\bfx| + |\bfx'| \big)}  \\ 
& \ & \qquad \qquad \quad + \frac{\tilde{g}^2}{32 \pi^3 \beta  |\bfx| |\bfx'|}\, \delta\big(|\bfx| - |\bfx'|\big) + \frac{\tilde{g}^2}{16 \pi^4 \big(  |\bfx|^2 - |\bfx'|^2 \big)^2 } + \cO(\beta) \,.\nn
\eea
Notice that the leading term in this expansion indeed matches the Schrodinger-picture calculation. This comparison reveals more explicitly the long-wavelength domain of validity inherent in the Markovian limit,\footnote{At least in the way it is derived here. A Markovian limit with a broader domain of validity might also be possible, such as if only $\sigma_\ssS(t-s)$ is expanded in powers of $s$ without also expanding $\phi(t-s)$. Examples along these lines are seen in some simpler examples involving qubits interacting with fields \cite{Kaplanek:2019vzj}.} revealing it to depend on both an expansion in powers of $\tilde g^2$ and on taking long wavelengths compared to the thermal length scale $\beta$.

\subsection{Decoherence}
\label{ssec:Decoherence}

Earlier sections explore the behaviour of the $\phi$-field state by computing its reduced density matrix once the hotspot fields $\chi^a$ are all integrated out. It was there argued that the kernel $\cB(\bfx,\bfx';t)$ --- or equivalently $\mfb(\bfk,\bfq;t)$ --- provides a diagnostic of whether the state is pure or mixed (because pure states seem to require $\cB = \mfb = 0$.)

That is what makes eqs.~\pref{mfbEvo} and \pref{mfbkk} so interesting; they quantify the rate with which decoherence accumulates when hotspot couplings are turned on with $\phi$ prepared in its vacuum state. These  equations show in particular the growth of decoherence at early times has a momentum-independent universal rate, set by $\tilde g^2/(2\pi \beta)$. Since all momenta start to decohere with the same rate, the accumulated decoherence should be dominated by the highest momenta, for which there is the most available phase space. It is for this reason that measures of integrated decoherence -- such as the purity diagnostic computed explicitly below -- tend to diverge in the ultraviolet, at least so long as the cutoff remains below the characteristic hotspot temperature, as assumed for the Markovian approximation above. 

\subsubsection{Purity}

To pin down the decoherence process more precisely it is useful to have a practical diagnostic for state purity. A standard choice for this is often $\TrA [ \sigma_\ssS^2 ]$ since this is unity if and only if the state is pure (in which case $\sigma_\ssS^2 = \sigma_\ssS$). 

This trace can be computed as a function of the state kernels $\cA_j$ and $\cB$, as follows:
\bea \label{PurityCalc}
\mathrm{Tr}[\sigma_{\ssS}^2] & = & \int \cD \varphi_1 \int \cD \varphi_2 \; \bra{ \varphi_1 } \sigma_{\ssS}(t) \ket{ \varphi_2 } \bra{\varphi_2} \sigma_{\ssS}(t) \ket{ \varphi_1 } \nn\\
 &=& \cN(t)^2 \int \cD \varphi_1 \int \cD \varphi_2 \; \exp \left\{-  \frac{1}{2} \int \mathrm{d}^3 \bfx \int \mathrm{d}^3 \bfy\; \bigg[ 2 \,\mathrm{Re}\Bigl[ \cA_1(\bfx, \bfy ; t) \Bigr] \Bigl( \varphi_1(\bfx) \varphi_1(\bfy) + \varphi_2(\bfx) \varphi_2(\bfy)  \Bigr)  \right.   \nn\\
 && \qquad\qquad\qquad\qquad\qquad\qquad\qquad\qquad \left.   + 4\, \mathrm{Re}\Bigl[\cB(\bfx, \bfy ; t)\Bigr] \varphi_1(\bfx) \varphi_2(\bfy)\bigg] \right\}   \\ 
 &= & \cN(t)^2 \left\{ \det\left[ \frac{1}{2\pi} \left( \begin{matrix} 2\,\mathrm{Re}[\cA_1(t)] & 2\,\mathrm{Re}[\cB(t)]  \\ 2\,\mathrm{Re}[\cB(t)]  & 2\,\mathrm{Re}[\cA_1(t)] \end{matrix} \right) \right] \right\}^{-1/2} \,.\nn
\eea
In the last line we write quantities $\cB(\bfx,\bfy)$ as matrices, with rows and columns labelled by position. In this notation the matrix $\mathrm{Re}[\cB(t)]$ is symmetric, since the identity $\cB^{\ast}(\bfx,\bfy;t) = \cB(\bfy, \bfx ; t)$ means
\be
\mathrm{Re}[ \cB ](\bfx, \bfy ; t) = \frac{\cB(\bfx,\bfy;t) + \cB^{\ast}(\bfx,\bfy;t)}{2} = \frac{\cB(\bfx,\bfy;t) + \cB(\bfy,\bfx;t)}{2} \,,
\ee
a fact that has been used in writing the last line of \pref{PurityCalc}. 

This result can be further simplified using the following identity\footnote{To see this, take the determinant of both sides of $\left[ \begin{smallmatrix} I & I \\ 0 & I \end{smallmatrix}  \right] \left[ \begin{smallmatrix} X & Y \\ Y & X \end{smallmatrix}  \right]\left[ \begin{smallmatrix} I & -I \\ 0 & I \end{smallmatrix}  \right] = \left[ \begin{smallmatrix} X+Y & 0 \\ Y & X-Y \end{smallmatrix}  \right]$.} that applies for any two square matrices $X$ and $Y$: 
\be
   \det\left( \begin{matrix} X & Y \\ Y & X \end{matrix}  \right) = \det(X - Y) \det(X+Y) \,.
\ee
This leads to
\bea  \label{purity1}
\mathrm{Tr}[\sigma_{\ssS}^2] &= & \cN(t)^2 \left\{ \det\left[ \frac{1}{\pi}   \Bigl(\mathrm{Re}[ \cA_1(t) ] - \mathrm{Re}[ \cB(t) ] \Bigr) \right]   \det\left[ \frac{1}{\pi} \Bigl( \mathrm{Re}[ \cA_1(t) ] + \mathrm{Re}[ \cB(t) ] \Bigr) \right] \right\}^{-1/2} \\
&= & \frac{ \det\left[ \frac{1}{\pi} \Bigl( \mathrm{Re}[\cA_1] + \cB(t) \Bigr) \right] }{ \sqrt{ \det\left[ \frac{1}{\pi} \Bigl(\mathrm{Re}[ \cA_1(t) ] - \mathrm{Re}[ \cB(t) ] \Bigr) \right]   \det\left[ \frac{1}{\pi} \Bigl(\mathrm{Re}[ \cA_1(t) ] + \mathrm{Re}[ \cB(t) ] \Bigr) \right] }}
\eea
which uses 
\be
   \cN(t)^2 = \det\left[  \frac{1}{\pi} \Bigl( \mathrm{Re}[\cA_1] + \cB(t) \Bigr) \right] = \det  \cM(t)   \,.
\ee
Notice that this last relation verifies that $\cB(\bfx,\bfy) = 0$ implies $\mathrm{Tr}[\sigma_{\ssS}^2] = 1$, so $\cB =0$ is sufficient to ensure that $\sigma_{\ssS}$ describes a pure state.

These expressions can be made even more explicit by switching to momentum space and evaluating them using the above perturbative solution near the free vacuum. In this case repeated use of the identity $\det  = \exp\, \mathrm{Tr} \log$ leads to
\bea \label{PurityFalling}
\mathrm{Tr}[\sigma_{\ssS}(t)^2] 
& = & \exp\left[ \int \exd^3 \bfk \; \frac{\pi}{|\bfk|} \cdot \frac{2\tilde g^2 \mfb(\bfk,\bfk;t)}{(2\pi)^4} + \cO(\tilde g^4) \right] \nn\\
& = & \exp\left[ - \frac{\tilde g^2 t}{16\pi^4 \beta} \int \frac{\exd^3 \bfk}{|\bfk|}+ \cO(\tilde g^4) \right] \\
& = & \exp\left[ - \frac{\tilde g^2 \Lambda^2 t}{8\pi^3 \beta} + \cO(\tilde g^4) \right] \,,\nn
\eea
revealing the divergence described above. These ultimately arise because the Markovian derivation required $\Lambda \ll 1/\beta$. Presumably the same calculation would not have diverged if $\Lambda$ could have been taken larger than the temperature, though we have not succeeded yet in capturing this evolution in a fuller  calculation.

Notice that the purity starts out at $1$ since the initial vacuum state is pure, and then drops monotonically as time passes. A strictly perturbative calculation would only have been able to capture the leading contribution in powers of $\tilde g^2$, and so would have given  
\be
\mathrm{Tr}[\sigma_{\ssS}(t)^2] \simeq 1  - \frac{\tilde g^2 \Lambda^2 t}{8\pi^3 \beta} + \cO(\tilde g^4)  \qquad \hbox{(perturbative)} \,,
\ee
with the linear secular growth with $t$ eventually causing the perturbative calculation to break down. It is the deviation of Nakajima-Zwanzig equation from straight-up perturbation theory that allows the resummation of the secularly growing terms to all orders in $\tilde g^2 t$ into the exponential form visible in \pref{PurityFalling}, along the lines also seen in simpler examples (such as in \cite{Kaplanek:2019dqu, Kaplanek:2019vzj, Kaplanek:2020iay, Burgess:2015ajz}). This allows us to see that it approaches $0$ for late times, corresponding to a maximally mixed state (in $D$-dimensional quantum mechanics, a maximally mixed state has purity $1/D$ and in the present instance $D \to \infty$). 

\section{Mean-field methods}
\label{sec:meanfieldH}

One of our goals is to explore the nature of nonlocality in open systems, in hopes that this can shed light on whether nonlocality can also arise near horizons for black holes. Nonlocality in this context traditionally means the extent to which the effective action (or Hamiltonian) is not simply the integral of a Hamiltonian density that depends only on fields and their derivatives at a single point (a property normally held by Wilsonian actions as a consequence of cluster decomposition and microcausality \cite{Weinberg:1995mt}).

Locality is a question for which the language of open quantum systems used above is less well adapted because it frames its predictions in terms of the evolution of the reduced density matrix that is obtained after tracing out any dependence on unmeasured sectors of the Hilbert space, as given by the Nakajima-Zwanzig equation \cite{Nak, Zwan}. Although entanglement and information exchange ensure this equation naturally involves nonlocality (particularly in time), the absence of an effective Hamiltonian or action in this approach makes its lessons relatively obscure.

To help understand the issues that are at play this section sets up the mean-field approximation for the hotspot system, since the mean-field limit provides a natural definition of an effective hamiltonian (and effective action) that can provide a good approximation to the system's evolution in some circumstances even for open systems \cite{EFTBook}. Part of the purpose is to identify the kinds of nonlocality that can emerge for the mean-field Hamiltonian, but at the same time also to identify precisely when the mean-field description is a good approximation to the full dynamics. The hotspot provides a relatively simple test laboratory for exploring these ideas.
 
\subsection{Definitions}
\label{sec:Definitions}

The essence of the mean-field approximation is that averages in the unmeasured (or `environment') part of the open system dominate the fluctuations in this sector, allowing an informative expansion in powers of small deviations from the mean. There is some freedom in how to set up this expansion when working beyond the leading order, which we first summarize before computing how things look specifically in the hotspot example. 

We use, as before, the `double-bracket' notation 
\be
    \avg{ \cO } := \TrB \Bigl[\rho_\beta \, \cO \Bigr] \quad \hbox{where} \quad
    \rho_\beta := \cI_+ \otimes \varrho_\beta
\ee
with $\varrho_\beta$ being the thermal state appearing in \pref{initialstate+-} and the partial trace running only over the environmental sector of the Hilbert space (which in the hotspot example is the sector spanned at the initial time by the $\chi^a$ fields). For example, for $\cO = \sum_n \cA_n \otimes \cB_n$ expanded in terms of a basis of operators acting in the $\phi$ and $\chi^a$ sectors,
\be
  \avg{\cO} = \sum_n \cA_n\; \TrB\Bigl[\varrho_\beta \, \cB_n \Bigr] = \sum_n \cA_n \, \bigl\langle \cB_n \bigr\rangle_\beta \,.
\ee
In particular, $\avg{\cO}$ is an operator that acts only in the measured sector ({\it i.e.}~the $\phi$ sector of the hotspot example). 

\subsubsection{Mean-field evolution}

For a perturbative analysis it is worth specializing to the interaction picture, for which expectation values for observables evolve according to
\be
  \cA(t) := \mathrm{Tr} \Bigl[ \rho_\ssS(t) \,\cO_\ssS \Bigr] = \Tr \Bigl[ \rho_\ssI(t) \cO_\ssI(t) \Bigr] = \Tr \Bigl[ \rho_\ssI(0) V^\star(t) \cO_\ssI(t) V(t) \Bigr]\,,
\ee
where $V(t) = U_0^\star(t) U(t)$ is the interaction-picture evolution operator for the state $\rho_\ssI(t)$. In what follows we drop the subscript `$I$' for interaction-picture quantities.

For the observables that only measure the $\phi$ sector the Schr\"odinger picture operator has the factorized form $\cO_\ssS = \cO_{\ssS +} \otimes \cI_-$. This factorization remains true in the interaction picture provided that the free part of the Hamiltonian does not couple the two sectors to one another. That is, if $H = H_0 + H_{{\rm int}}$ where $H_0 = \cH_+ \otimes \cI_- + \cI_+ \otimes \cH_-$, then
\be \label{freeevo}
  U_0(t) := \exp \Bigl[ -i H_0 t \Bigr] = U_+(t) \otimes U_-(t) \,,
\ee
and so in the interaction picture
\be \label{IntPicObs}
   \cO(t) := U_0^\star(t) \, \cO_\ssS \, U_0(t) = \cO_+(t) \otimes \cI_-
\ee
with $\cO_+(t) = U_+^\star(t) \, \cO_{\ssS +} \, U_+(t)$.

Now comes the main point:  in the mean-field approximation the interaction-picture evolution operator is well-approximated within the observable sector by the operator
\bea \label{VavgDef}
  \ol V(t) &:=& \avg{V(t)} \\
  &=& I - i \int_0^t \exd \tau \, \avg{ H_{\rm int}(\tau) } + \frac12(-i)^2 \int_0^t \exd \tau_1 \int_0^{\tau_1} \exd \tau_2  \, \avg{H_{\rm int}(\tau_1) \, H_{\rm int}(\tau_2)} + \cdots \,,\nn
\eea
where $H_{\rm int}(t)$ denotes the usual interaction-picture interaction Hamiltonian $H_{\rm int}(t) = U^\star_0(t) H_{\rm int} U_0(t)$. Writing the full evolution operator as a mean-field part plus the rest (with the rest here called the `diffuse' evolution, in analogy to optics)
\be \label{VbarVcV}
   V(t) =: \ol V(t) \otimes \cI_-  + \cV(t) \,,
\ee
the mean-field approximation is a good one when contributions of the diffuse evolution operator, $\cV(t)$ are parametrically small. 

The point of defining the mean-field evolution using $\ol V(t)$ (as opposed to simply averaging the Hamiltonian, say) is that this ensures that for any observable of the form \pref{IntPicObs} there is no cross-interference -- to all orders in perturbation theory -- between the mean field evolution and diffuse evolution. That is, computing the expectation of $\cO(t)$ using \pref{VbarVcV} gives
\be \label{DecoupVbarVdiff}
  \mathrm{Tr} \Bigl[ \rho(t) \,\cO(t) \Bigr] = \mathrm{Tr} \Bigl[ V(t) \rho(0) V^\star(t) \,\cO(t) \Bigr] 
  = \Tr_+ \Bigl[ \ol V(t) \rho_+ \ol V^\star(t)\, \cO_{ +}(t) \Bigr] + \Tr \Bigl[ \cV(t) \rho(0) \, \cV^\star(t) \cO(t) \Bigr] \,,
\ee
with the first trace being only over the measured ($\phi$) sector, and we assume the uncorrelated initial conditions \pref{initialstate} and \pref{initialstate+-}. Notice the absence of any cross terms involving both $\ol V$ and $\cV$, which vanish because \pref{VavgDef} and \pref{VbarVcV} together imply $\avg{\cV(t)} = 0$.

\subsubsection{Mean-field Hamiltonian}

Given the mean-field evolution, the mean-field interaction Hamiltonian, $\ol H_{\rm int}(t)$, is defined as the operator that generates $\ol{V}(t)$, or equivalently is related to $\ol V(t)$ by the usual iterative expression
\be
  \ol V(t) = I - i \int_0^t \exd \tau \, \ol H_{\rm int}(\tau)  + \frac12(-i)^2 \int_0^t \exd \tau_1 \int_0^{\tau_1} \exd \tau_2  \, \ol H_{\rm int}(\tau_1) \, \ol H_{\rm int}(\tau_2) + \cdots \,.
\ee
Comparing this with \pref{VavgDef} we read off
\be\label{MFHam}
  \ol H_{\rm int} (t)= \avg{H_{\rm int}(t) } -i \int_0^t \exd \tau \; \avg{ \delta H_{\rm int}(t) \, \delta H_{\rm int}(\tau) } + \cdots \,,
\ee
where the ellipses represent terms at least third order in $\delta H_{\rm int}$ and 
\be
  \delta H_{\rm int}(t) := H_{\rm int}(t) - \avg{ H_{\rm int}(t) } \,.
\ee

Notice that starting at second order in $\delta H_{\rm int}$ the mean-field Hamiltonian need not be hermitian, since $\ol H_{\rm int}(t) = \mfH(t) + i \,\mfI(t)$ with
\bea
   \mfH(t) &=&  \avg{H_{\rm int}(t) } - \frac{i}{2} \int_0^t \exd \tau \; \avg{ \bigl[ \delta H_{\rm int}(t) \,, \delta H_{\rm int}(\tau) \bigr]} + \cdots \nn\\
 \hbox{and} \quad
   \mfI(r) &=&  - \frac{1}{2} \int_0^t \exd \tau \; \avg{ \bigl\{ \delta H_{\rm int}(t) \,, \delta H_{\rm int}(\tau) \bigr\} } + \cdots\,.
\eea
The failure of unitarity for $\ol V(t)$ that arises in this way merely reflects the relatively artificial nature of the mean-field/diffuse split, since `diffuse' interactions with fluctuations in the environment can deplete probability from the mean-field description. 

These diffuse interactions are themselves described perturbatively by 
\be \label{cVdefPtbn}
   \cV(t) \simeq -i \int_0^t \exd \tau \; \delta H_{\rm int}(\tau) + \cdots \,,
\ee
where ellipses now represent terms second-order in $H_{\rm int}$. (Recall that knowing $\cV$ at linear order suffices to compute observables to second order because expressions like \pref{DecoupVbarVdiff} depend quadratically on $\cV$.) It is only the total evolution (with mean-field and diffuse contributions combined) that must be unitary, and this can be expressed as a generalization of the optical theorem, relating the imaginary part of $\ol H_{\rm int}$ to the rate of diffuse scattering \cite{EFTBook}. This makes the relative size of the imaginary and real parts of $\ol H_{\rm int}$ a proxy for the relative importance of diffuse and mean-field evolution.

\subsubsection{Domain of validity}

Broadly speaking, mean-field descriptions arise as good approximations for real systems in two common ways, depending on whether or not $\avg{H_{\rm int}}$ is zero. The simplest case is when $\avg{H_{\rm int}} \neq 0$, because then perturbation theory alone can justify the mean-field approximation. This can be seen because the leading (linear) contribution in powers of $H_{\rm int}$ is necessarily a mean-field result (because $\cV$ first enters expressions like \pref{DecoupVbarVdiff} at second order). Neutrino interactions with matter inside the Sun or Earth provide practical examples of this type, for which the mean-field description follows as a consequence of the extreme feebleness of the weak interactions \cite{Burgess:1996mz, Bamert:1997jj}.

Things are more subtle when $\avg{H_{\rm int}} = 0$, however, because then the leading contribution of $\ol V(t)$  to eq.~\pref{DecoupVbarVdiff} arises at the same order in $\delta H_{\rm int}$ as does the diffuse contribution $\cV(t)$. This is typically what happens for the interactions of photons with transparent dielectric materials, for example, and in this case perturbation theory in $\delta H_{\rm int}$ itself is insufficient for mean-field methods to dominate.\footnote{For photons it is instead a large-$N$ argument based on coherence that justifies mean-field methods (see {\it e.g.}~\cite{EFTBook}).} This is also the regime appropriate to the hotspot, and in what follows we identify possible control parameters for mean-field methods by comparing for the hotspot the relative size of the real and imaginary parts of $\ol H_{\rm int}$.

\subsection{Application to the hotspot}

The above is made concrete by specializing to the specific hotspot interactions of the previous sections. Consider first the case of non-negligible hotspot size, $\xi$, where the underlying coupling has the form given in \pref{Hint(t)}:
\be
   H_{\rm int}(t) = \int_{S_\xi} \exd^2 x \; G_a \, \chi^a(t, \bfx) \, \phi(t, \bfx ) \,,
\ee 
with $S_\xi$ the 2-sphere of radius $\xi$ centred on the origin. For the purposes of the present discusion the coupling $G_\phi \, \phi^2$ can be regarded as part of the unperturbed Hamiltonian because it does not couple the $\phi$ and $\chi^a$ sectors to one another.

\subsubsection{Nonlocal mean-field Hamiltonian}

Since $\langle \chi^a \rangle = 0$ in the thermal bath the mean of the interaction Hamiltonian vanishes, $\avg{H_{\rm int}(t) } = 0$, and so the leading term in \pref{MFHam} is the second-order contribution, giving 
\be \label{MFHeffHS}
  \ol H_{\rm int}(t) \simeq  -i \widetilde G^2 \int_0^t \exd \tau \int_{S_\xi}\exd^2x\int_{S_\xi} \exd^2x' \;  \mathscr{W}_\beta(t, \bfx; \tau, \bfx') \, \phi(t,\bfx) \, \phi(\tau,\bfx')   \,,
\ee
where (as before) $\avg{ \chi^a(t,\bfx) \, \chi^b(\tau,\bfx') } =  \delta^{ab} \, \mathscr{W}_\beta(t, \bfx; \tau, \bfx')$ and $\widetilde G^2 := G_a G_b \, \delta^{ab} = N G^2$ is the coupling after summing over environmental fields. Evaluating the correlator using \pref{thermalcorrelatorab} gives the explicit form
\bea \label{ThermalCorrelator}
\mathscr{W}_\beta(t, \bfx; \tau, \bfx') &=&  \frac{1}{8 \pi \beta |\mathbf{x} - \mathbf{x}'|}\left\{ \coth\left[  \dfrac{\pi}{\beta} \left( t - \tau + |\mathbf{x} - \mathbf{x}'| - i \delta \right) \right] \right. \\
&& \qquad\qquad \qquad\qquad \qquad \left. - \coth\left[ \dfrac{\pi}{\beta} \left( t - \tau - |\mathbf{x} - \mathbf{x}'| - i \delta \right) \right] \right\} \,, \nn
\eea

Expression \pref{MFHeffHS}  for the mean-field interaction Hamiltonian is explicitly nonlocal, in two different ways. First, it is nonlocal in space because the correlator $\langle \chi^a(t,\bfx) \, \chi^b(t',\bfx') \rangle$ has support for arbitrary pairs of points $\bfx$ and $\bfx'$ on the localized interaction region $\cS_\xi$. No interactions at all (local or nonlocal) arise in $\ol{H}_{\rm int}(t)$ away from $\cS_\xi$ as a consequence of the absence of $\chi^a$ fields anywhere in $\cR_+$ away from the interaction surface. This does not preclude the field $\phi$ from acquiring nontrivial autocorrelations away from $\cS_\xi$ in response to these interactions, however, such as those seen in \pref{Wpertfull}. The spatial nonlocality has a relatively simple form in the limit where the times of interest are more widely separated than the light-crossing time for the hotspot itself: $|t - \tau|\geq |\bfx - \bfx'|  \gg 2\xi $. In this case eq.~\pref{ThermalCorrelator} shows that the effective interaction \pref{MFHeffHS} becomes
\be \label{MFHeffHS2}
  \ol H_{\rm int}(t) \simeq  \frac{i \tilde g^2}{4\beta^2} \int_0^t \exd \tau   \;  \hat\phi_{\ell = 0}(t,\xi) \, \hat\phi_{\ell =0}(\tau,\xi) \; \csch^2\left[ \dfrac{\pi}{\beta} \left( t - \tau - i \delta \right) \right]    \,,
\ee
which uses the relation \pref{gavsGa} ({\it i.e.}~$g_a = 4\pi \xi^2 G_a$) and defines the projector onto the $\ell = 0$ spherical harmonic of the field $\phi(\bfx,t)$
\be
   \hat\phi_{\ell = 0}(t,\xi) := \frac{1}{4\pi} \int_0^{4\pi} \exd^2\Omega \; \phi(t,r=\xi,\theta,\varphi) \,,
\ee
with the integration $\exd^2\Omega$ being over $4\pi$ solid angle. This form, local in angular-momentum space (and so nonlocal in position space), is also seen in other applications \cite{PPEFTDis}.

The second source of nonlocality is in time, although this dies off exponentially quickly once $\pi |t-\tau| \gg \beta$. This nonlocality is only consistent with the approximation that led to \pref{MFHeffHS2} if $\beta \gg \xi$ since otherwise it is impossible to satisfy both $|t - \tau| \lsim \beta/\pi$ and $|t-\tau| \gg 2\xi$. It should be noticed in this context that in the black hole analogy $\beta = 4\pi \xi$ and so choosing $|t-\tau| \gg 2\xi$ would also imply $|t-\tau| \gg \beta/\pi$. 

Both of these sources of nonlocality have their roots in the fluctutation of $\chi^a$ and so exist only in regions where the $\chi^a$'s have support. This is why all of the nonlocality mentioned above is restricted to the world-tube swept out by the interaction surface $\cS_\xi$, from the point of view of an external observer in $\cR_+$.  

The leading contribution to the diffuse evolution describing deviations from the mean-field limit is given by \pref{cVdefPtbn}, which (because $\avg{H_{\rm int}} = 0$ in the present instance), becomes
\be \label{cVdefPtbn2}
   \cV(t) \simeq -i \int_0^t \exd \tau \;  H_{\rm int}(\tau) = -i \int_0^t \exd \tau \; \int_{S_\xi} \exd^2 x \; G_a \, \chi^a(\tau,\bfx) \, \phi(\tau,\bfx) \,.
\ee
As mentioned above, second-order contributions from this generically compete with first-order contributions from \pref{MFHeffHS} since both arise at second order in $H_{\rm int}$.

\subsubsection{The local limit}

The above mean-field interaction simplifies and becomes approximately local in the special case that $\xi$ and $\beta$ are both microscopic scales, showing how locality ultimately re-emerges in the long-wavelength limit. To examine this limit explicitly we use small $\xi$ to integrate out the hotspot's spatial size, leading to the effective point-like coupling given in \pref{Hint(t)pt}. Using this to compute the mean-field Hamiltonian then gives 
\be \label{HmLocal}
{\ol{H}}_{\mathrm{int}}(t) = \frac{\lambda}{2} \, {\phi}^2(t,\mathbf{0}) - i \tilde g^2 {\phi}(t,\mathbf{0}) \int_0^t\; \exd s\; \mathscr{W}_\beta(s)\, {\phi}(t-s,\mathbf{0})\,,
\ee
where (as before) $\langle \chi^a(t,\mathbf{0}) \, \chi^b(t-s,\mathbf{0}) \rangle = \delta^{ab} \, \mathscr{W}_\beta(s)$, with
\be \label{Wbeta00}
   \mathscr{W}_\beta(s) := - \frac{1}{4\beta^2}\; \csch^2\left[ \frac{\pi}{\beta} (s - i \delta) \right] \,.
\ee

Two simplifications follow from the observation that $\mathscr{W}_\beta(s)$ is peaked exponentially sharply around $s = 0$, with width of order $\beta/\pi$. First, the upper integration limit can be taken to infinity at the expense of errors that are $\sim e^{-2 \pi t/\beta}$ and so are exponentially small in the regime $t \gg \beta/\pi$. Second, for fields varying on scales long compared with $\beta$ we can expand $\phi(t-s,\mathbf{0}) \simeq \phi(t,\mathbf{0}) - s \, \partial_t \phi(t,\mathbf{0}) + \cdots$ inside the integrand to get
\be
{\ol{H}}_{\mathrm{int}}(t) = \frac{\lambda}{2} \, {\phi}^2(t,\mathbf{0}) + \cA \,  {\phi}^2(t,\mathbf{0}) + \cB \, {\phi} \, \partial_t  {\phi}(t,\mathbf{0}) + \ldots \,,
\ee
where  
\be
 \cA  = - i\tilde g^2 \int_0^\infty\; \exd s\; \mathscr{W}_\beta(s) \quad\hbox{and} \quad
\cB = i \tilde g^2\int_0^\infty\; \exd s\; s\, \mathscr{W}_\beta(s)  \,,
\ee
and so on.

These reveal the coefficient $\cA$ to be a renormalization of the effective coupling $\lambda$, while the coefficient $\cB$ multiplies a new effective interaction proportional to $\phi \, \partial_t \phi$. This last interaction plays no role in the physics to follow because (depending on the operator ordering) it either involves the commutator of $\phi$ with its canonical momentum (and so is a divergent contribution to an irrelevant field-independent piece in $\ol H_{\rm int}$) or it involves a total time derivative, $\partial_t \phi^2$ (and so can be eliminated using an appropriate canonical transformation).  

The integrals giving the coefficients $\cA$ and $\cB$ can be evaluated explicitly to give
\be \label{cAcBresults}
 \cA  = - \tilde g^2 \left[\frac{1}{4\pi^2\delta}+  \frac{i}{4\pi\beta}  \right] + \cO(\delta) \quad \hbox{and}\quad
\cB  = i \tilde g^2 \left[ \frac{1}{4\pi^2} \log\left( \frac{2 \pi \delta}{\beta} \right) - \frac{i}{8 \pi} \right] + \cO(\delta)\,, 
\ee
where the infinitesimal $\delta$ is meant to be taken to zero. The divergences arise because $\mathscr{W}_\beta \sim s^{-2}$ as $s \to 0$, and so can be regarded as being ultraviolet in origin. They arise here as divergences when $\delta \to 0$, which just means that this infinitesimal -- which was introduced for other reasons -- is playing double duty; providing here also a near-hotpsot regularization for the singular integration (which could indeed have been regulated in other ways). To the extent that they contribute to non-redundant interactions these divergences can be renormalized into effective couplings, such as $\lambda$, thereby underlining that such self-couplings are generically always present in the effective theory. 

What is {\it not} simply a small change to the effective description, even in the local limit, is the generation of an imaginary part of $\cA$ seen in \pref{cAcBresults}. This cannot be absorbed into $\lambda$ without changing the reality properties of $\lambda$, and its appearance is a manifestation of the general probability loss away from the mean-field sector into its `diffuse' complement. Since the relative size of $\tilde g^2/\beta$ to $\lambda$ provides a measure of the relative importance of the real and imaginary parts of this coupling we should also only expect the mean-field description to be a good approximation when $\lambda \gg \tilde g^2/\beta$. The above discussion also suggests that if the $\phi$ self-coupling $\lambda$ at the hotspot is ultimately induced by the microscopic $\phi$--$\chi$ coupling, then its natural size is $\lambda \sim \tilde g^2/\xi$. If true, this would suggest the mean-field limit should appear to work best in the regime $\xi \ll \beta$. 

A more explicit expression for the real part of $\ol H_{\rm int}$ is given by
\bea
\mfH(t) & := & \frac12 \Bigl[ \ol H_{\rm int}(t) + \ol H_{\rm int}^{\star}(t) \Bigr] \nn \\
& \simeq & \frac{\lambda}{2} \, {\phi}^2(t,\mathbf{0}) - \frac{i \tilde g^2}{2} \int_0^t \exd s  \; \Bigl(  \mathscr{W}_\beta( s ) \,  \phi(t,\mathbf{0})  \, \phi( t-s, \mathbf{0}) - \mathscr{W}^{\ast}_\beta( s ) \, \phi( t-s, \mathbf{0}) \, \phi(t,\mathbf{0})  \Bigr) \nn\\
& = & \frac{\lambda}{2} \, {\phi}^2(t,\mathbf{0}) - \frac{i \tilde g^2}{2} \int_0^t \exd s \; \mathrm{Re}[\mathscr{W}_{\beta}(s)] \; \Bigl[ \phi(t,\mathbf{0})  \,, \phi( t-s, \mathbf{0})  \Bigr] \\
&& \qquad\qquad\qquad\qquad\qquad + \frac{ \tilde g^2}{2} \int_0^t \exd s \; \mathrm{Im}[\mathscr{W}_{\beta}(s)] \; \Bigl\{ \phi(t,\mathbf{0})  \,, \phi( t-s, \mathbf{0})  \Bigr\} \nn \\
& = & \frac{\lambda}{2} \, {\phi}^2(t,\mathbf{0}) + \frac{ \tilde g^2}{4\pi} \cI_{+} \int_0^t \exd s \; \mathrm{Re}[\mathscr{W}_{\beta}(s)] \; \delta'(s) + \frac{ \tilde g^2}{8\pi} \int_0^t \exd s \; \delta'(s) \; \Bigl\{ \phi(t,\mathbf{0})  \,, \phi( t-s, \mathbf{0})  \Bigr\} \nn 
\eea
where the last equality uses the free-field commutator,  $[ \phi(t,\bfx) \,, \phi(t',\bfx) ] = i \delta'(t-t') \cI_+ /(2\pi)$, computed in \pref{CCRneqtimeeqpos}, and that the imaginary part of the free thermal Wightman function is $\mathrm{Im}[\mathscr{W}_{\beta}(t)] = \delta'(t) / (4\pi)$, as computed in (\ref{ImWbeta}). 

Integrating by parts, the above formula becomes (noting that the $\delta(t)$ factors vanish for $t>0$)
\bea
\mfH(t) & = & \frac{\lambda}{2} \, {\phi}^2(t,\mathbf{0}) + \frac{ \tilde g^2}{4\pi}\, \cI_{+} \bigg( - \mathrm{Re}[\mathscr{W}_{\beta}(0)] \; \delta'(0) - \mathrm{Re}[\mathscr{W}'_{\beta}(0)] \bigg) \\
& \ & \qquad \qquad \qquad \qquad - \frac{ \tilde g^2}{8\pi} \bigg(  \delta(0) \cdot 2 \phi^2(t,\mathbf{0}) + \phi(t,\mathbf{0}) \partial_t \phi(t,\mathbf{0}) + \partial_t  \phi(t,\mathbf{0}) \phi(t,\mathbf{0}) \bigg) \,.\nn
\eea
For the present purposes we may drop any terms that are proportional to $\cI_{+}$, since these do not contribute to the dynamics because they drop out of commutators with fields. We can also (as always) omit redundant operators like $\phi \, \partial_t \phi$ --- see the logic given below equation (\ref{cAcBresults}) --- and after doing so we have
\be \label{realMFH}
\mfH(t) \simeq \frac{1}{2} \left[ \lambda - \frac{\tilde{g}^2}{2\pi} \, \delta(0) \right] \, {\phi}^2(t,\mathbf{0}) \,,
\ee
showing once more how the real part of the mean-field Hamiltonian serves to renormalize\footnote{Note that the shift shown in \pref{realMFH} matches the shift in (\ref{cAcBresults}), when one interprets $\delta(0) = 1/(\pi\delta)$ (which follows from writing $\frac{1}{x- i\delta}  = 1/x + i \pi \delta(x)$.} the self-interaction parameter $\lambda$, with
\be \label{lambdaRdef}
\lambda \; \to \; \lambda_{\ssR} := \lambda - \frac{\tilde{g}^2}{2\pi} \, \delta(0) \,.
\ee
As usual we henceforth suppress the subscript `$R$'. 

The imaginary part of $H_{\rm int}(t)$ is similarly given by
\bea \label{imaginaryMFH}
\mfI(t) & := & \frac{1}{2i} \Bigl[ \ol H_{\rm int}(t) - \ol H_{\rm int}^{\star}(t) \Bigr] \nn \\
& = & - \frac{\tilde g^2}{2} \int_0^t \exd s  \; \Bigl(  \mathscr{W}_\beta( t-s )  \,\phi(t,\mathbf{0}) \, \phi( s, \mathbf{0}) + \mathscr{W}^{\ast}_\beta( t-s ) \,\phi( s, \mathbf{0}) \,\phi(t,\mathbf{0})  \Bigr) \,.
\eea
Although this can also be written in terms of commutators and anti-commutators of $\phi$, it turns out to be less useful to do so. 
After renormalizing the self-interaction parameter using \pref{lambdaRdef} and combining terms, the complete mean-field Hamiltonian can therefore be written as
\bea  \label{HIsum}
\ol H_{\rm int}(t)  &=&  \mfH(t) + i\, \mfI(t)  \\
&=&  \frac{\lambda}{2} \phi^2(t,\mathbf{0}) - \frac{i \tilde g^2}{2} \int_0^t \exd s  \; \Bigl(  \mathscr{W}_\beta( t-s ) \, \phi(t,\mathbf{0}) \, \phi( s, \mathbf{0}) + \mathscr{W}^{\ast}_\beta( t-s ) \,\phi( s, \mathbf{0})\, \phi(t,\mathbf{0})  \Bigr)\ . \nn
\eea

\subsection{Mean-field $\phi$ correlation function}
\label{sec:MFcorinv}

The virtue of the hotspot model is that it can be solved exactly, making a comparison with mean-field predictions instructive about the latter's domain of validity. This comparison is most easily made using the $\langle \phi  \, \phi \rangle$ correlation function, since this has a known form \cite{Hotspot} --- given explicitly at late times by \pref{pertcorr} in the perturbative limit. To make this comparison we now evaluate the $\langle \phi  \, \phi \rangle$ two-point function within the mean-field limit. 

\subsubsection{Mean-field contribution}

Keeping in mind that the similarity transformation relating the Heisenberg and interaction pictures is $\phi_\ssH(t,\bfx) = V^{\star}(t) \, \phi(t,\bfx) \, V(t)$ where $V(t) = U_0^{-1}(t) \, U(t)$ (as before), the correlation function can be written
\be \label{HHvsIntPic}
\Tr\Bigl[ {\phi}_\ssH(t,\bfx) \,{\phi}_\ssH(t',\bfx') \rho_\ssH \Bigr] = \Tr\Bigl[ {V}^{\star}(t) \, {\phi}(t,\bfx) \, {V}(t) \, {V}^{\star}(t') \, {\phi}(t',\bfx') \, {V}(t') \rho(0)   \Bigr] \,,
\ee
in which we also use that the two pictures agree at the initial time, so $\rho_\ssH = \rho(0)$ given by (\ref{initialstate}). The mean-field result is obtained by using in this expression the approximate form
\be \label{VvsolV}
{V}(t) \simeq \ol{V}(t) \otimes \cI_-
\ee
with $\ol{V}(t)$ given by \pref{VavgDef}.

With this replacement -- and using the initial conditions \pref{initialstate} and \pref{initialstate+-} -- the mean-field correlation function reduces to an in-in expectation in the observed $\phi$ sector, of the form
\bea  \label{correlatorMF}
\TrAB\Bigl[ {\phi}_{\ssH}(t,\bfx) {\phi}_{\ssH}(t',\bfx') \rho_\ssH \Bigr]_\MF &:= & \TrA\bigg[ \Vm^\star(t){\phi}(t,\bfx) \Vm(t)  \Vm^\star(t') {\phi}(t',\bfx')  \Vm(t') \, \rho_+ \bigg] \nn \\
& = & \bra{ \vac } \Vm^\star(t){\phi}(t,\bfx) \Vm(t)  \Vm^\star(t') {\phi}(t',\bfx')  \Vm(t') \ket{\vac} 
\eea
where the interaction-picture state is evaluated at $t = 0$. Evaluating $\ol{H}_{\rm int}$ using \pref{HmLocal},  we have
\be  \label{MFVexp}
\Vm(t) \simeq \cI_+ - i \int_0^{t} \exd \tau \; {\ol {H}}_{\mathrm{int}}(\tau) \simeq  \cI_+ - i \int_0^t \exd \tau  \Bigl[ \mfH(\tau) + i\, \mfI(\tau) \Bigr] \,,
\ee
where real and imaginary parts of $\ol H_{\rm int}(t)$ are given by \pref{HIsum}. Inserting (\ref{MFVexp}) into (\ref{correlatorMF}) and working to leading nontrivial order in $\ol H_{\rm int}$ then yields the quantity to be evaluated:
\bea \label{MFcorr}
&&\TrAB\Bigl[ {\phi}_{\ssH}(t,\bfx) {\phi}_{\ssH}(t',\bfx') \rho_\ssH \Bigr]_\MF   \simeq    \bra{ \vac } {\phi}(t,\bfx) {\phi}(t',\bfx') \ket{\vac} \\
&& \qquad\qquad + i \int_0^{t} \exd \tau\; \bra{ \vac } \Bigl[ \mfH(\tau), \phi(t,\bfx) \Bigr] {\phi}(t',\bfx') \ket{\vac} + i \int_0^{t'} \exd \tau\; \bra{ \vac } {\phi}(t,\bfx) \Bigl[ \mfH(\tau) , {\phi}(t',\bfx') \Bigr] \ket{\vac} \nn \\
&&\qquad\qquad\qquad  + \int_0^{t} \exd \tau\; \bra{ \vac } \Bigl\{ \mfI(\tau), \phi(t,\bfx) \Bigr\} {\phi}(t',\bfx') \ket{\vac} + \int_0^{t'} \exd \tau\; \bra{ \vac } {\phi}(t,\bfx) \Bigl\{ \mfI(\tau) , {\phi}(t',\bfx') \Bigr\} \ket{\vac} \,.\nn
\eea

\subsubsection{Aside: $\ol V^\star$ vs $\ol V^{-1}$}

We pause the main line of development here to settle a side issue that might bother the reader at this point. The issue is this: equation \pref{MFcorr} is derived by substituting the mean-field expression \pref{VvsolV} into the general correlator definition \pref{HHvsIntPic}. For the full theory unitarity ensures $V^\star = V^{-1}$ but the same is {\it not} true for the mean-field limit, since we have seen $\ol V^\star \neq \ol V^{-1}$. So although nothing changes if we replace $V^\star \to V^{-1}$ in \pref{HHvsIntPic}, making the replacement \pref{VvsolV} in the result instead leads to 
\be  \label{correlatorMFwrong}
\cU(t,\bfx;t',\bfx') := \bra{ \vac } \Vm^{-1}(t)\, {\phi}(t,\bfx)\, \Vm(t)  \Vm^{-1}(t')\, {\phi}(t',\bfx') \, \Vm(t') \ket{\vac}
\ee
which differs from the right-hand side of \pref{correlatorMF}. Using \pref{correlatorMFwrong} would change \pref{MFcorr} by replacing the anticommutators $\{ \mfI , (\cdot) \}$ with commutators $[ \mfI , (\cdot) ]$ -- an important difference in practice because the commutator is much easier to evaluate (as we do for completeness in Appendix \ref{App:correlatorwrong}).   

Which is right? This is partially a matter of definition, since it hinges on how the full result gets spit into mean-field and diffuse parts. The guiding principle in \S\ref{sec:Definitions} is to make this split so that observables like \pref{DecoupVbarVdiff} break into a sum of mean-field and diffuse pieces, with no interference terms. The same principle tells us to define the mean-field correlator using \pref{correlatorMF} rather than \pref{correlatorMFwrong}. Specialized to $t = t'$ both equations have the same form as \pref{DecoupVbarVdiff}, and it is only for \pref{correlatorMF} that mean-field and diffuse parts cleanly split, because $V^\star = \ol V^\star + \cV^\star$ divides linearly while $V^{-1}$ does not.

\subsubsection{Equal-time Limits}

With the mean-field/diffuse split in mind, we next specialize the correlation function to equal times ($t = t'$), so that \pref{HHvsIntPic} agrees with (\ref{DecoupVbarVdiff}) with the choice $\cO(t) = \phi(t,\bfx) \phi(t,\bfx')$. As described above, this ensures the equal-time correlation function nicely splits into the sum of mean-field and diffuse parts
\be \label{totalsum}
\TrAB\Bigl[ {\phi}_{\ssH}(t,\bfx) {\phi}_{\ssH}(t,\bfx') \rho_\ssH \Bigr] = \TrAB\Bigl[ {\phi}_{\ssH}(t,\bfx) {\phi}_{\ssH}(t,\bfx') \rho_\ssH \Bigr]_{\MF} + \TrAB\Bigl[ {\phi}_{\ssH}(t,\bfx) {\phi}_{\ssH}(t,\bfx') \rho_\ssH \Bigr]_{\mathrm{diff}}
\ee
where we define the (equal-time) mean-field correlations as\footnote{Note that this is {\it not} the same as taking $t=t'$ in the formula (\ref{MFVexp}), since $\ol V(t) \ol V^{\star}(t) \neq \cI_{+}$.} 
\bea \label{MFequal1}
\TrAB\Bigl[ {\phi}_{\ssH}(t,\bfx) {\phi}_{\ssH}(t,\bfx') \rho_\ssH \Bigr]_\MF & := & \bra{ \vac } \ol V^{\star}(t) \phi(t,\bfx) \phi(t,\bfy) \ol V(t) \ket{\vac} \\
& = & \bra{ \vac } {\phi}(t,\bfx) \,{\phi}(t,\bfx') \ket{\vac} \ + i \int_0^{t} \exd \tau\; \bra{ \vac } \Bigl[ \mfH(\tau) , {\phi}(t,\bfx){\phi}(t,\bfx')  \Bigr]  \ket{\vac} \nn \\
&& \qquad  + \int_0^{t} \exd \tau\; \bra{ \vac } \Bigl\{ \mfI(\tau) , {\phi}(t,\bfx){\phi}(t,\bfx')  \Bigr\}  \ket{\vac} \nn
\eea
{\it c.f.} \pref{MFVexp}, while the diffuse part of the correlations is
\be \label{diffcorr}
\TrAB\Bigl[ {\phi}_{\ssH}(t,\bfx) {\phi}_{\ssH}(t,\bfx') \rho_\ssH \Bigr]_\mathrm{diff}  := \TrAB \Bigl[ \cV^{\star}(t) \phi(t,\bfx) \phi(t,\bfy)  \cV(t) \rho_\ssH \Bigr] \ .
\ee

\subsubsection{Equal-time mean-field correlation}

This section explicitly evaluates the equal-time mean-field correlation \pref{MFequal1}, ending with a final form that reduces the mode sums to a single integration over explicit elementary functions. 

The terms involving the real part $\mfH$ of the mean-field Hamiltonian involve commuators are simplest and so all integrals  can be evaluated explicitly. The quantity to be evaluated is
\be
i \int_0^{t} \exd \tau\; \bra{ \vac } \Bigl[ \mfH(\tau) , {\phi}(t,\bfx){\phi}(t,\bfx')  \Bigr]  \ket{\vac} = \frac{i\lambda}{2} \int_0^{t} \exd \tau\; \bra{ \vac } \Bigl[ \phi^2(\tau,\mathbf{0}), {\phi}(t,\bfx){\phi}(t,\bfx')  \Bigr]  \ket{\vac}
\ee
(which is why we could ignore terms in $\mfH(t)$ proportional to the $\cI_{+}$  in \pref{realMFH}). Using the single-field commutator
\be \label{commtaux}
\Bigl[ \phi(\tau,\mathbf{0}),\phi(t,\bfx) \Bigr] = \frac{i}{4\pi |\bfx|} \bigg( \delta\big[\tau - (t - |\bfx|) \big] -  \delta\big[\tau - (t + |\bfx|) \big] \bigg) \cI_{+} \ ,
\ee
evaluated in Appendix \ref{sec:freeequalcorrelator} allows this term to be written 
\be
\Bigl[ \phi^2(\tau,\mathbf{0}),\phi(t,\bfx) \Bigr] = \frac{i}{2\pi |\bfx|} \bigg( \delta\big[\tau - (t - |\bfx|) \big] -  \delta\big[ \tau - (t + |\bfx|) \big] \bigg) \phi(\tau,\mathbf{0}) \ ,
\ee
which in turn implies
\bea
\Bigl[ \phi^2(\tau,\mathbf{0}),\phi(t,\bfx)\phi(t,\bfx') \Bigr] & = & \Bigl[ \phi^2(\tau,\mathbf{0}),\phi(t,\bfx) \Bigr] \phi(t,\bfx') + \phi(t,\bfx) \Bigl[ \phi^2(\tau,\mathbf{0}),\phi(t,\bfx') \Bigr] \nn \\
& = & \frac{i}{2\pi |\bfx|} \bigg( \delta\big[\tau - (t - |\bfx|) \big] -  \delta\big[\tau - (t + |\bfx|) \big] \bigg) \phi(\tau,\mathbf{0}) \phi(t,\bfx') \\
&\ & \qquad \qquad + \frac{i}{2\pi |\bfx'|} \bigg( \delta\big[\tau - (t - |\bfx'|) \big] -  \delta\big[\tau - (t + |\bfx'|) \big] \bigg) \phi(t,\bfx) \phi(\tau,\mathbf{0}) \ . \nn
\eea

Only $\delta$-functions with singularities at the retarded times $t - |\bfx|$ and $t - |\bfx'|$ contribute in the regime of interest, so
\bea
&& i \int_0^{t} \exd \tau\; \bra{ \vac } \Bigl[ \mfH(\tau) , {\phi}(t,\bfx){\phi}(t,\bfx')  \Bigr]  \ket{\vac} \nn \\
&& \quad = - \frac{\lambda}{4\pi} \int_0^{t} \exd \tau\; \bigg( \frac{\delta\big[\tau - (t - |\bfx|) \big]}{|\bfx|} \bra{ \vac } \phi(\tau,\mathbf{0}) \phi(t,\bfx') \ket{\vac} + \frac{\delta\big[\tau - (t - |\bfx'|) \big]}{|\bfx'|} \bra{ \vac } \phi(t,\bfx) \phi(\tau,\mathbf{0}) \ket{\vac} \bigg) \nn \\
&& \quad = - \frac{\lambda }{4\pi |\bfx|}\, \Theta(t - |\bfx|) \bra{ \vac } \phi(t-|\bfx|,\mathbf{0}) \phi(t,\bfx') \ket{\vac} - \frac{\lambda}{4\pi |\bfx'|}\, \Theta(t - |\bfx'|) \bra{ \vac } \phi(t,\bfx) \phi(t-|\bfx'|,\mathbf{0})  \ket{\vac} \nn \\
&& \quad = - \frac{\lambda \Theta(t - |\bfx|)}{16\pi^3 |\bfx| \big( - ( - |\bfx| - i \delta )^2 + |\bfx'|^2 \big) } - \frac{\lambda \Theta(t - |\bfx'|)}{16\pi^3 |\bfx'| \big( - ( |\bfx'| - i \delta )^2 + |\bfx|^2 \big) }
\eea
In the limit that the transients have passed -- {\it i.e.}~ once $t - |\bfx| > 0$ and $t - |\bfx'| >0$ -- this simplifies to 
\bea
&& i \int_0^{t} \exd \tau\; \bra{ \vac } \Bigl[ \mfH(\tau) , {\phi}(t,\bfx){\phi}(t,\bfx')  \Bigr]  \ket{\vac} = - \frac{\lambda}{16\pi^3 |\bfx| |\bfx'| \big( |\bfx| + |\bfx'| ) }
\eea
which agrees with the $\lambda$-dependent part of equal-time correlation functions computed in \pref{largeNcorrteqs0} and \pref{eqtimecorrMarkovianLT}, (evaluated using the renormalized coupling $\lambda = \lambda_{\ssR}$). 

The more complicated contribution involves the imaginary part of $\ol H_{\rm int}$, in which we evalaute $\mfI$ using (\ref{imaginaryMFH}) to get
\bea \label{imanti}
&& \int_0^{t} \exd \tau\; \bra{ \vac } \Bigl\{ \mfI(\tau) , {\phi}(t,\bfx){\phi}(t,\bfx')  \Bigr\}  \ket{\vac} \\
&& \quad =  - \int_0^{t} \exd \tau \int_0^{\tau} \exd \tau'\; \bra{ \vac } \bigg\{ \sfrac{\tilde{g}^2 \Bigl(  \mathscr{W}_\beta( \tau-\tau' )  \phi(\tau,\mathbf{0})  \phi( \tau', \mathbf{0}) + \mathscr{W}^{\ast}_\beta( \tau - \tau' ) \phi( \tau', \mathbf{0}) \phi(\tau,\mathbf{0})  \Bigr)}{2} , {\phi}(t,\bfx){\phi}(t,\bfx') \bigg\}  \ket{\vac} \nn \\
&& \quad =  - \frac{\tilde g^2}{2} \int_0^{t} \exd \tau \int_0^{t} \exd \tau'\;  \mathscr{W}_\beta( \tau-\tau' ) \bra{ \vac } \Bigl\{  \phi(\tau,\mathbf{0})  \phi( \tau', \mathbf{0}) , {\phi}(t,\bfx){\phi}(t,\bfx')  \Bigr\}  \ket{\vac} \,,\nn
\eea
where the last equality changes integration variables $\tau \leftrightarrow \tau'$ in one of the two terms and uses the property $\mathscr{W}_{\beta}^{\ast}(t) = \mathscr{W}_{\beta}(-t)$ of the Wightman function. The formula (\ref{MFequal1}) therefore takes the final form
\bea \label{MFequal2}
\TrAB\Bigl[ {\phi}_{\ssH}(t,\bfx) {\phi}_{\ssH}(t,\bfx') \rho_\ssH \Bigr]_\MF & \simeq & \frac{1}{4\pi^2 |\bfx -\bfx'|^2} - \frac{\lambda}{16\pi^3 |\bfx| |\bfx'| \big( |\bfx| + |\bfx'| ) } \\
&& \quad  - \frac{\tilde g^2}{2} \int_0^{t} \exd \tau \int_0^{t} \exd \tau'\;  \mathscr{W}_\beta( \tau-\tau' ) \bra{ \vac } \Bigl\{  \phi(\tau,\mathbf{0})  \phi( \tau', \mathbf{0}) , {\phi}(t,\bfx){\phi}(t,\bfx')  \Bigr\}  \ket{\vac} \nn
\eea
in the regime $t-|\bfx| > 0$ and $t-|\bfx'|>0$ (after transients have passed from the switch-on of couplings at $t = |\bfx| = 0$).

The matrix element in this last expression is evaluated in Appendix \ref{App:eqtimeInts} with the result
\bea \label{MFequal3}
&&\TrAB\Bigl[ {\phi}_{\ssH}(t,\bfx) {\phi}_{\ssH}(t,\bfx') \rho_\ssH \Bigr]_\MF   \simeq   \frac{1}{4\pi^2 |\bfx -\bfx'|^2} \left[1  - \frac{\tilde g^2}{4 \pi^2} \left( \dfrac{\zeta(3) t}{\pi \beta^3} - {\displaystyle \int_0^\infty } \exd p \; p\; \dfrac{\exd \cD_{\beta}(p,\delta)}{\exd p} \right) \right]  \nn \\
&& \qquad\qquad - \frac{\lambda}{16\pi^3 |\bfx| |\bfx'| \big( |\bfx| + |\bfx'| ) }  - \frac{\tilde g^2}{16\pi^4 |\bfx| |\bfx'|} \int_0^\infty \exd p\; \sin(k|\bfx| )  \int_0^\infty \exd k\; \sin(k|\bfx'| ) \\
&& \qquad\qquad\qquad \times \left\{ \frac{\cC_{\beta}(p) + \cC_{\beta}(k)}{p+k} \sin\big[(p+k)t \big] - \frac{\cD_{\beta}(p,\delta) + \cD_{\beta}(k,\delta)}{p+k} \Bigl[ 1 + \cos\big[(p+k)t \big] \Bigr] \right\} \nn
\eea
where the Riemann-Zeta function evaluates to $\zeta(3) \simeq 1.202$ and the functions $\cC_{\beta}$ and $\cD_{\beta}$ are given by
\be
\cC_{\beta} = \frac{p}{4 \pi} \coth\left( \frac{ \beta p }{2}  \right) \quad \hbox{and} \quad
\cD_{\beta} = \frac{p}{2\pi^2} \log\left( \frac{2\pi e^{\gamma} \delta}{\beta} \right) + \frac{p}{2\pi^2} \mathrm{Re}\left[ \psi^{(0)}\left( - i \; \frac{\beta p}{2\pi} \right) \right] \ ,
\ee
where $\psi^{(0)}$ is the digamma function defined by $\psi^{(0)}(z) = \frac{\mathrm{d}}{\mathrm{d} z} \log \Gamma(z)$, and where $\delta$ is (as usual) to be taken to zero at the end (after renormalization). 

Although the $\tilde g^2$-independent terms in \pref{MFequal3} agree with the perturbative and Markovian results \pref{largeNcorrteqs0} and \pref{eqtimecorrMarkovianLT}, those that include $\tilde g^2$ do not. This difference is due to the contributions of the diffuse evolution first entering at this order, as we now show. 

\subsubsection{Including diffuse correlations}
\label{ssec:IncDiff}

The above calculation omits the diffuse correlations \pref{diffcorr}, and to the order we work it suffices to use the lowest-order expression \pref{cVdefPtbn} for $\cV$:
\be
  \cV(t) \simeq  -i \int_0^t \exd \tau \; \delta H_{\rm int}(\tau) = -i \int_0^t \exd \tau \; g_a \, \phi(\tau , \mathbf{0}) \otimes \chi^a(\tau, \mathbf{0}) \,.
\ee
Using this in \pref{diffcorr} then gives
\be \label{diffcorr2}
\Tr\Bigl[ {\phi}_\ssH(t,\bfx) {\phi}_\ssH(t,\bfx') \rho_\ssH \Bigr]_{\rm diff}  \simeq  \tilde g^2 \int_0^t \exd \tau  \int_0^t \exd \tau' \; \mathscr{W}_{\beta}(\tau - \tau') \; \bra{\vac} {\phi}(\tau,\mathbf{0}) {\phi}(t,\bfx) {\phi}(t,\bfx') {\phi}(\tau', \mathbf{0})  \ket{\vac}  \,.
\ee

We now show that adding \pref{diffcorr2} to the mean-field result  \pref{MFequal2} reproduces the perturbative expression for the full correlator given in \pref{largeNcorrteqs0}. Summing these mean-field and diffuse contributions gives the result
\be
\TrAB\Bigl[ {\phi}_{\ssH}(t,\bfx) {\phi}_{\ssH}(t,\bfx') \rho_\ssH \Bigr]  \simeq  \frac{1}{4\pi^2  |\bfx -\bfx'|^2 } - \frac{\lambda}{16\pi^3 |\bfx| |\bfx'| \big( |\bfx| + |\bfx'| ) }+ \cI_{\beta}(t,\bfx,\bfx') \label{totalsum2}
\ee
where the last term is given by the following combination of matrix elements
\bea
\cI_{\beta}(t,\bfx,\bfx') & := & - \frac{\tilde g^2}{2} \int_0^{t} \exd \tau \int_0^{t} \exd \tau'\;  \mathscr{W}_\beta( \tau-\tau' ) \bra{ \vac } \Bigl\{  \phi(\tau,\mathbf{0})  \phi( \tau', \mathbf{0}) , {\phi}(t,\bfx){\phi}(t,\bfx')  \Bigr\}  \ket{\vac} \\
&& \qquad \qquad + \tilde g^2 \int_0^t \exd \tau  \int_0^t \exd \tau' \; \mathscr{W}_{\beta}(\tau - \tau') \; \bra{\vac} {\phi}(\tau,\mathbf{0}) {\phi}(t,\bfx) {\phi}(t,\bfx') {\phi}(\tau', \mathbf{0})  \ket{\vac} \nn \ .
\eea

These can be usefully (but tediously) re-written as a double commutator plus a remainder,
\bea
\cI_{\beta}(t,\bfx,\bfx') & = & \cP_{\beta}(t,\bfx,\bfx') + \cQ_{\beta}(t,\bfx,\bfx')
\eea
where the double commutator is
\be \label{Nbetadef}
\cP_{\beta}(t,\bfx,\bfx') := - \tilde g^2 \int_0^{t} \exd \tau \int_0^{t} \exd \tau'\;  \mathscr{W}_\beta( \tau-\tau' ) \bra{ \vac } \Bigl[ \phi(\tau,\mathbf{0}) , {\phi}(t,\bfx) \Bigr] \Bigl[ \phi(\tau',\mathbf{0}) , {\phi}(t,\bfx') \Bigr] \ket{\vac}
\ee
while the remainder becomes
\bea  \label{Ebetadef}
\cQ_{\beta}(t,\bfx,\bfx') & : = & \frac{\tilde g^2}{2} \int_0^{t} \exd \tau \int_0^{t} \exd \tau'\;  \mathscr{W}_\beta( \tau-\tau' ) \bra{ \vac }  \bigg( \Bigl[ \phi(\tau,\mathbf{0}) , {\phi}(t,\bfx) \Bigr] \phi(\tau',\mathbf{0}) {\phi}(t,\bfx') \nn\\
&& \qquad - \phi(\tau,\mathbf{0}) \Bigl[ \phi(\tau',\mathbf{0}) , {\phi}(t,\bfx) \Bigr]  {\phi}(t,\bfx')  + {\phi}(t,\bfx) \Bigl[ \phi(\tau,\mathbf{0}) , {\phi}(t,\bfx') \Bigr] \phi(\tau',\mathbf{0}) \\
&& \qquad\qquad\qquad  - {\phi}(t,\bfx) \phi(\tau,\mathbf{0}) \Bigl[ \phi(\tau',\mathbf{0}) , {\phi}(t,\bfx') \Bigr] \bigg) \ket{\vac} \ .\nn
\eea

These integrals evaluate (see Appendix \ref{App:DiffuseInts}) in the regime $t - |\bfx| > 0$ and $t - |\bfx'| > 0$ to
\be \label{Nbetaanswer}
\cP_{\beta}(t,\bfx,\bfx') = - \frac{\tilde g^2}{64 \pi^2 \beta^2 |\bfx| |\bfx'| \sinh^2\left[ \frac{\pi}{\beta} ( - |\bfx| + |\bfx'| - i \delta ) \right] } \ , 
\ee
and
\be \label{Ebetaanswer}
\cQ_{\beta}(t,\bfx,\bfx') = \frac{\tilde g^2}{64 \pi^4 |\bfx| |\bfx'|} \bigg[ \frac{1}{( |\bfx| - |\bfx'| + i \delta )^2} - \frac{1}{( |\bfx| + |\bfx'| )^2} \bigg] \ .
\ee
When we use these in \pref{totalsum2} the overall correlation function is therefore
\bea  \label{totalsum3}
\TrAB\Bigl[ {\phi}_{\ssH}(t,\bfx) {\phi}_{\ssH}(t,\bfx') \rho_\ssH \Bigr]  & \simeq &  \frac{1}{4\pi^2  |\bfx -\bfx'|^2 } - \frac{\lambda}{16\pi^3 |\bfx| |\bfx'| \big( |\bfx| + |\bfx'| ) }  \\
&& \qquad \qquad - \frac{\tilde g^2}{64 \pi^2 \beta^2 |\bfx| |\bfx'| \sinh^2\left[ \frac{\pi}{\beta} ( - |\bfx| + |\bfx'| - i \delta )\right] }+ \frac{\tilde g^2}{16 \pi^4 } \;  \frac{1}{( |\bfx|^2 - |\bfx'|^2 )^2}    \ .\nn
\eea
This last expressions agrees perfectly with the perturbative Heisenberg-picture result computed in \cite{Hotspot}, once this is evaluated in the equal-time limit -- see eq.~\pref{largeNcorrteqs0} -- inside the future light-cone of the event at $t = |\bfx| = 0$ where the coupling switches on.

\subsubsection{Domain of validity of mean-field methods}

We see from these calculations that the mean-field correlator does {\it not} in general agree with the Heisenberg-picture result, even if this comparison is only made at leading order in $\lambda$ and $\tilde g^2$.  The comparison of the previous section shows that the difference between the mean-field and Heisenberg-picture answers is  precisely given by the diffuse contribution that must be small for mean-field methods to apply.

The difference between mean-field result \pref{MFequal3} and the corresponding Heisenberg-picture answer \pref{totalsum3} lies completely in their $\tilde g^2$ dependence; the term involving the self-coupling $\lambda$ is identical in both cases. Since $\tilde g$ and $\lambda$ both have dimensions of length, another scale must appearin the comparison of $\lambda$ and $\tilde g^2$, and the explicit evaluation -- {\it e.g.}~\pref{totalsum3} -- shows this scale to be either $\beta$ or a combination of $|\bfx|$ and $|\bfx'|$. 

Because our interest is typically where $|\bfx|$ and $|\bfx'|$ are much larger than $\beta$ the relative size of the $\lambda$-dependent term and the largest of $\tilde g^2$ corrections is set by the relative size of $\lambda$ and $\tilde g^2/\beta$, suggesting that mean-field methods provide a reliable approximation in the regime $\lambda \gg \tilde g^2/\beta$.  

In summary, we see that mean-field methods can apply to the hotspot problem, but only in some parts of parameters space such as when $\lambda \gg \tilde g^2/\beta$. Where it does apply the resulting effective Hamiltonian can be nonlocal, both in the angular directions of $\cS_\xi$ and in time, due to the nonlocality of the $\chi$ correlations with which the external $\phi$ field interacts.

\section*{Acknowledgements}
We thank Sarah Shandera for useful conversations, and KITP Santa Barbara for hosting the workshop (during a pandemic) that spawned this work. (Consequently this research was supported in part by the National Science Foundation under Grant No. NSF PHY-1748958.)  CB's research was partially supported by funds from the Natural Sciences and Engineering Research Council (NSERC) of Canada. Research at the Perimeter Institute is supported in part by the Government of Canada through NSERX and by the Province of Ontario through MRI.

\appendix

\section{Useful intermediate steps}

This appendix gathers together many intermediate steps not given in the main text, including the evaluation of several of the integrals encountered there. Our goal is to be as explicit as possible.

\subsection{Kernel equations for the gaussian ansatz}
\label{App:KernelPosSpace}

This section evaluations the implications of the master equation \pref{Markeq} for the gaussian kernels in the ansatz \pref{firstAnsatz}.

First we compute the time-derivative of the above ansatz
\bea
\frac{ \pd \sigma_{\ssS}( t, \varphi_1, \varphi_2 )}{\pd t} & = & \bigg[ \frac{\partial_t \cN(t)}{\cN(t)}  - \int \exd^3 \bfx \int \exd^3 \bfy\; \bigg\{ \frac{1}{2} \partial_t \mathcal{A}_1(\bfx,\bfy ; t ) \varphi_1(\bfx) \varphi_{1}(\bfy) \\
& \ & \quad \quad \quad \quad + \frac{1}{2} \partial_t \mathcal{A}_2(\bfx,\bfy ; t )  \varphi_2(\bfx) \varphi_{2}(\bfy) + \partial_t \cB(\bfx, \bfy ; t) \varphi_1(\bfx) \varphi_{2}(\bfy) \bigg\} \bigg]  \sigma_{\ssS}( t, \varphi_1, \varphi_2 ) \ . \notag
\eea
and we note the RHS of equation (\ref{Markeq})
\bea
\text{RHS of (\ref{Markeq})} & = & - \frac{i}{2} \int \exd^3 \bfx \bigg[  -  \frac{\delta^2}{\delta \varphi_1(\bfx)^2 } + \big| \boldsymbol{\nabla} \varphi_1(\bfx) \big|^2 + \frac{\delta^2}{\delta \varphi_2(\bfx)^2 } - \big| \boldsymbol{\nabla} \varphi_2(\bfx) \big|^2 \bigg] \sigma_{\ssS}( t, \varphi_1, \varphi_2 ) \\
 & \ & \qquad \qquad - \frac{\tilde g^2}{4\pi\beta}  \big(  \varphi_{1}(\mathbf{0}) - \varphi_{2}(\mathbf{0}) \big)^2 \sigma_{\ssS}( t, \varphi_1, \varphi_2 ) - \frac{i \lambda }{2} \big(  \varphi_{1}(\mathbf{0})^2 - \varphi_{2}(\mathbf{0})^2 \big) \sigma_{\ssS}( t, \varphi_1, \varphi_2 )  \ . \notag
\eea
We first need to compute the functional derivative
\be
\frac{\delta \sigma_{\ssS}[t, \varphi_1, \varphi_2]}{\delta \varphi_{1}(\bfx)}  =   \left( - \int d^{3} \bfy \; \big[ \cA_1(\bfx,\bfy;t) \varphi_1(\bfy) + \cB(\bfx, \bfy ; t) \varphi_{2}(\bfy) \big]  \right) \sigma_{\ssS}[t, \varphi_1, \varphi_2]
\ee
which assumes the symmetry $\cA_j(\bfx,\bfy;t) = \cA_j(\bfy,\bfx;t)$. From there we have
\bea
\frac{\delta^2 \sigma_{\ssS}[t, \varphi_1, \varphi_2]}{\delta \varphi_{1}(\bfx)^2} & = & \left( -  \cA_1(\bfx,\bfx;t) + \bigg\{ \int d^{3} \bfy \; \big[ \cA_1(\bfx,\bfy;t) \varphi_1(\bfy) + \cB(\bfx, \bfy ; t) \varphi_{2}(\bfy) \big] \bigg\}^2 \right) \sigma_{\ssS}[t, \varphi_1, \varphi_2] \nn  \\
& = & \bigg( -  \cA_1(\bfx,\bfx;t) + \int d^{3} \bfy  \int d^{3} \bfz \; \big[ \cA_1(\bfx,\bfy;t) \varphi_1(\bfy) + \cB(\bfx, \bfy ; t) \varphi_{2}(\bfy) \big] \label{var1der} \\ 
& \  & \qquad \qquad \qquad \qquad \qquad \qquad \qquad \times \big[ \cA_1(\bfx,\bfz;t) \varphi_1(\bfz) + \cB(\bfx, \bfz ; t) \varphi_{2}(\bfz) \big] \bigg) \sigma_{\ssS}[t, \varphi_1, \varphi_2] \ .\nn
\eea
Since $\cB$ is {\it not} symmetric, the other $\varphi_2$-derivative differs slightly from \pref{var1der} (note the variable being integrated in $\cB$ here) where
\bea
\frac{\delta^2 \sigma_{\ssS}[t, \varphi_1, \varphi_2]}{\delta \varphi_{2}(\bfx)^2}
& = & \bigg( -  \cA_2(\bfx,\bfx;t) + \int d^{3} \bfy  \int d^{3} \bfz \; \big[ \cA_2(\bfx,\bfy;t) \varphi_2(\bfy) + \cB(\bfy, \bfx ; t) \varphi_{1}(\bfy) \big]  \\ 
& \  & \qquad \qquad \qquad \qquad \qquad \qquad \qquad \times \big[ \cA_2(\bfx,\bfz;t) \varphi_2(\bfz) + \cB(\bfz, \bfx ; t) \varphi_{1}(\bfz) \big] \bigg) \sigma_{\ssS}[t, \varphi_1, \varphi_2] \ , \nn
\eea
which implies that
\bea
\sfrac{ \mathrm{RHS\ of\ }(\ref{Markeq}) }{  \sigma_{\ssS}( t, \varphi_1, \varphi_2 ) } & = & - \frac{i}{2} \int \exd^3 \bfx \bigg[ \cA_1(\bfx,\bfx;t) -  \cA_2(\bfx,\bfx;t) + |\boldsymbol{\nabla} \varphi_1(\bfx)|^2 - |\boldsymbol{\nabla} \varphi_2(\bfx)|^2 \bigg]  \\
& \ &  \ \ + \int \exd^3 \bfx \int d^{3} \bfy \int d^{3} \bfz \; \bigg( \tfrac{i \; \big[ \cA_1(\bfx,\bfy;t) \varphi_1(\bfy) + \cB(\bfx, \bfy ; t) \varphi_{2}(\bfy) \big] \big[ \cA_1(\bfx,\bfz;t) \varphi_1(\bfz) + \cB(\bfx, \bfz ; t) \varphi_{2}(\bfz) \big]}{2} \notag \\
& \ & \qquad \qquad \qquad \qquad \qquad \qquad - \; \tfrac{i \big[ \cA_2(\bfx,\bfy;t) \varphi_2(\bfy) + \cB(\bfy, \bfx ; t) \varphi_{1}(\bfy) \big] \big[ \cA_2(\bfx,\bfz;t) \varphi_2(\bfz) + \cB(\bfz, \bfx ; t) \varphi_{1}(\bfz) \big] }{2} \bigg) \nn  \\
& \ & \qquad \qquad + \left( - \frac{\tilde g^2}{4\pi \beta} - \frac{i \lambda}{2} \right) \varphi_{1}(\mathbf{0})^2 + \frac{\tilde g^2}{2 \pi \beta} \varphi_{1}(\mathbf{0}) \varphi_{2}(\mathbf{0}) + \left( - \frac{\tilde g^2}{4\pi \beta} + \frac{i \lambda}{2} \right) \varphi_{2}(\mathbf{0})^2 \ . \nn
\eea
We next need to collect the terms that are proportional to the various possible powers of $\varphi_1(\mathbf{x})$ and $\varphi_{2}(\mathbf{x})$ and so on (note that we re-label some integration variables here):
\bea
\sfrac{ \mathrm{RHS\ of\ }(\ref{Markeq})  }{  \sigma_{\ssS}( t, \varphi_1, \varphi_2 ) } & = & - \bigg( \frac{\tilde g^2}{4\pi \beta} + \frac{i \lambda}{2} \bigg) \varphi_{1}(\mathbf{0})^2 + \frac{\tilde g^2}{2\pi \beta} \varphi_{1}(\mathbf{0}) \varphi_{2}(\mathbf{0}) - \bigg( \frac{\tilde g^2}{4\pi \beta} - \frac{i \lambda}{2} \bigg) \varphi_{2}(\mathbf{0})^2 \\
& \ & \qquad  - \frac{i}{2} \int \exd^3 \bfx \bigg[ \cA_1(\bfx,\bfx;t) -  \cA_2(\bfx,\bfx;t) \bigg] - \frac{i}{2} \int \d^3 \bfx \; \bigg( |\boldsymbol{\nabla} \varphi_1(\bfx)|^2 - |\boldsymbol{\nabla} \varphi_2(\bfx)|^2 \bigg) \nn \\
& \ & \qquad + \frac{i}{2} \int \exd^{3} \bfx \int \exd^{3}\bfy \int \exd^3 \bfz \; \big[ \cA_{1}(\bfz,\bfx ; t) \cA_{1}(\bfz,\bfy ; t)  - \cB(\bfx,\bfz ; t) \cB(\bfy,\bfz ; t) \big] \varphi_{1}(\bfx) \varphi_{1}(\bfy) \notag \\
& \ & \qquad + i  \int \exd^{3} \bfx \int \exd^{3}\bfy \int \exd^3 \bfz \; \big[ \cA_{1}(\bfz,\bfx ; t) \cB(\bfz,\bfy ; t) - \cB(\bfx,\bfz ; t) \cA_{2}(\bfz,\bfy ; t) \big]  \varphi_{1}(\bfx) \varphi_{2}(\bfy) \notag \\
& \ & \qquad + \frac{i}{2} \int \exd^{3} \bfx \int \exd^{3}\bfy \int \exd^3 \bfz \; \big[ - \cA_{2}(\bfx,\bfz ; t) \cA_{2}(\bfx,\bfz ; t) + \cB(\bfz,\bfx ; t) \cB(\bfz,\bfy ; t) \big] \varphi_{2}(\bfx) \varphi_{2}(\bfy) \notag 
\eea
Note that the above is equal to the quantity (with the time-derivative we computed above)
\bea
\sfrac{ \mathrm{LHS\ of\ }(\ref{Markeq}) }{  \sigma_{\ssS}( t, \varphi_1, \varphi_2 ) }  & = & \frac{\partial_t \cN(t)}{\cN(t)} + \int \exd^3 \bfx \int \exd^3 \bfy\; \bigg\{ - \frac{1}{2} \partial_t \mathcal{A}_1(\bfx,\bfy ; t ) \varphi_1(\bfx) \varphi_{1}(\bfy) \\
& \ & \qquad \qquad \qquad \qquad \qquad \qquad \quad  - \frac{1}{2} \partial_t \mathcal{A}_2(\bfx,\bfy ; t )  \varphi_2(\bfx) \varphi_{2}(\bfy) - \partial_t \cB(\bfx, \bfy ; t) \varphi_1(\bfx) \varphi_{2}(\bfy) \bigg\} \notag
\eea
and so we need to get the RHS into this form. We then use integration by parts (twice) to write
\bea
 \int \d^3 \bfx \; |\boldsymbol{\nabla} \varphi_1(\bfx)|^2 & = & -  \int \d^3 \bfx \;  \varphi_1(\bfx) \nabla_\bfx^2 \varphi_1(\bfx) \nn\\
& = & -  \int \d^3 \bfx  \int \d^3 \bfy  \; \delta^3(\bfx - \bfy) \varphi_1(\bfy) \nabla_\bfx^2 \varphi_1(\bfx) \\
& = & - \int \d^3 \bfx  \int \d^3 \bfy \; \varphi_1(\bfx)  \varphi_1(\bfy) \;  \nabla_\bfx^2 \delta^3(\bfx - \bfy) \nn
\eea
which gives
\bea
&\ & \sfrac{ \mathrm{RHS\ of\ }(\ref{Markeq}) }{ \sigma_{\ssS}( t, \varphi_1, \varphi_2 ) } \ = \ - \frac{i}{2} \int \exd^3 \bfx \bigg[ \cA_1(\bfx,\bfx;t) -  \cA_2(\bfx,\bfx;t) \bigg] \\
& \ & \quad+ \int \d^3 \bfx \int \d^3 \bfy \; \varphi_1(\bfx)  \varphi_1(\bfy) \bigg\{ \frac{i}{2} \nabla_\bfx^2 \delta^3(\bfx - \bfy) - \bigg( \sfrac{\tilde g^2}{4\pi \beta} + \sfrac{i \lambda}{2} \bigg) \delta^3(\bfx) \delta^3(\bfy) \notag \\
& \ & \qquad \qquad \qquad \qquad \qquad \qquad \qquad \qquad \qquad + \frac{i}{2} \int \exd^3 \bfz \; \big[ \cA_{1}(\bfz,\bfx ; t) \cA_{1}(\bfz,\bfy ; t) - \cB(\bfx,\bfz ; t) \cB(\bfy,\bfz ; t) \big] \bigg\} \notag \\
& \ & \quad  + \int \d^3 \bfx \int \d^3 \bfy \; \varphi_2(\bfx)  \varphi_2(\bfy) \bigg\{ - \frac{i}{2} \nabla_\bfx^2 \delta^3(\bfx - \bfy) - \bigg(  \sfrac{\tilde g^2}{4\pi \beta} - \sfrac{i \lambda}{2} \bigg) \delta^3(\bfx) \delta^3(\bfy) \notag \\
& \ & \qquad \qquad \qquad \qquad \qquad \qquad \qquad \qquad \qquad + \frac{i}{2} \int \exd^3 \bfz \; \big[ - \cA_{2}(\bfx,\bfz ; t) \cA_{2}(\bfy,\bfz ; t) + \cB(\bfz,\bfx ; t) \cB(\bfz,\bfy ; t) \big] \bigg\} \notag \\
& \ & \quad +  \int \d^3 \bfx \int \d^3 \bfy \; \varphi_1(\bfx)  \varphi_2(\bfy) \bigg[ \sfrac{\tilde g^2}{2\pi \beta} \delta^3(\bfx) \delta^3(\bfy) + i \int \exd^3 \bfz \; \big[ \cA_{1}(\bfz,\bfx ; t) \cB(\bfz,\bfy ; t) - \cB(\bfx,\bfz ; t) \cA_{2}(\bfz,\bfy ; t) \big] \bigg]  \ .\notag
\eea
Setting $\mathrm{LHS}=\mathrm{RHS}$ gives four equations. The constant piece gives
\be \label{App:Neq}
\frac{1}{\cN(t)} \frac{\pd \cN}{\pd t}  =   - \frac{i}{2} \int \exd^3 \bfx \bigg[ \cA_1(\bfx,\bfx;t) -  \cA_2(\bfx,\bfx;t) \bigg] \ , 
\ee
while coefficient of $\varphi_1(\bfx)  \varphi_1(\bfy)$ gives
\bea\label{App:11eq}
 \frac{\pd \cA_{1}(\bfx, \bfy ; t)}{\pd t} & = & - i \nabla_\bfx^2 \delta^3(\bfx - \bfy) + \left(  \frac{\tilde g^2}{2\pi \beta} + i \lambda \right) \delta^3(\bfx) \delta^3(\bfy)  \\
& \ & \quad  \quad  \quad  \quad  \quad  \quad + \int \exd^3 \bfz \; \big[ - i \cA_{1}(\bfz,\bfx ; t) \cA_{1}(\bfz,\bfy ; t) + i \cB(\bfx,\bfz ; t) \cB(\bfy,\bfz ; t) \big] \notag \ ,
\eea
the coefficient of $\varphi_2(\bfx)  \varphi_2(\bfy)$ gives (the first term in the integral has used the symmetry of $\cA_2$)
\bea \label{App:22eq}
 \frac{\pd \cA_{2}(\bfx, \bfy ; t)}{\pd t} & = &  i \nabla_\bfx^2 \delta^3(\bfx - \bfy) + \left(  \frac{\tilde g^2}{2\pi \beta} - i \lambda \right) \delta^3(\bfx) \delta^3(\bfy)  \\
& \ & \quad  \quad  \quad  \quad  \quad  \quad + \int \exd^3 \bfz \; \big[  i \cA_{2}(\bfx,\bfz ; t) \cA_{2}(\bfy,\bfz ; t) - i \cB(\bfz,\bfx ; t) \cB(\bfz,\bfy ; t) \big] \ , \notag
\eea
and the coefficient of $\varphi_1(\bfx) \varphi_2(\bfy)$ gives
\bea  \label{App:12eq} 
\frac{\pd \cB(\bfx, \bfy ; t)}{\pd t} & = & - \frac{\tilde g^2}{2\pi \beta} \delta^3(\bfx) \delta^3(\bfy) + \int \exd^3 \bfz \; \big[ - i  \cA_{1}(\bfz,\bfx ; t) \cB(\bfz,\bfy ; t) + i \cB(\bfx,\bfz ; t) \cA_{2}(\bfz,\bfy ; t) \big] \ . \qquad \qquad
\eea

\subsection{The Calculation of $\cM^{-1}(\bfx,\bfx';t)$}
\label{MinvCalc}

We here compute the matrix $\cM^{-1}(\bfx,\bfx';t)$ appearing in the correlator (\ref{cMinvcorrelator}). We do so by going to momentum space and perturbing in the interactions. Defining the momentum-space version of $\cM$ using the expression
\be
\cM(\bfx, \bfx' ; t)  : =  \int \frac{\exd^3 \bfk}{(2\pi)^3} \int \frac{\exd^3 \bfq}{(2\pi)^3} \;e^{+ i \bfk \cdot \bfx} M(\bfk,\bfq ;t) e^{ - i \bfq \cdot \bfx'}
\ee
for which the inverse-matrix condition 
\be
\int \exd^{3} \bfz\; \cM^{-1}(\bfx, \bfz ; t) \cM(\bfz, \bfx' ; t) =  \delta^3(\bfx - \bfx' ) 
\ee
takes the form
\be \label{doubleinversion}
\int \frac{\exd^{3} \bfp}{(2\pi)^3} \; M^{-1}(\bfk, \bfp;t) M(\bfp,\bfq;t) =  (2\pi)^3 \delta^3(\bfk - \bfq) \ , 
\ee
where $M^{-1}(\bfk, \bfq ; t)$ denotes the momentum-space components of $\cM^{-1}(\bfx,\bfx';t)$. 

To solve for $M^{-1}(\bfk, \bfq ; t)$ we perturb about the free-vacuum solution, writing
\bea \label{minvpert}
M(\bfp,\bfq; t) &=& 2  (2\pi)^3 |\bfp| \delta^3(\bfp - \bfq)  + \mfm(\bfp,\bfq;t) \\
M^{-1}(\bfk,\bfq ; t) &=&  \frac{(2\pi)^3}{2|\bfk|} \delta^3(\bfk - \bfq) + \mathfrak{i}(\bfk,\bfq ; t) \,, \nn
\eea
where in both lines the first term is just the free-field result --- see Appendix \ref{sec:freeequalcorrelator} --- and the second term is the perturbation that is to be solved to linear order in $\tilde g^2$ and $\lambda$. Inserting these into the relation (\ref{doubleinversion}) gives at linear order
\be
\frac{1}{2|\bfk|} \cdot \mfm(\bfk, \bfq;t) + \mathfrak{i}(\bfk,\bfq;t) \cdot 2 |\bfq|  \ \simeq \ 0
\ee
and so $\mathfrak{i}(\bfk,\bfq;t) \simeq  - {\mfm(\bfk, \bfq;t)}/({4 \; |\bfk| \; |\bfq|})$. Using expression \pref{MvsAB} giving $\cM$ in terms of $\cA_j$ and $\cB$, together with the solutions \pref{mfa1kk} through \pref{mfbkk}, then implies
\be
\mathfrak{i}(\bfk,\bfq;t) = - \frac{1}{2|\bfk|\; |\bfq|} \mathrm{Re}\left[ \left( \lambda - \frac{i \tilde g^2 }{2\pi\beta} \right) \sfrac{1 - e^{- i \big( |\bfk| + |\bfq| \big) t}}{|\bfk| + |\bfq|} \right]  - \frac{1}{2|\bfk|\; |\bfq|} \mathrm{Re}\left[ \frac{i \tilde g^2}{2\pi\beta} \cdot \sfrac{1 - e^{- i \big( |\bfk| - |\bfq| \big) t} }{|\bfk| - |\bfq|} \right] \,.
\ee

The desired position-space inverse is now found by Fourier transforming:
\bea \label{corrwithmfi}
 \cM^{-1}(\bfx, \bfx' ; t) 
& = & \int \frac{\exd^3 \bfk}{(2\pi)^3} \int \frac{\exd^3 \bfq}{(2\pi)^3} \; e^{+ i \bfk \cdot \bfx} M^{-1}(\bfk,\bfq ;t) \, e^{ - i \bfq \cdot \bfx'} \nn \\
& = &  \frac{1}{4\pi^2|\bfx - \bfx'|^2} + \int \frac{\exd^3 \bfk}{(2\pi)^3} \int \frac{\exd^3 \bfq}{(2\pi)^3} \; e^{+ i \bfk \cdot \bfx}\, \mathfrak{i}(\bfk,\bfq;t) \, e^{ - i \bfq \cdot \bfx'} 
\eea
where the first term is the free result computed in \S\ref{sec:freeequalcorrelator}. The angular integrals are simple because $\mathfrak{i}(\bfk,\bfq;t)$ depends only on $|\bfk|$ and $|\bfq|$, and so
\bea \label{corrwithI12}
&& \int \frac{\exd^3 \bfk}{(2\pi)^3} \int \frac{\exd^3 \bfq}{(2\pi)^3} \; e^{+ i \bfk \cdot \bfx} \, \mathfrak{i}(\bfk,\bfq;t) \, e^{ - i \bfq \cdot \bfx'} \nn \\
&& \qquad \ = \ \frac{1}{4 \pi^4 |\bfx| |\bfx'|} \int_0^\infty\exd k \int_0^\infty \exd q\; \bigg( - \frac{1}{2 k q} \mathrm{Re}\left[ \left( \lambda - \frac{i \tilde g^2}{2\pi \beta} \right) \frac{1 - e^{- i ( k + q ) t } }{k+q} \right] \nn \\
&& \qquad \qquad \qquad \qquad \qquad \qquad \qquad \qquad \qquad \qquad - \frac{1}{2 k q } \mathrm{Re}\left[ \frac{i \tilde g^2}{2\pi\beta} \cdot \frac{1 - e^{- i ( k-q) t} }{k-q} \right] \bigg) k \sin(k |\bfx|  ) q \sin(q|\bfx'| ) \nn \\
&& \qquad  =  - \frac{\lambda}{8 \pi^4 |\bfx| |\bfx'|} \;\mathrm{Re}\big[ I_{1}(\bfx, \bfx', t) \big] + \frac{\tilde g^2}{16 \pi^5 \beta |\bfx| |\bfx'|} \; I_{2}(\bfx,\bfx', t) 
\eea
where we define the integrals
\be \label{I1def}
I_{1}(\bfx,\bfx', t) := \int_0^\infty\exd k \int_0^\infty \exd q\;  \frac{1 - e^{- i ( k + q) t } }{k+q} \; \sin(k|\bfx| ) \sin(q|\bfx'| ) 
\ee
and
\be \label{I2def}
I_{2}(\bfx,\bfx', t) := \int_0^\infty\exd k \int_0^\infty \exd q\; \bigg( - \frac{\sin\big[ (k+q) t \big] }{k+q} + \frac{\sin\big[ (k-q) t \big] }{k-q} \bigg) \sin(k|\bfx| ) \sin(q|\bfx'| ) \ . 
\ee

To compute $I_1$ we use the Schwinger parametrization trick, which uses the identity 
\be
 \frac{1}{p} =  \int_0^\infty \exd \zeta\; e^{- p \zeta}   \,,
\ee
for any parameter $p>0$ to rewrite the factor of $(k+q)^{-1}$ in the integrand. This gives
\bea
I_{1}(\bfx,\bfx', t) & = & \int_0^\infty \exd \zeta \int_0^\infty\exd k \int_0^\infty \exd q\; ( 1 - e^{- i ( k + q) t } ) \sin(k|\bfx| ) \sin(q|\bfx'| ) e^{ - (k+q) \zeta } \nn\\
& = & \int_0^\infty \exd \zeta \; \bigg\{ \int_0^\infty \exd k \; e^{- k \zeta} \sin(k|\bfx|) \bigg\} \bigg\{ \int_0^\infty \exd q\; e^{- q \zeta} \sin(q|\bfx'|) \bigg\}  \\
&\ & \qquad \qquad \qquad -  \int_0^\infty \exd \zeta \; \bigg\{ \int_0^\infty \exd k\; e^{- k (\zeta+ i t)} \sin(|\bfx|k) \bigg\} \bigg\{ \int_0^\infty \exd q\; e^{- q ( \zeta+ i t )} \sin(|\bfx'|q) \bigg\} \nn \\
& = & \int_0^\infty \exd \zeta \; \bigg[ \bigg\{ \frac{|\bfx|}{\zeta^2 + |\bfx|^2 } \bigg\} \bigg\{ \frac{|\bfx'|}{\zeta^2 + |\bfx'|^2 } \bigg\} - \bigg\{ \frac{|\bfx|}{(\zeta+it)^2 + |\bfx|^2 } \bigg\} \bigg\{ \frac{|\bfx'|}{(\zeta+it)^2 + |\bfx'|^2 } \bigg\} \bigg] \nn
\eea
leaving an elementary integral over $\zeta$. Performing this integral we find that $I_1$ evaluates to 
\be \label{I1answer}
I_{1}(\bfx,\bfx', t) = \frac{|\bfx| |\bfx'|}{|\bfx|^2 - |\bfx'|^2} \bigg( \frac{i}{|\bfx'|}\log\left| \sfrac{1 + t/|\bfx'|}{1 - t/|\bfx'|} \right| + \frac{\pi \Theta(t - |\bfx'| )}{2 |\bfx'|}  - \frac{i}{|\bfx|}\log\left| \sfrac{1 + t/|\bfx|}{1 - t/|\bfx|} \right|  - \frac{\pi \Theta(t - |\bfx| )}{2 |\bfx|} \bigg) \ . 
\ee
Only the real part of this expression 
\be \label{I1answerR}
\hbox{Re} \big[  I_{1}(\bfx,\bfx', t) \big] = \frac{|\bfx| |\bfx'|}{|\bfx|^2 - |\bfx'|^2} \bigg[  \frac{\pi}{2 |\bfx'|} \,  \Theta(t - |\bfx'| ) - \frac{\pi }{2 |\bfx|}\, \Theta(t - |\bfx| ) \bigg]  
\ee
appears in \pref{corrwithI12}.

To compute $I_2$ it proves easier to first differentiate with respect to $t$, leading to
\bea
\frac{\pd I_2(\bfx,\bfx',t)}{\pd t} & = & 2 \int_0^\infty\exd k \int_0^\infty \exd q\; \sin( t k ) \sin( t q ) \sin(|\bfx| k) \sin(|\bfx'| q) \nn\\
& = & \frac{1}{2} \int_0^\infty \exd k \; \bigg( \cos\big[(t-|\bfx|)k\big]- \cos\big[(t+|\bfx|)k\big] \bigg) \\
&& \qquad\qquad\qquad\qquad \times \int_0^\infty \exd q\; \bigg( \cos\big[(t-|\bfx'|)q\big]- \cos\big[(t+|\bfx'|)q\big] \bigg) \nn \\
& = & \frac{\pi^2}{2} \bigg( \delta(t-|\bfx|) - \delta(t+|\bfx|) \bigg) \bigg( \delta(t-|\bfx'|)- \delta(t+|\bfx'|) \bigg) \nn
\eea
where the last line uses the real part of the Fourier transform of a Heaviside step function. Since $\delta(t-|\bfx|)= \delta(t-|\bfx'|)=0$ for $t>0$, $|\bfx|>0$ and $|\bfx'|>0$ this simplifies to  
\be
\frac{\pd I_2(\bfx,\bfx',t)}{\pd t} = \frac{\pi^2}{2} \delta(t-|\bfx|) \delta(t-|\bfx'|) \ = \ \frac{\pi^2}{2} \delta(t-|\bfx|) \delta(|\bfx|-|\bfx'|) \,.
\ee
Integrating with respect to $t$ with the initial condition $I_2(0,\bfx,\bfx')=0$ (from the definition \pref{I2def}) then gives 
\be \label{I2answer}
I_2(\bfx,\bfx',t)= \frac{\pi^2}{2} \Theta(t-|\bfx|) \delta(|\bfx|-|\bfx'|) \ . 
\ee

Putting everything together gives
\bea \label{App:corrwithmfians}
 \cM^{-1}(\bfx, \bfx' ; t) 
& = & \frac{1}{4\pi^2|\bfx - \bfx'|^2}   - \frac{\lambda}{8 \pi^4 |\bfx| |\bfx'|} \;\mathrm{Re}\big[ I_{1}(\bfx, \bfx', t) \big] + \frac{\tilde g^2}{16 \pi^5 \beta |\bfx| |\bfx'|} \; I_{2}(\bfx,\bfx', t)  \nn\\
& = &  \frac{1}{4\pi^2|\bfx - \bfx'|^2}  - \frac{\lambda}{16 \pi^3 (|\bfx|^2 - |\bfx'|^2)}   \bigg[  \frac{1}{ |\bfx'|} \,  \Theta(t - |\bfx'| ) - \frac{1 }{ |\bfx|}\, \Theta(t - |\bfx| ) \bigg]   \\
&&\qquad\qquad\qquad\qquad\qquad\qquad \qquad \qquad \qquad + \frac{\tilde g^2}{32 \pi^3 \beta |\bfx| |\bfx'|} \;   \Theta(t-|\bfx|) \delta(|\bfx|-|\bfx'|) \,.\nn
\eea
This is the expression quoted in \pref{corrwithmfians} of the main text.

\subsection{Schr\"odinger-picture equal-time free-field correlator}
\label{sec:freeequalcorrelator}

In this Appendix we compute the $\langle \phi \, \phi \rangle$ correlator for free fields at equal times $t=t'$, as a check on Schr\"odinger picture methods. Using the field basis and the vacuum wave-functional the equal-time Wightman function is given by the functional integral
\bea\label{App:beforeN0}
\bra{\mathrm{vac}} \phi_{\ssS}(\mathbf{x}) \phi_{\ssS}(\mathbf{x}') \ket{\mathrm{vac}} 
& = & \int \cD [\varphi\sb] \; \bra{\varphi \sb} \phi_{\ssS}(\mathbf{x}) \phi_{\ssS}(\mathbf{x'}) \ket{\mathrm{vac}} \langle \mathrm{vac} | \varphi \sb \rangle \nn \\
& = & \int \cD [\varphi\sb] \;  \varphi(\mathbf{x}) \varphi(\mathbf{x}') \la \varphi \sb | \mathrm{vac} \ra \langle \mathrm{vac} | \varphi \rangle  \\
& = & \cN_0 \int \cD \varphi \; \varphi(\mathbf{x}) \varphi(\bfx')\; e^{- \tfrac{1}{2} \int \mathrm{d}^{3}\bfx \int \mathrm{d}^3 \bfy\; 2 \cE(\bfx - \bfy) \varphi(\bfx)  \varphi(\bfy) } \nn \\
& = & [2 \cE]^{-1}(\bfx - \bfx') \nn
\eea
where we use the free Gaussian solution in the second-last line, where $\cE(\bfx - \bfx') = \int \frac{\mathrm{d}^3 \bfk}{(2\pi)^3} |\bfk| e^{i (\bfx - \bfx') \cdot \bfk}$ given in (\ref{freekernel}). In order to perform the Gaussian integral we use the standard gaussian results
\bea
\int_{-\infty}^{\infty} \prod \exd \xi_{r}\; \xi_{r_1} \xi_{r_2} e^{ - \tfrac{1}{2} \sum_{r,s} K_{rs} \xi_{r} \xi_{s} } & = & \det\left( \sfrac{K}{2\pi} \right)^{-1/2} (K^{-1})_{r_1r_2} \label{WeinbergGauss1} \\
\int_{-\infty}^{\infty} \prod \exd \xi_{r}\; e^{ - \tfrac{1}{2} \sum_{r,s} K_{rs} \xi_{r} \xi_{s} } & = & \det\left( \sfrac{K}{2\pi} \right)^{-1/2}  \label{WeinbergGauss2}
\eea
The latter formula determines $\cN_0 =  \sqrt{\det( \frac{2\mathcal{E}}{2\pi} ) }$. Now we need to invert the ``matrix'' $2\cE$ here, where the matrix $[2\cE]^{-1}$ is defined by
\bea
\int \exd^3 \bfz\; [ 2 \cE ]^{-1}(\bfx - \bfz) 2 \cE(\bfz - \bfx') \ = \ \delta^3(\bfx - \bfx') \ ,
\eea
We solve the above in Fourier space, by writing $[ 2 \cE ]^{-1}(\bfx - \bfx') = \int \frac{\mathrm{d}^3 \bfk}{(2\pi)^3} \mathfrak{I}_{\bfk} e^{i (\bfx - \bfx') \cdot \bfk}$ for some function $\mathfrak{I}_{\bfk}$ in momentum space which we solve for here. The above equation then implies
\bea
2 |\bfk| \mathfrak{I}_{\bfk} \ = \ 1 \ ,
\eea
which means that $\mathfrak{I}_{\bfk} = (2 |\bfk| )^{-1}$. We Fourier transform this to position space to find that 
\bea
[2 \cE]^{-1}(\bfx - \bfx') = \int \frac{\mathrm{d}^3 \bfk}{(2\pi)^3}\; \bigg[ \frac{1}{2|\bfk|} \bigg] e^{i (\bfx - \bfx') \cdot \bfk} & = & \frac{1}{4\pi^2 |\bfx - \bfx'|} \int_0^\infty \exd k\; \sin( k |\bfx-\bfx'| ) \\
& = & -  \frac{1}{4\pi^2 |\bfx - \bfx'|} \cdot \mathrm{Im}\left[ \int_{-\infty}^\infty \exd k\;  \Theta(k) e^{- i |\bfx - \bfx'| k} \right]  \nn
\eea
From here we use Weinberg's formula (6.2.15) in \cite{Weinberg:1995mt} for the Fourier transform of a Heaviside step function, where (in the limit $\delta \to 0^{+}$)
\bea
\Theta(x) = - i \int_{-\infty}^{\infty} \frac{\exd y}{2\pi} \cdot  \frac{e^{+ i y x}}{y - i \delta} \qquad \iff \qquad  \int_{-\infty}^{\infty} \exd x\; \Theta(x) e^{- i y x} = \frac{-i}{y - i \delta} \label{WeinbergTheta}
\eea
which gives us 
\be
[2 \cE]^{-1}(\bfx - \bfx') = -  \frac{1}{4\pi^2 |\bfx - \bfx'|} \left( - \frac{1}{|\bfx - \bfx' |} \right)\ ,
\ee
which tells us that the free correlation function is 
\bea
\bra{\mathrm{vac}} \phi_{\ssS}(\mathbf{x}) \phi_{\ssS}(\mathbf{x}') \ket{\mathrm{vac}} & = & + \frac{1}{4\pi^2 |\bfx - \bfx'|^2}  \ , 
\eea
which is the correct answer for the Wightman function (for equal times $t=t'$).

\section{Mean-field details}

This appendix collects various intermediate steps  encountered in the mean-field calculations of \S\ref{sec:meanfieldH}. 

\subsection{Correlators using $\ol V^{-1}$}
\label{App:correlatorwrong}

We first compute the mean-field correlator of eq.~\pref{correlatorMFwrong} that would have been obtained if the transition to mean field methods had been done using $V^{-1}$ rather than $V^\star$ in \pref{HHvsIntPic}. 

Starting with \pref{correlatorMFwrong} leads to the following expression at leading nontrivial order in $\ol H_{\rm int}$:
\bea \label{MFcor1} 
 \cU(t,\bfx; t',\bfx')  
& \simeq &  \bra{ \vac } {\phi}(t,\bfx) {\phi}(t',\bfx') \ket{\vac} + i \int_0^{t} \exd \tau\; \bra{ \vac } \Bigl[ {\ol{H}}_{\mathrm{int}}(\tau) ,  {\phi}(t,\bfx) \Bigr] {\phi}(t',\bfx') \ket{\vac} \nn \\
&& \qquad \qquad \qquad + i \int_0^{t'} \exd \tau\; \bra{ \vac } {\phi}(t,\bfx) \Bigl[ {\ol{H}}_{\mathrm{int}}(\tau) , {\phi}(t',\bfx') \Bigr] \ket{\vac} 
\eea
where only terms linear in ${\ol{H}}_{\mathrm{int}}$ are kept. The first term in \pref{MFcor1} is simply the free Wightman function
\be \label{freeWightman} 
\bra{\vac} {\phi}(t,\bfx) {\phi}(t',\bfx') \ket{\vac} = \frac{1}{4 \pi^2 \big[ - (t - t' - i \delta)^2 + |\bfx - \bfx'|^2 \big]} \,,
\ee
while the commutator in the subleading terms is evaluated in Appendix \S\ref{App:CommutatorHbarint}, giving
\bea \label{commresult}
  \Bigl[ {\ol{H}}_{\mathrm{int}}(\tau) ,  {\phi}(t,\bfx) \Bigr] & = & \frac{i \abare  }{4 \pi |\bfx|} \bigg[ \delta\big( \tau - (t - |\bfx| ) \big) - \delta\big( \tau  - ( t + |\bfx| ) \big)  \bigg] {\phi}(\tau,\mathbf{0}) \nn\\ 
&& \qquad + \frac{\tilde g^2}{4 \pi |\bfx|} \bigg[  \Theta(t - | \bfx|) \Theta\big( \tau - [ t - |\bfx| ] \big) \mathscr{W}_{\beta}\big( \tau - [ t - |\bfx| ] \big) \\
&& \qquad\qquad\qquad\qquad\qquad\qquad \qquad - \Theta\big( \tau - [t + \bfx] \big) \mathscr{W}_{\beta}\big( \tau - [t + \bfx] \big) \bigg] {\phi}(\tau,\mathbf{0}) \notag \\
&& \qquad + \frac{\tilde g^2}{4 \pi |\bfx|} \bigg[ \delta\big( \tau - (t - |\bfx| ) \big) - \delta\big( \tau  - ( t + |\bfx| ) \big)  \bigg] \int_{0}^{\tau} \exd \tau' \; \mathscr{W}_{\beta}(\tau') {\phi}(\tau - \tau',\mathbf{0}) \,. \notag
\eea
Using this in (\ref{MFcor1}) we get, after some manipulations,
\bea
  &&i \int_0^{t} \exd \tau\; \bra{ \vac } \Bigl[ {\ol{H}}_{\mathrm{int}}(\tau), {\phi}(t,\bfx) \Bigr] {\phi}(s,\bfx') \ket{\vac} = - \frac{\abare \Theta(t - |\bfx| )}{4 \pi |\bfx|} \bra{ \vac } {\phi}(t - |\bfx| ,\mathbf{0}) {\phi}(s,\bfx') \ket{\vac}\nn \\
&& \qquad\qquad   + \frac{i \tilde g^2 \Theta(t - | \bfx|)}{4 \pi |\bfx|}  \int_{0}^{|\bfx|} \exd \tau'\; \mathscr{W}_{\beta}\big( \tau'  \big)  \bra{ \vac } {\phi}(\tau' + t - |\bfx|,\mathbf{0}) {\phi}(s,\bfx') \ket{\vac}  \\
&& \qquad \qquad\qquad + \frac{i\tilde  g^2 \Theta(t - |\bfx|)}{4 \pi |\bfx|} \int_{0}^{t - |\bfx|} \exd \tau' \; \mathscr{W}_{\beta}(\tau') \bra{ \vac } {\phi}(t - |\bfx| - \tau',\mathbf{0}) {\phi}(s,\bfx') \ket{\vac} \notag
\eea
and so the mean-field correlation function becomes
\bea  \label{VinvabareCOMP}
 && \cU(t,\bfx; t',\bfx')  \simeq  \frac{1}{4 \pi^2 \big[ - (t - t' - i \delta)^2 + |\bfx - \bfx'|^2 \big]} \nn\\
 &&\qquad - \frac{\abare \Theta(t - |\bfx| )}{4 \pi |\bfx|} \bra{ \vac } {\phi}(t - |\bfx| ,\mathbf{0}) {\phi}(t',\bfx') \ket{\vac} - \frac{\abare \Theta(t' - |\bfx'| )}{4 \pi |\bfx'|} \bra{ \vac } {\phi}(t,\bfx)  {\phi}(t' - |\bfx'| ,\mathbf{0}) \ket{\vac} \nn \\
&& \qquad\qquad   + \frac{i \tilde g^2 \Theta(t - | \bfx|)}{4 \pi |\bfx|}  \int_{0}^{|\bfx|} \exd \tau\; \mathscr{W}_{\beta}( \tau )  \bra{ \vac } {\phi}( t - |\bfx| + \tau ,\mathbf{0}) {\phi}(t',\bfx') \ket{\vac} \notag \\
&& \qquad\qquad\qquad + \frac{i \tilde g^2 \Theta(t - |\bfx|)}{4 \pi |\bfx|} \int_{0}^{t - |\bfx|} \exd \tau \; \mathscr{W}_{\beta}(\tau) \bra{ \vac } {\phi}(t - |\bfx| - \tau,\mathbf{0}) {\phi}(t',\bfx') \ket{\vac}  \\
& & \qquad\qquad\qquad \qquad + \frac{i \tilde g^2 \Theta(t' - | \bfx'|)}{4 \pi |\bfx'|} \int_{0}^{|\bfx'|} \exd \tau\; \mathscr{W}_{\beta}( \tau )  \bra{ \vac } {\phi}(t,\bfx) {\phi}(t' - |\bfx'| + \tau,\mathbf{0}) \ket{\vac} \notag \\
&& \qquad \qquad\qquad\qquad\qquad + \frac{i \tilde g^2 \Theta(t' - |\bfx'|)}{4 \pi |\bfx'|} \int_{0}^{t' - |\bfx'|} \exd \tau \; \mathscr{W}_{\beta}(\tau) \bra{ \vac } {\phi}(t,\bfx) {\phi}(t' - |\bfx'| - \tau,\mathbf{0})  \ket{\vac} \ . \notag
\eea

This expression simplifies further in the regime where all of $t - |\bfx|$, $|\bfx|$, $t' - | \bfx'|$ and $|\bfx'|$ are much greater than $\beta$, because in this case the narrowness of the Wightman function --- $\mathscr{W}_{\beta}(\tau) \propto e^{ - 2 \pi \tau /  \beta}$ for $\tau \gg \beta$ --- makes it a good approximation to approximate the upper integration limits by $\infty$ (with only exponentially small error). Under these assumptions we also know $\Theta(t - |\bfx|) = \Theta(t' - |\bfx'|) = 1 $ and so get
\bea  \label{MFcor2}
&& \cU(t,\bfx; t',\bfx') \;  \simeq \; \frac{1}{4 \pi^2 \big[ - (t - t' - i \delta)^2 + |\bfx - \bfx'|^2 \big]}  \nn\\
&& \quad - \frac{\abare}{4 \pi } \bigg[ \frac{\bra{ \vac } {\phi}(t - |\bfx| ,\mathbf{0}) {\phi}(t',\bfx') \ket{\vac}  }{|\bfx|} + \frac{\bra{ \vac } {\phi}(t,\bfx)  {\phi}(t' - |\bfx'| ,\mathbf{0}) \ket{\vac} }{|\bfx'|} \bigg]   \\
&& \quad\; + \frac{i\tilde g^2}{4 \pi |\bfx|}  \int_{0}^{\infty} \exd \tau\; \mathscr{W}_{\beta}( \tau ) \bigg[ \bra{ \vac } {\phi}( t - |\bfx| + \tau ,\mathbf{0}) {\phi}(t',\bfx') \ket{\vac} + \bra{ \vac } {\phi}(t - |\bfx| - \tau,\mathbf{0}) {\phi}(t',\bfx') \ket{\vac}  \bigg] \notag \\
&& \quad\; \;  + \frac{i \tilde g^2}{4 \pi |\bfx'|} \int_{0}^{\infty} \exd \tau\; \mathscr{W}_{\beta}( \tau ) \bigg[ \bra{ \vac } {\phi}(t,\bfx) {\phi}(t' - |\bfx'| + \tau,\mathbf{0}) \ket{\vac} + \bra{ \vac } {\phi}(t,\bfx) {\phi}(t' - |\bfx'| - \tau,\mathbf{0})  \ket{\vac} \bigg] \ . \notag
\eea

The integrals involving $\mathscr{W}_{\beta}( \tau ) $ above are computed in Appendix \ref{App:ThermalWInt}. These contain a divergence from the $\tau \to 0$ limit, but this has the same structure as does the second line of \pref{MFcor2} so the divergence can be absorbed into $\lambda$. Once this is done, and using the explicit form \pref{freeWightman} for the free Wightman function, we find 
\bea  \label{MFcor4} 
&& \cU(t,\bfx; t',\bfx') \; \simeq \; \frac{1}{4 \pi^2 \big[ - (t - t' - i \delta)^2 + |\bfx - \bfx'|^2 \big]} \nn\\
&&\quad + \frac{\lambda}{32 \pi^3 |\bfx| |\bfx'| } \bigg[ - \frac{1}{t - t' + |\bfx| + |\bfx'| - i \delta } + \frac{1}{t - t' - |\bfx| - |\bfx'| - i \delta } \bigg]  \\
&& \quad\;   - \frac{\tilde  g^2 }{128 \pi^2 \beta^2 |\bfx| |\bfx'| \sinh^2 \left[ \frac{\pi}{\beta} (   t - t' + |\bfx| + |\bfx'| - i \delta )\right]} - \frac{i \tilde g^2}{64\pi^4 \beta^2  |\bfx| |\bfx'|  } \;  \mathrm{Im} \; \psi^{(1)}\left[ 1 + \sfrac{i ( t - t' + |\bfx| + |\bfx'| ) }{\beta} \right] \notag \\
&&  \quad\;\;    + \frac{\tilde g^2}{128 \pi^2 \beta^2 |\bfx| |\bfx'|  \sinh^2 \left[ \frac{\pi }{\beta}(  t - t' - |\bfx| - |\bfx'| - i \delta ) \right]} + \frac{i \tilde g^2}{64 \pi^4 \beta^2 |\bfx| |\bfx'| } \;  \mathrm{Im}\;  \psi^{(1)}\left[ 1 + \sfrac{i (  t - t' - |\bfx| - |\bfx'| )}{\beta} \right] \ , \notag
\eea
where $\psi^{(1)}(z) := (\exd/\exd z)^2 \ln \Gamma(z)$ is the polygamma function of order 1. Recall that the derivation of \pref{MFcor4} assumes $|\bfx| \ , t - |\bfx| \ , |\bfx'| \ , t' - |\bfx'| \gg \beta$.

The equal-time limit of eq.~\pref{MFcor4} is given by
\be  \label{MFcoreqtime} 
 \cU(t,\bfx; t,\bfx') \simeq \frac{1}{4 \pi^2 |\bfx - \bfx'|^2} -  \left(\lambda - \frac{i \tilde g^2}{2\pi \beta} \right) \frac{1}{ 16 \pi^3|\bfx| |\bfx'| \big( |\bfx| + |\bfx'| \big) }  
\ee
where the assumption $|\bfx| + |\bfx'| \gg \beta$ used in the above derivation allows use of the large-$z$ approximation $\mathrm{Im}[ \psi^{(1)}(1+ i z)  ] \simeq - 1/z$. Notice that the $\tilde g^2$ term in this expression does not satisfy the hermiticity condition $\cU^{\ast}(t,\bfx; t,\bfx') = \cU(t,\bfx'; t,\bfx)$ because the density matrix is evaluated at $t=0$ and the $\tilde g^2$ contributions ensure the effective mean-field hamiltonian that evolves to general $t$ is also not hermitian.

We next evaluate the commutators required in the above, starting with the unequal-time commutator of the field itself. 

\subsubsection{Field commutators at unequal times}
\label{App:UnequalTime}

For later use in Appendix \ref{App:CommutatorHbarint}, we here compute the commutator ${ [ } {\phi}(t,\bfx) , {\phi}(t',\bfy) ]$ of interaction-picture fields at unequal times. This can be done using the standard field expansion in terms of creation and annihilation operators, but it is simpler to obtain it directly from the Wightman function given in \pref{freeWightman}. This can be done because the commutator of two free fields is a $c$-number, and so is equal to its expectation value in the vacuum, giving
\bea \label{unequal2}
{ \Bigl[ } {\phi}(t,\bfx) , {\phi}(t',\bfx') \Bigr] & = & \Bigl( \bra{ \mathrm{vac} } {\phi}(t,\bfx) {\phi}(t',\bfx') \ket{ \mathrm{vac} } - \bra{ \mathrm{vac} } {\phi}(t',\bfx') {\phi}(t,\bfx) \ket{ \mathrm{vac} } \Bigr)  \nn \\
& = & \frac{1}{4\pi^2} \left[ \frac{1}{-(t - t'- i \delta)^2 + |\bfx - \bfx'|^2 } - \frac{1}{-(t - t' + i \delta)^2 + |\bfx - \bfx'|^2 } \right]  \\
& = & \frac{1}{8 \pi^2 |\bfx - \bfx'|} \bigg[ \frac{1}{ (t - t') + |\bfx - \bfx'| - i \delta}  -  \frac{1}{ (t - t') - |\bfx - \bfx'| - i \delta} \nn\\
&& \quad \quad \quad \quad \quad \quad \quad \quad \quad \quad - \frac{1}{ (t - t') + |\bfx - \bfx'| + i \delta} + \frac{1}{ (t - t') - |\bfx - \bfx'| + i \delta} \bigg]   \nn
\eea
where a factor of the unit operator, $\cI_+$, is implicit everywhere on the right-hand side.

Using the Sochocki-Plemelj identity $({z - i 0^+})^{-1} - ({z + i 0^+})^{-1} = 2 i \pi \delta(z)$ for infinitesimal and positive $0^{+}$, the above becomes
\be \label{phiphicomm}
{ \Bigl[ } {\phi}(t,\bfx) , {\phi}(t',\bfx') \Bigr] = \frac{i}{4 \pi |\bfx - \bfx'|} \bigg[ \delta\big( t - t' +  |\bfx - \bfx'|  \big) -  \delta\big( t - t' -  |\bfx - \bfx'|  \big)  \bigg] \cI_+ \,,
\ee
which reduces when $t = t'$ to the standard equal-time commutator when $\bfx \neq \bfx'$:
\be \label{ordinaryCCR1}
{ \Bigl[ } {\phi}(t,\bfx) , {\phi}(t,\bfx') \Bigr]  =  0 \ .
\ee
Specializing (\ref{unequal2}) to the commutator for two fields at different times but with $\bfx = \bfx'$, instead gives
\be \label{CCRneqtimeeqpos}
{ \Bigl[ } {\phi}(t,\bfx) , {\phi}(t',\bfx) \Bigr] = \frac{1}{4\pi^2} \left[ - \frac{1}{(t - t'- i \delta)^2} + \frac{1}{(t - t' + i \delta)^2} \right] \cI_+ =  \frac{i}{2\pi} \; \delta'(t-t') \; \cI_+
\ee
where the limit $\delta \to 0$ is taken in the last equality and $\delta'(x)$ denotes the derivative of the Dirac delta function with respect to its argument.

\subsubsection{The Commutator $\left[ {\ol{H}}_{\mathrm{int}}(\tau), {\phi}(t,\bfx) \right]$}
\label{App:CommutatorHbarint}

The next intermediate step required is the commutator with $\phi$ of the local mean-field Hamiltonian ${\ol{H}}_{\mathrm{int}}(t)$ defined in (\ref{HmLocal}):
\be \label{CommHbarint1}
\Bigl[ {\ol{H}}_{\mathrm{int}}(\tau) , {\phi}(t,\bfx) \Bigr] = \frac{1}{2} \abare \, \Bigl[ {\phi}^2(\tau,\mathbf{0}) , {\phi}(t,\bfx) \Bigr] - i \tilde  g^2 \int_{0}^{\tau} \exd s \; \mathscr{W}_{\beta}(\tau') \Bigl[ {\phi}(\tau,\mathbf{0}) {\phi}(\tau-s, \mathbf{0}) ,  {\phi}(t,\bfx) \Bigr] \ . 
\ee
Using
\be \label{commaterm}
\Bigl[ {\phi}^2(\tau,\mathbf{0}) , {\phi}(t,\bfx) \Bigr]  
= - {\phi}(\tau,\mathbf{0}) \Bigl[ {\phi}(t,\bfx) , {\phi}(\tau,\mathbf{0}) \Bigr] - \Bigl[ {\phi}(t,\bfx) ,  {\phi}(\tau,\mathbf{0}) \Bigr] {\phi}(\tau,\mathbf{0}) 
\ee
with the result (\ref{phiphicomm}) we have
\be
\Bigl[ {\phi}^2(\tau,\mathbf{0}) , {\phi}(t,\bfx) \Bigr] = \frac{i}{2 \pi |\bfx|} \bigg[ \delta\big( \tau - (t - |\bfx| ) \big) - \delta\big( \tau  - ( t + |\bfx| ) \big)  \bigg] {\phi}(\tau,\mathbf{0}) \ . \label{phiSQphicomm}
\ee
Similarly
\bea \label{nonloccomm}
\Bigl[ {\phi}(\tau,\mathbf{0}) {\phi}(\tau-s, \mathbf{0}), {\phi}(t,\bfx) \Bigr] 
& = & \frac{i}{4 \pi |\bfx|} \bigg[ \delta\big( s - [ \tau - (t - |\bfx| ) ] \big) - \delta\big( s - [ \tau - (t + |\bfx| ) ] \big) \bigg] {\phi}(\tau,\mathbf{0})  \\
& \ & \qquad \qquad \qquad  + \frac{i}{4 \pi |\bfx|} \bigg[ \delta\big( \tau - (t - |\bfx| ) \big) - \delta\big( \tau  - ( t + |\bfx| ) \big)  \bigg] {\phi}(\tau - s,\mathbf{0}) \nn
\eea
and so
\bea
\left[ {\ol{H}}_{\mathrm{int}}(\tau) ,  {\phi}(t,\bfx) \right] & = & \frac{i \abare   }{4 \pi |\bfx|} \bigg[ \delta\big( \tau - (t - |\bfx| ) \big) - \delta\big( \tau  - ( t + |\bfx| ) \big)  \bigg] {\phi}(\tau,\mathbf{0}) \\
&& \quad  + \frac{\tilde g^2}{4 \pi |\bfx|} \,{\phi}(\tau,\mathbf{0}) \int_{0}^{\tau} \exd s  \; \mathscr{W}_{\beta}(s ) \bigg[ \delta\big( s  - [ \tau - (t - |\bfx| ) ] \big) - \delta\big( s  - [ \tau - (t + |\bfx| ) ] \big) \bigg]  \notag \\
&& \quad \quad  + \frac{\tilde g^2}{4 \pi |\bfx|} \bigg[ \delta\big( \tau - (t - |\bfx| ) \big) - \delta\big( \tau  - ( t + |\bfx| ) \big)  \bigg] \int_{0}^{\tau} \exd s  \; \mathscr{W}_{\beta}(s ) \,{\phi}(\tau - s ,\mathbf{0}) \ .\nn
\eea

Performing the $s$-integrals using the delta functions gives
\be
\int_{0}^{\tau} \exd s \; \mathscr{W}_{\beta}(s) \, \delta\big( s - [ \tau - (t - |\bfx| ) ] \big)  =  \Theta(t - | \bfx|) \Theta\big( \tau - [ t - |\bfx| ] \big) \mathscr{W}_{\beta}\big( \tau - [ t - |\bfx| ] \big) \ ,
\ee
where the step functions express the conditions under which the delta function has support within the integration range: $0 < \tau - (t - |\bfx|) < \tau$, which in turn implies $\tau > t - |\bfx| > 0$. Similarly
\be
\int_{0}^{\tau} \exd s \; \mathscr{W}_{\beta}(s) \delta\big( s - [ \tau - (t + |\bfx| ) ] \big)= \Theta\big( \tau - [t + \bfx] \big) \mathscr{W}_{\beta}\big( \tau - [t + \bfx] \big) \,.
\ee
Putting the above terms together gives the result (\ref{commresult}).

\subsubsection{The $\mathscr{W}_{\beta}$ Integrals}
\label{App:ThermalWInt}

This section computes the integrals
\be  \label{IbetaINT}
\cJ_{1}(t, \bfx, t', \bfx')   : =   \int_{0}^{\infty} \exd \tau\; \mathscr{W}_{\beta}( \tau ) \bigg[ \bra{ \vac } {\phi}( t - |\bfx| + \tau ,\mathbf{0}) {\phi}(t',\bfx') \ket{\vac} + \bra{ \vac } {\phi}(t - |\bfx| - \tau,\mathbf{0}) {\phi}(t,\bfx') \ket{\vac}  \bigg]\ ,
\ee
and
\be  \label{JbetaINT}
 \cJ_{2}(t, \bfx, t', \bfx')  :=   \int_{0}^{\infty} \exd \tau\; \mathscr{W}_{\beta}( \tau ) \bigg[ \bra{ \vac } {\phi}(t,\bfx) {\phi}(t' - |\bfx'| + \tau,\mathbf{0}) \ket{\vac} + \bra{ \vac } {\phi}(t,\bfx) {\phi}(t' - |\bfx'| - \tau,\mathbf{0})  \ket{\vac} \bigg] \ ,
\ee
which appear in eq.~(\ref{MFcor2}) above. Writing the free Wightman function (\ref{freeWightman}) as a mode sum yields
\bea  \label{freeW_FT} 
\bra{\vac} {\phi}(t,\bfx) {\phi}(t',\bfx') \ket{\vac} & = & \frac{1}{4 \pi^2 \big[ - (t - t' - i \delta)^2 + |\bfx - \bfx'|^2 \big]} \nn\\
& = & \frac{1}{8 \pi^2 |\bfx - \bfx'|} \bigg[  \frac{1}{t - t' + |\bfx - \bfx'| - i \delta} - \frac{1}{t - t' - |\bfx - \bfx'| - i \delta} \bigg] \\
& = & \sfrac{i}{8 \pi^2 |\bfx - \bfx'|} \int_0^\infty \exd p \; \bigg[ e^{- ip ( t - t' + |\bfx - \bfx'| - i \delta)  } - \int_0^\infty \exd p \; e^{- ip ( t - t' - |\bfx - \bfx'|- i \delta )   } \bigg] \qquad \nn\\
& = & \frac{1}{4 \pi^2 |\bfx -\bfx'| } \int_0^\infty \exd p \; e^{ - ip (t - t' - i \delta) } \sin\big( p|\bfx - \bfx'|  \big) \,,\nn
\eea
which allows (\ref{IbetaINT}) to be written as
\bea \label{IbetaINT2}
\cJ_{1}(t, \bfx, t, \bfx') & = &  \int_{0}^{\infty} \exd \tau\; \frac{ \mathscr{W}_{\beta}( \tau ) }{ 4 \pi^2 |\bfx'|} \int_0^\infty \exd p \; \bigg[ e^{ - i p(t - |\bfx| + \tau - t'- i \delta)   }  \sin( p|\bfx'|  )  + e^{ - ip (t - |\bfx| - \tau - t' - i \delta)   }  \sin(p |\bfx'| ) \bigg]  \nn \\
& = & \frac{ 1 }{ 4 \pi^2 |\bfx'| }  \int_0^\infty \exd p \; e^{ - ip (t - |\bfx| - t' - i \delta)  } \sin(p |\bfx'| ) \bigg[ \mathcal{C}_{\beta}(p) + i \mathcal{K}_{\beta} \bigg] 
\eea
with the definitions
\be
\cC_{\beta}(p) := 2 \int_0^\infty \exd \tau \; \mathrm{Re}\left[ \mathscr{W}_{\beta}(\tau) \right] \cos( p \tau )\qquad \mathrm{and} \qquad \mathcal{K}_{\beta} := 2 \int_0^\infty \exd \tau \; \mathrm{Im}\left[ \mathscr{W}_{\beta}(\tau) \right] \cos( p \tau ) \ .
\ee
The first integral $\cC_{\beta}$ was computed in \cite{Kaplanek:2019dqu,Kaplanek:2020iay} and gives 
\be \label{ReWcosFT}
\cC_{\beta}(p) = \frac{p}{4 \pi} \coth\left( \frac{ \beta p }{2} \right) \ .
\ee
Meanwhile to compute $\mathcal{K}_{\beta}$ (which turns out to be divergent as well as $p$-independent), note that the imaginary part of $\mathscr{W}_{\beta}$ actually vanishes if $\tau$ is fixed but nonzero as $\delta \to 0$, since it can be written as
\be
\mathrm{Im}\left[ \mathscr{W}_{\beta}(\tau) \right]   =   \frac{i}{8 \beta^2 } \bigg\{ \frac{1}{\sinh^2\left[ \frac{\pi}{\beta}  (\tau - i \delta )\right]} - \frac{1}{\sinh^2\left[ \frac{\pi}{\beta}  ( \tau + i \delta )\right]} \bigg\}  
\ee
and so the complete contribution comes only from the regime near $\tau \to 0$, for which
\be
\mathrm{Im}\left[ \mathscr{W}_{\beta}(\tau) \right]   \simeq   \frac{i}{8\pi^2} \left[ \frac{1}{( \tau - i \delta )^2} - \frac{1}{( \tau + i \delta )^2} \right] \to \frac{1}{4\pi} \; \delta'(\tau) \,,  \label{ImWbeta}
\ee
as $\delta \to 0$, which follows from the the Sochocki-Plemelj identity. The required integral then is
\be \label{Kbetaintresult}
\mathcal{K}_{\beta} = 2 \int_0^\infty \exd \tau \; \frac{\delta'(\tau)}{4\pi} \, \cos( p \tau ) = - \frac{1}{2\pi} \, \delta(0)
\ee
which displays a divergence that ultimately gets absorbed into the coupling parameter $\abare$. 

Combining the above into (\ref{IbetaINT2}) yields
\be
\cJ_{1}(t, \bfx, t', \bfx') = \frac{ 1 }{ 4 \pi^2 |\bfx'| }  \int_0^\infty \exd p \; e^{ - i p(t - |\bfx| - t' -i \delta)   } \sin( p|\bfx'|  ) \bigg[ \frac{p}{4 \pi} \coth\left( \frac{ \beta p }{2}  \right) - \frac{i}{2\pi}\; \delta(0) \bigg] 
\ee
The divergent term simplifies using (\ref{freeW_FT}), leading to
\bea \label{IbetaINT3}
\cJ_{1}(t, \bfx, t', \bfx') & = & - \frac{i}{2\pi}\, \delta(0) \bra{ \vac } {\phi}( t - |\bfx| , \mathbf{0} ) {\phi}(t,\bfx') \ket{ \vac } \\
&& \qquad - \frac{ i }{ 32 \pi^3 |\bfx'| }  \int_0^\infty \exd p \; p \coth\left( \frac{ \beta p }{2}  \right) \bigg[ e^{ - i p(t - t' - |\bfx| + |\bfx'| -i \delta)   } -e^{ - ip (t - t' - |\bfx| - |\bfx'| -i \delta)   } \bigg] \notag
\eea
which can be integrated using
\be \label{FTcoth}
\int_0^\infty \exd p\; p \coth\left( \frac{\beta p}{2} \right) e^{ - i p \tau - i \delta  } =  - \frac{\pi^2}{\beta^2 \sinh^2 \left[ \frac{\pi }{\beta} ( \tau - i \delta )\right]} - \frac{2 i}{\beta^2} \;  \mathrm{Im} \bigg[ \psi^{(1)}\left( 1 + \frac{i\tau}{\beta} \right) \bigg] \,,
\ee
where $\psi^{(1)}(z) := \frac{\mathrm{d}}{\mathrm{d}z} \log\big( \Gamma(z) \big)$ is the Polygamma function of order 1 (for a derivation of this integral see Appendix \ref{App:FTcoth}). The final result found by inserting (\ref{FTcoth}) into (\ref{IbetaINT3}) is then
\bea \label{IbetaINT4} 
&&\cJ_{1}(t, \bfx, t', \bfx') = - \frac{i}{2\pi}\, \delta(0) \bra{ \vac } {\phi}( t - |\bfx| , \mathbf{0} ) {\phi}(t',\bfx') \ket{ \vac } \\
&& \qquad + \frac{i \pi^2}{32 \pi^3 \beta^2 |\bfx'|  \sinh^2 \left[ \frac{\pi }{\beta}(  t - t' - |\bfx| + |\bfx'| - i \delta ) \right]} - \frac{1}{16\pi^3 \beta^2  |\bfx'| } \;  \mathrm{Im} \bigg[ \psi^{(1)}\left( 1 + \frac{i ( t - t' - |\bfx| + |\bfx'| ) }{\beta} \right) \bigg] \notag \\
&& \qquad - \frac{i \pi^2}{32 \pi^3 \beta^2 |\bfx'|  \sinh^2 \left[ \frac{\pi  }{\beta} ( t - t' - |\bfx| - |\bfx'| - i \delta )\right]} + \frac{1}{16\pi^3 \beta^2  |\bfx'| } \;  \mathrm{Im} \bigg[ \psi^{(1)}\left( 1 + \frac{i (  t - t' - |\bfx| - |\bfx'| )}{\beta} \right) \bigg] \notag 
\eea
and in an almost identical calculation the integral (\ref{JbetaINT}) evaluates to
\bea
&&\cJ_{2}(t, \bfx, t', \bfx') = - \frac{i}{2\pi}\, \delta(0) \bra{ \vac } {\phi}( t , \bfx ) {\phi}(t'-|\bfx'|, \mathbf{0}) \ket{ \vac } \\
&& \qquad + \frac{i \pi^2}{32 \pi^3 \beta^2 |\bfx|  \sinh^2 \left[ \frac{\pi }{\beta}( t - t' + |\bfx| + |\bfx'| - i \delta ) \right]} - \frac{1}{16\pi^3 \beta^2  |\bfx| } \;  \mathrm{Im} \bigg[ \psi^{(1)}\left( 1 + \frac{i ( t - t' + |\bfx| + |\bfx'| ) }{\beta} \right) \bigg] \notag \\
&& \qquad- \frac{i \pi^2}{32 \pi^3 \beta^2 |\bfx|  \sinh^2 \left[ \frac{\pi }{\beta}(  t - t' - |\bfx| + |\bfx'| - i \delta ) \right]} + \frac{1}{16\pi^3 \beta^2  |\bfx| } \;  \mathrm{Im} \bigg[ \psi^{(1)}\left( 1 + \frac{i (  t - t' - |\bfx| + |\bfx'| )}{\beta} \right) \bigg] \,.\notag 
\eea
Using these in (\ref{MFcor2}) then gives the result (\ref{MFcor4}).

\subsubsection{One-Sided Fourier Transform of $p \coth(\beta p /2)$}
\label{App:FTcoth}

Here we derive the integral \pref{FTcoth}. To this end we use the identity $\coth\big( \frac{\beta p }{2} \big) = 1 + \frac{2}{e^{\beta p } - 1}$ to write the LHS of (\ref{FTcoth}) as
\bea
\int_0^\infty \exd p\; p \coth\left( \sfrac{\beta p }{2} \right) e^{ - i p (\tau - i \delta )  } & = & \int_0^\infty \exd p\; p e^{ - i p ( \tau - i \delta ) } +  2 \int_0^\infty \exd p\; \sfrac{ p \cos( \tau p ) }{ e^{\beta p} - 1 } + 2 i \int_0^\infty \exd p\; \sfrac{ p \sin ( \tau  p) }{ e^{\beta p} - 1 } \ . \qquad \quad \  
\eea
where the limit $\delta \to 0$ can be safely taken in the latter two integrals (since they are both convergent at $\tau = 0$). The first integral evaluates to 
\be
\int_{0}^{\infty} \exd p\; p  e^{- i p (\tau - i \delta) } \ = \ - \frac{1 }{( \tau - i \delta )^2 } \ ,
\ee
and the second integral is given in equation (3.951.5) of \cite{grad} (which converges for any $\mathrm{Re}[\beta] > 0$)
\bea
\int_0^\infty \exd p\; \frac{ p \cos( \tau  p ) }{ e^{\beta p} - 1 } \ = \ \frac{1}{2\tau^2} - \frac{\pi^2}{2\beta^2 \sinh^2\big( \frac{\pi \tau}{\beta} \big)} \ , 
\eea
and the third integral can be exactly evaluated as\footnote{Note the integral representation $\psi^{(1)}(z) = \int_0^{\infty} \exd q\; {q\; e^{ - z q} }/(1 - e^{-q})$ which follows from formula (5.9.12) of \cite{NIST}. This implies $\psi^{(1)}(1 + i y) = \int_0^{\infty} \exd q\; {q\; e^{ - i y q} }/(e^q - 1)$, and then taking the imaginary part of this gives \pref{sincothFT}.}
\bea
\int_0^\infty \exd p\; \frac{ p \sin( \tau  p ) }{ e^{\beta p} - 1 } \ = \  - \frac{1}{\beta^2} \mathrm{Im} \bigg[ \psi^{(1)}\left( 1 + \frac{i\tau}{\beta} \right) \bigg] \label{sincothFT}
\eea
where $\zeta(3) \simeq 1.202$. Putting the above altogether (in the limit $\delta \to 0$) gives formula (\ref{FTcoth}).

\subsection{Integrals appearing in the Equal-Time Correlator}
\label{App:eqtimeInts}

Here we simplify the integral
\be
\cM_{\beta}(t,\bfx,\bfx') := - \frac{\tilde g^2}{2} \int_0^{t} \exd \tau \int_0^{t} \exd \tau'\;  \mathscr{W}_\beta( \tau-\tau' ) \bra{ \vac } \left\{  \phi(\tau,\mathbf{0})  \phi( \tau', \mathbf{0}) , {\phi}(t,\bfx){\phi}(t,\bfx')  \right\}  \ket{\vac}
\ee
appearing in the equal-time mean-field correlator \pref{MFequal2}.

\subsubsection{Four-Point Wightman Functions}

First we note the functional form of the four-point Wightman functions appearing in the above, where 
\bea
&\ &\bra{\vac} \phi(t_1,\bfx_1) \phi(t_2,\bfx_2) \phi(t_3,\bfx_3) \phi(t_4,\bfx_4) \ket{\vac} \label{4ptW} \\
& \ & \qquad = \int \frac{\exd^3 \bfk}{(2\pi)^3 2 E_k } \int \frac{\exd^3 \bfp}{(2\pi)^3 2 E_p } \bigg[ e^{ - i E_{k} (t_1 - t_4) + i \bfk \cdot ( \bfx_1 - \bfx_4 ) } e^{ - i E_{p} (t_2 - t_3) + i \bfp \cdot ( \bfx_2 - \bfx_3 ) }  \nn \\
&& \qquad + e^{ - i E_{k} (t_1 - t_3) + i \bfk \cdot ( \bfx_1 - \bfx_3 ) } e^{ - i E_{p} (t_2 - t_4) + i \bfp \cdot ( \bfx_2 - \bfx_4 ) } + e^{ - i E_{k} (t_1 - t_2) + i \bfk \cdot ( \bfx_1 - \bfx_2 ) } e^{ - i E_{p} (t_3 - t_4) + i \bfp \cdot ( \bfx_3 - \bfx_4 ) } \bigg] \notag
\eea
where the commutation relations $[ \mfa_{\bfk}, \mfa_{\bfp} ] = [ \mfa^{\ast}_{\bfk}, \mfa^{\ast}_{\bfp} ] = 0$ and $[\mfa_{\bfk}, \mfa_{\bfp}] = \delta^3(\bfk - \bfp)$ have been used, as well as the expectation values
\bea
\bra{\vac} \hat{\mfa}_{\bfk} \hat{\mfa}_{\bfl} \hat{\mfa}_{\bfp}^{\ast} \hat{\mfa}_{\bfq}^{\ast} \ket{\vac} & = & \delta^3(\bfk - \bfq) \delta^3(\bfp - \bfl) + \delta^3(\bfk - \bfp) \delta^3(\bfq - \bfl) \\
\bra{\vac} \hat{\mfa}_{\bfk} \hat{\mfa}_{\bfl}^{\ast} \hat{\mfa}_{\bfp} \hat{\mfa}_{\bfq}^{\ast} \ket{\vac} & = & \delta^3(\bfk - \bfl) \delta^3(\bfp - \bfq) \nn
\eea
In terms of free (two-point) Wightman functions, the above has the simple form
\bea
\bra{\vac} \phi(x_1) \phi(x_2) \phi(x_3) \phi(x_4) \ket{\vac} & = & \bra{\vac} \phi(x_1) \phi(x_4) \ket{\vac} \bra{\vac} \phi(x_2) \phi(x_3) \ket{\vac} \\
& \ & \qquad + \bra{\vac} \phi(x_1) \phi(x_3) \ket{\vac} \bra{\vac} \phi(x_2) \phi(x_4) \ket{\vac} \nn \\
& \ & \qquad \qquad + \bra{\vac} \phi(x_1) \phi(x_2) \ket{\vac} \bra{\vac} \phi(x_3) \phi(x_4) \ket{\vac} \nn
\eea
using the shorthand $x_{j} = (t_j, \bfx_j)$.

\subsubsection{One-Sided Fourier Transform of $\mathscr{W}_{\beta}(t)$}
\label{App:WbetaFT}

Here we compute the integrals
\be
\int_0^\infty \exd \sigma\; \mathscr{W}_{\beta}(\sigma) e^{\pm i p \sigma} = \frac{1}{2} \bigg( \big[ \cC_{\beta} + i \mathcal{K}_{\beta} \big] \pm \big[ - \cS_{\beta} + i \cD_{\beta} \big] \bigg) \label{1sided}
\ee
where the function $\cC_{\beta}(p) = \frac{p}{4\pi} \coth\big( \frac{\beta p}{2} \big)$ is given in \pref{ReWcosFT} and the divergent constant $\mathcal{K}_{\beta} = -\delta(0) / (2\pi) $ is given in \pref{Kbetaintresult}, and we furthermore define
\bea \label{Ibetaint}
\cS_{\beta}(p) & := & 2 \int_0^\infty \exd \sigma \; \mathrm{Im}\left[ \mathscr{W}_{\beta}(\sigma) \right] \sin( p \sigma) \nn\\
\cD_{\beta}(p,\delta) & := & 2 \int_0^\infty \exd \sigma \; \mathrm{Re}\left[ \mathscr{W}_{\beta}(\sigma) \right] \sin( p \sigma) \nn
\eea
The functions $\cS_{\beta}$ and $\cD_{\beta}$ have also been computed in \cite{Kaplanek:2019dqu,Kaplanek:2020iay} (where $\beta$ is replaced by either the Unruh or Hawking temperatures). These functions take the form
\bea
\cS_{\beta}(p) & = & - \frac{p}{4\pi} \ , \\
\cD_{\beta}(p,\delta) & = & \frac{p}{2\pi^2} \log\left( \frac{2\pi e^{\gamma} \delta}{\beta} \right) + \frac{p}{2\pi^2} \mathrm{Re}\left[ \psi^{(0)}\left( - i \; \frac{\beta p}{2\pi} \right) \right] \ , \nn
\eea
where $\gamma$ is the Euler-Mascheroni constant and $\psi^{(0)}(z) = {\Gamma'(z)}/{\Gamma(z)}$ is the digamma function. Note that the function $\cD_{\beta}$ has a $\delta$-divergence, where $\delta >0$ is the regulator appearing the correlation function
\be
\mathscr{W}_{\beta}(\sigma) = - \frac{1}{4 \beta^2 \sinh^2\left( \frac{\pi ( \sigma - i \delta )}{\beta} \right)} \ .
\ee
Putting this all together in \pref{1sided} we find that 
\bea
\int_0^\infty \exd \sigma\; W_{\beta}(\sigma) e^{\pm i p \sigma} & = & \bigg[ \frac{p}{8 \pi} \coth\left( \frac{ \beta p }{2}  \right) - \frac{i}{4\pi^2 \delta} \bigg] \\
& \ & \quad \quad \quad \quad \pm \bigg[ \frac{p}{8\pi} + i \left( \frac{p}{4\pi^2} \log\left( \frac{2\pi e^{\gamma} \delta}{\beta} \right) + \frac{p}{4 \pi^2} \mathrm{Re}\left[ \psi^{(0)}\left( - i \; \frac{\beta p}{2\pi} \right) \right]  \right) \bigg] \ . \notag
\eea

\subsubsection{The Integral $\cM_{\beta}$}

It turns out that it is easiest to express the integral $\cM_{\beta}$ in the nested-integral form of \pref{imanti}, where
\bea
\cM_{\beta}(t,\bfx,\bfx') & = & - \frac{\tilde g^2}{2} \int_0^{t} \exd \tau \int_0^{\tau} \exd \tau'\;  \mathscr{W}_\beta( \tau-\tau' ) \bra{ \vac } \left\{ \phi(\tau,\mathbf{0}) \phi( \tau', \mathbf{0}) , {\phi}(t,\bfx){\phi}(t,\bfx') \right\} \ket{\vac} \\
& \ & \qquad - \frac{\tilde g^2}{2} \int_0^{t} \exd \tau \int_0^{\tau} \exd \tau'\;  \mathscr{W}^{\ast}_\beta( \tau-\tau' ) \bra{ \vac } \left\{ \phi(\tau',\mathbf{0}) \phi( \tau, \mathbf{0}) , {\phi}(t,\bfx){\phi}(t,\bfx') \right\} \ket{\vac} \ . \nn
\eea
By expanding the anti-commutators above and also using $[\phi(t,\bfx),\phi(t,\bfx')]=0$, the above can be manipulated into the form
\bea
\cM_{\beta}(t,\bfx,\bfx') & = & - \tilde g^2 \int_0^{t} \exd \tau \int_0^{\tau} \exd \sigma\; \mathrm{Re}\bigg[ \mathscr{W}_\beta( \sigma ) \bigg( \bra{ \vac } \phi(\tau,\mathbf{0}) \phi( \tau - \sigma, \mathbf{0}) {\phi}(t,\bfx){\phi}(t,\bfx') \ket{\vac} \\
& \ & \qquad \qquad \qquad \qquad \qquad \qquad \qquad \qquad \quad + \bra{ \vac } {\phi}(t,\bfx){\phi}(t,\bfx') \phi(\tau,\mathbf{0}) \phi( \tau - \sigma, \mathbf{0}) \ket{\vac} \bigg) \bigg] \nn
\eea
where the change of variable $\tau' = \tau  - \sigma$ has also been made. Using the formula \pref{4ptW} the four-point correlators can be written in momentum space as 
\bea
&& \bra{ \vac } \phi(\tau,\mathbf{0}) \phi( \tau - \sigma, \mathbf{0}) {\phi}(t,\bfx){\phi}(t,\bfx') \ket{\vac} + \bra{ \vac } {\phi}(t,\bfx){\phi}(t,\bfx') \phi(\tau,\mathbf{0}) \phi( \tau - \sigma, \mathbf{0}) \ket{\vac} \\
&& \qquad = \int \frac{\exd^3 \bfp}{(2\pi)^3 2 E_p } \int \frac{\exd^3 \bfk}{(2\pi)^3 2 E_k } \bigg( 2 e^{- i E_p \sigma} e^{+ i \bfk \cdot (\bfx - \bfx')} \nn \\
&& \qquad \qquad\qquad \qquad\qquad \qquad \qquad \quad + 2 \mathrm{Re}\left[ e^{- i E_p t + i \bfp \cdot \bfx} e^{- i E_k t + i \bfk \cdot \bfx'} e^{+ i (E_p + E_k) \tau} (e^{- i E_p \sigma} + e^{ - i E_{k} \sigma } )  \right] \bigg) \nn \ .
\eea
With this, the integral $\cM_{\beta}$ splits into two pieces
\bea
&& \cM_{\beta}(t,\bfx,\bfx') =  \cM^{(1)}_{\beta}(t,\bfx,\bfx') +  \cM^{(2)}_{\beta}(t,\bfx,\bfx')
\eea
where
\be
\cM^{(1)}_{\beta}(t,\bfx,\bfx') := - 2 \tilde g^2 \int \frac{\exd^3 \bfp}{(2\pi)^3 2 E_p } \int \frac{\exd^3 \bfk}{(2\pi)^3 2 E_k }  \int_0^{t} \exd \tau \int_0^{\tau} \exd \sigma\; \mathrm{Re}\bigg[ \mathscr{W}_\beta( \sigma ) e^{- i E_p \sigma} e^{+ i \bfk \cdot (\bfx - \bfx')} \bigg]
\ee
and
\bea
\cM^{(2)}_{\beta}(t,\bfx,\bfx') & := & - 2 \tilde g^2 \int \frac{\exd^3 \bfp}{(2\pi)^3 2 E_p } \int \frac{\exd^3 \bfk}{(2\pi)^3 2 E_k }  \int_0^{t} \exd \tau \int_0^{\tau} \exd \sigma\; \mathrm{Re}[ \mathscr{W}_\beta( \sigma ) ] \\
&\ & \qquad \qquad \qquad \qquad \times\mathrm{Re}\left[ e^{- i E_p t + i \bfp \cdot \bfx} e^{- i E_k t + i \bfk \cdot \bfx'} e^{+ i (E_p + E_k) \tau} (e^{- i E_p \sigma} + e^{ - i E_{k} \sigma } )  \right] \ . \nn
\eea
First we focus on simplifying $\cM^{(1)}$ above. The $\bfk$-integration is easily done, and then integrating the $\bfp$-angles away yields
\be
\cM^{(1)}_{\beta}(t,\bfx,\bfx') = - \frac{\tilde g^2}{8 \pi^4 |\bfx - \bfx'|^2} \int_0^\infty \exd p \; p \int_0^{t} \exd \tau \int_0^{\tau} \exd \sigma\; \mathrm{Re}\bigg[ \mathscr{W}_\beta( \sigma ) e^{- i p \sigma} \bigg] \ .
\ee
By switching the order of integration in the $(\tau,\sigma)$-plane the above integral can be written as 
\bea
\cM^{(1)}_{\beta}(t,\bfx,\bfx') &  = & - \frac{\tilde g^2}{8 \pi^4 |\bfx - \bfx'|^2} \int_0^\infty \exd p \; p \int_0^{t} \exd \sigma \int_{\sigma}^{t} \exd \tau\; \mathrm{Re}\bigg[ \mathscr{W}_\beta( \sigma ) e^{- i p \sigma} \bigg] \nn \\
& = & - \frac{\tilde g^2}{8 \pi^4 |\bfx - \bfx'|^2} \int_0^\infty \exd p \; p \int_0^{t} \exd \sigma\; (t - \sigma) \mathrm{Re}\bigg[ \mathscr{W}_\beta( \sigma ) e^{- i p \sigma} \bigg] \\
& = & - \frac{\tilde g^2}{8 \pi^4 |\bfx - \bfx'|^2} \int_0^\infty \exd p \; p  \; \bigg[ t \int_0^{t} \exd \sigma\; \bigg( \mathrm{Re}[ \mathscr{W}_\beta( \sigma ) ] \cos(p \sigma) + \mathrm{Im}[ \mathscr{W}_\beta( \sigma ) ] \sin(p \sigma) \bigg) \nn \\
& \ & \qquad \qquad \qquad \qquad \qquad \quad - \frac{\exd}{\exd p}\int_0^{t} \exd \sigma\; \bigg( \mathrm{Re}[ \mathscr{W}_\beta( \sigma ) ] \sin(p \sigma) - \mathrm{Im}[ \mathscr{W}_\beta( \sigma ) ] \sin(p \sigma) \bigg) \bigg] \qquad \nn
\eea
To simplify the integrals, we next assume that we probe times
\be
t \gg \beta\ ,
\ee
so that the upper limit on the $\sigma$-integrals can be taken to be $\simeq \infty$ (since $\mathscr{W}_{\beta}(\sigma) \propto e^{ - 2 \pi \sigma / \beta} $). Upon doing so the $\sigma$-integrals in the above may be expressed in terms of the functions given in \pref{1sided}, where
\be
\cM^{(1)}_{\beta}(t,\bfx,\bfx') \simeq - \frac{\tilde g^2}{16 \pi^4 |\bfx - \bfx'|^2} \int_0^\infty \exd p \; p  \; \bigg[ t \big( \cC_{\beta}(p) +  \cS_{\beta}(p) \big) - \frac{\exd \cD_{\beta}(p,\delta)}{\exd p} + \frac{\exd \mathcal{K}_{\beta}(p,\delta)}{\exd p} \bigg] \ . \qquad 
\ee
Note that $\partial_{p} \mathcal{K}_{\beta}(p,\delta) = 0$ from the functional form \pref{Kbetaintresult}. Furthermore, using the functional forms of $\cC_{\beta}$ and $\cS_{\beta}$ we note the value of the integral 
\be
\int_0^\infty \exd p \; p \; \big( \cC_{\beta}(p) +  \cS_{\beta}(p) \big) \ = \ \int_0^\infty \frac{\exd p\; p^2}{2\pi(e^{\beta p} - 1)} \ = \ \frac{\zeta(3)}{\pi \beta^3}
\ee
where $\zeta(3) \simeq 1.202$ (with $\zeta$ the Riemann-Zeta function). With this we get 
\be
\cM^{(1)}_{\beta}(t,\bfx,\bfx') \simeq - \frac{\tilde g^2 \left( \dfrac{\zeta(3) t}{\pi \beta^3} - {\displaystyle \int_0^\infty } \exd p \; p\; \dfrac{\exd \cD_{\beta}(p,\delta)}{\exd p} \right)}{16 \pi^4 |\bfx - \bfx'|^2}
\ee
where we note that the remaining momentum-integral appears to be ultraviolet divergent in the momentum $p$. 

\vspace{2mm}

Moving on to the second integral $\cM^{(2)}$, we first integrate away the angles in the momentum integrals and simplify to get 
\bea
\cM^{(2)}_{\beta}(t,\bfx,\bfx') & = & - \frac{\tilde g^2}{16\pi^4 |\bfx| |\bfx'|} \int_0^\infty \exd p \int_0^\infty \exd k \int_0^{t} \exd \tau \int_0^{\tau} \exd \sigma\; \sin(|\bfx| p) \sin(|\bfx'| k) \\
&\ & \qquad \qquad \qquad \qquad \times \mathrm{Re}\bigg[ e^{- i (p + k) t} e^{+ i (p + k) \tau}  ( e^{- i p \sigma} + e^{ - i k \sigma } )  \mathrm{Re}[ \mathscr{W}_\beta( \sigma ) ] \bigg] \ . \nn
\eea
We again can switch the order of integration in the $(\tau,\sigma)$-plane giving us 
\bea
\cM^{(2)}_{\beta}(t,\bfx,\bfx') & = & - \frac{\tilde g^2}{16\pi^4 |\bfx| |\bfx'|} \int_0^\infty \exd p \int_0^\infty \exd k \int_0^{t} \exd \sigma \int_{\sigma}^{t} \exd \tau\; \sin(|\bfx| p) \sin(|\bfx'| k) \\
&\ & \qquad \qquad \qquad \qquad \times \mathrm{Re}\bigg[ e^{- i (p + k) t} e^{+ i (p + k) \tau}  ( e^{- i p \sigma} + e^{ - i k \sigma } )  \mathrm{Re}[ \mathscr{W}_\beta( \sigma ) ] \bigg]  \nn \ . 
\eea
The $\tau$-integration can now be easily performed such that 
\bea
\cM^{(2)}_{\beta}(t,\bfx,\bfx') & =  & - \frac{\tilde g^2}{16\pi^4 |\bfx| |\bfx'|} \int_0^\infty \exd p\; \sin(|\bfx| p)  \int_0^\infty \exd k\; \sin(|\bfx'| k) \int_0^{t} \exd \sigma  \\
&\ & \qquad \times \mathrm{Re}\bigg[ \frac{i e^{- i (p + k) t}  }{k+p}  ( e^{+ i p \sigma} + e^{+ i k \sigma } ) \mathrm{Re}[ \mathscr{W}_\beta( \sigma ) ] -  \frac{i}{k+p} ( e^{- i p \sigma} + e^{ - i k \sigma } )  \mathrm{Re}[ \mathscr{W}_\beta( \sigma ) ] \bigg] \ . \nn
\eea
As noted above, we assume $t \gg \beta$ and so the $\sigma$-integrals can be expressed in terms of the functions $\cC_{\beta}$ and $\cD_{\beta}$ where
\bea
\cM^{(2)}_{\beta}(t,\bfx,\bfx') & \simeq & - \frac{\tilde g^2}{16\pi^4 |\bfx| |\bfx'|} \int_0^\infty \exd p\; \sin(|\bfx| p)  \int_0^\infty \exd k\; \sin(|\bfx'| k) \nn \\
&\ & \qquad \times \mathrm{Re}\bigg[ \frac{i e^{- i (p + k) t}  }{k+p}  \big( \cC_{\beta}(p) + i \cD_{\beta}(p,\delta)  + \cC_{\beta}(k) + i \cD_{\beta}(k,\delta)  \big) \nn \\
& \ & \qquad \qquad \qquad \qquad \qquad \qquad -  \frac{i}{k+p} \big( \cC_{\beta}(p) - i \cD_{\beta}(p,\delta)  + \cC_{\beta}(k) - i \cD_{\beta}(k,\delta)  \big)\bigg] \nn \\
& = &  - \frac{\tilde g^2}{16\pi^4 |\bfx| |\bfx'|} \int_0^\infty \exd p\; \sin(|\bfx| p)  \int_0^\infty \exd k\; \sin(|\bfx'| k) \\
&\ & \qquad \times \bigg( \frac{\cC_{\beta}(p) + \cC_{\beta}(k)}{p+k} \sin\big((p+k)t \big) - \frac{\cD_{\beta}(p,\delta) + \cD_{\beta}(k,\delta)}{p+k} \big[ 1 + \cos\big((p+k)t \big) \big]   \bigg) \ . \nn
\eea

\subsection{Quantities entering with the diffuse correlator}
\label{App:DiffuseInts}

This Appendix computes quantities that arise in \S\ref{ssec:IncDiff} where the diffuse contributions to the Wightman function are computed.

\subsubsection{The integral $\cP_{\beta}$}

First we compute the integral $\cP_{\beta}$ defined in \pref{Nbetadef}. Using the commutator \pref{commtaux} we easily find that \pref{Nbetadef} simplifies to
\bea
\cP_{\beta}(t,\bfx,\bfx') & = & - \tilde g^2 \int_0^{t} \exd \tau \int_0^{t} \exd \tau'\;  \mathscr{W}_\beta( \tau-\tau' ) \sfrac{i \delta\big(\tau  - [t - |\bfx|]\big)}{4\pi |\bfx|}\sfrac{i \delta\big(\tau'  - [t - |\bfx'|]\big)}{4\pi |\bfx'|} \bra{ \vac }  \cI^2_{+} \ket{\vac} \nn \\
& = & + \frac{\tilde g^2 }{16 \pi^2 |\bfx| |\bfx'| } \Theta(t - |\bfx|) \Theta(t - |\bfx'|) \mathscr{W}_\beta( - |\bfx| +|\bfx'| ) \nn \\
& = & - \left( \frac{\tilde g^2 }{64 \pi^2 \beta^2} \right) \frac{ \Theta(t - |\bfx|) \Theta(t - |\bfx'|)}{  |\bfx| |\bfx'| \sinh^2\left[ \frac{\pi}{\beta} ( - |\bfx| + |\bfx'| - i \delta ) \right] } 
\eea
This is the result quoted as \pref{Nbetaanswer}  in the main text, and it agrees precisely with the $\cO(\tilde g^2)$  part of the correlator given in  \pref{largeNcorrteqs0}.

\subsubsection{The integral $\cQ_{\beta}$}

Next we compute the integral $\cQ_{\beta}(t,\bfx,\bfy)$ defined in (\ref{Ebetadef}). Using the commutator \pref{commtaux} allows this integral to be rewritten as 
\bea
&& \cQ_{\beta}(t,\bfx,\bfx') = \frac{\tilde g^2}{2} \int_0^{t} \exd \tau \int_0^{t} \exd \tau'\;  \mathscr{W}_\beta( \tau-\tau' ) \bra{ \vac } \tfrac{i \delta[\tau - (t - |\bfx|)] \phi(\tau',\mathbf{0}) {\phi}(t,\bfx') - i \delta[\tau' - (t - |\bfx|)]\phi(\tau,\mathbf{0}) {\phi}(t,\bfx')  }{4 \pi |\bfx|} \ket{\vac} \nn \\
&& \quad  + \frac{\tilde g^2}{2} \int_0^{t} \exd \tau \int_0^{t} \exd \tau'\;  \mathscr{W}_\beta( \tau-\tau' ) \bra{ \vac } \tfrac{i \delta[\tau - (t - |\bfx'|)] {\phi}(t,\bfx) \phi(\tau',\mathbf{0})  - i \delta[\tau' - (t - |\bfx|) ] {\phi}(t,\bfx) \phi(\tau,\mathbf{0}) }{4 \pi |\bfx'|}  \ket{\vac} \ .
\eea
Relabelling $\tau$ and $\tau'$ in the first term of each of the lines above allows this to be rewritten as 
\bea
\cQ_{\beta}(t,\bfx,\bfx') & = & \frac{\tilde g^2}{4 \pi |\bfx|} \int_0^{t} \exd \tau \int_0^{t} \exd \tau'\;  \mathrm{Im}[\mathscr{W}_\beta( \tau-\tau' )] \delta\big[\tau' - (t - |\bfx|)\big] \bra{ \vac } \phi(\tau,\mathbf{0}) {\phi}(t,\bfx') \ket{\vac} \\
&& \qquad   + \frac{\tilde g^2}{4\pi |\bfx'|} \int_0^{t} \exd \tau \int_0^{t} \exd \tau'\;   \mathrm{Im}[\mathscr{W}_\beta( \tau-\tau' )] \delta\big[\tau' - (t - |\bfx'|) \big] \bra{ \vac } {\phi}(t,\bfx) \phi(\tau,\mathbf{0})  \ket{\vac} \notag \ .
\eea
Performing the integrations over $\tau'$ now gives
\bea
\cQ_{\beta}(t,\bfx,\bfx') & = & \frac{\tilde g^2 \Theta(t - |\bfx|) }{4 \pi |\bfx|} \int_0^{t} \exd \tau \; \mathrm{Im}\big[\mathscr{W}_\beta\big( \tau - [t - |\bfx|]  \big)\big]  \bra{ \vac } \phi(\tau,\mathbf{0}) {\phi}(t,\bfx') \ket{\vac} \\
&& \qquad   + \frac{\tilde g^2  \Theta(t - |\bfx'|) }{4\pi |\bfx'|} \int_0^{t} \exd \tau \; \mathrm{Im}[\mathscr{W}_\beta\big( \tau - [ t - |\bfx'| ] \big)] \bra{ \vac } {\phi}(t,\bfx) \phi(\tau,\mathbf{0})  \ket{\vac} \notag \ ,
\eea
and using the explicit form for the free Wightman functions as well as $\mathrm{Im}[\mathscr{W}_{\beta}(t)] = \delta'(t) / (4\pi)$ makes the integrand explicit:
\be
\cQ_{\beta}(t,\bfx,\bfx') = \frac{\tilde g^2 \Theta(t - |\bfx|) }{64 \pi^4 |\bfx|} \int_0^{t} \exd \tau \; \sfrac{\delta'\big( \tau - [t - |\bfx|]  \big)}{-(\tau - t - i \delta)^2 + |\bfx'|^2} + \frac{\tilde g^2  \Theta(t - |\bfx'|) }{64 \pi^4 |\bfx'|} \int_0^{t} \exd \tau \; \sfrac{ \delta'\big( \tau - [ t - |\bfx'| ] \big) }{ - (t - \tau - i \delta)^2 + |\bfx|^2 } \ .
\ee
Integrating this by parts yields
\bea
&& \cQ_{\beta}(t,\bfx,\bfx') = \frac{\tilde g^2 \Theta(t - |\bfx|) }{64 \pi^4 |\bfx|} \bigg[ \sfrac{\delta\big( |\bfx| \big)}{|\bfx'|^2} - \sfrac{\delta\big( t - |\bfx|  \big)}{-( |\bfx| + i \delta)^2 + |\bfx'|^2} - \int_0^{t} \exd \tau \; \delta\big( \tau - [t - |\bfx|]  \big) \frac{\exd}{\exd \tau} \bigg\{ \sfrac{1}{-(\tau - t - i \delta)^2 + |\bfx'|^2} \bigg\}  \bigg] \nn \\
&& \quad + \frac{\tilde g^2  \Theta(t - |\bfx'|) }{64 \pi^4 |\bfx'|} \bigg[\sfrac{ \delta(|\bfx'|) }{|\bfx|^2 } - \sfrac{ \delta(t - |\bfx'|) }{ - (|\bfx'| + i \delta)^2 + |\bfx|^2 } -  \int_0^{t} \exd \tau \; \delta\big( \tau - [ t - |\bfx'| ] \big) \frac{\exd}{\exd \tau} \bigg\{ \sfrac{ 1 }{ - (t - \tau - i \delta)^2 + |\bfx|^2 } \bigg\} \bigg]
\eea
where boundary terms with $\delta(|\bfx|)$ factors never contribute (since $|\bfx| >0$). In the regime of interest, $t - |\bfx| > 0$ and $t - |\bfx'| > 0$, the other boundary terms also do not contribute (since the Heaviside functions all turn on). 

All that is left is to perform the $\delta$-function integrations which gives the final result
\bea
\cQ_{\beta}(t,\bfx,\bfx') & = & \frac{\tilde g^2}{64 \pi^4 |\bfx|} \bigg[ \frac{1}{2 |\bfx'| \big( |\bfx| - |\bfx'| + i \delta \big)^2} - \frac{1}{2 |\bfx'| \big( |\bfx| + |\bfx'| + i \delta \big)^2} \bigg] \nn\\
&& \qquad\qquad  + \frac{\tilde g^2}{64 \pi^4 |\bfx'|} \bigg[ \frac{1}{2 |\bfx| \big( |\bfx| - |\bfx'| + i \delta \big)^2} - \frac{1}{2 |\bfx| \big( |\bfx| + |\bfx'| - i \delta \big)^2} \bigg] \\
&=&  \frac{\tilde g^2}{64 \pi^4 |\bfx| |\bfx'|} \bigg[ \frac{1}{( |\bfx| - |\bfx'| + i \delta )^2} - \frac{1}{( |\bfx| + |\bfx'| )^2} \bigg] \ ,\nn
\eea
where the last line safely takes $\delta \to 0$. This is the result quoted in (\ref{Ebetaanswer}).

\end{document}